\documentclass[aps,prf,preprint,groupedaddress]{revtex4-2}

\usepackage{amsmath,amssymb,graphicx,graphics,hyperref,xcolor}

\begin{document}

\title{On the Hinch-Kim dualism between singularity  and Faxén operators in the
hydromechanics of arbitrary bodies in Stokes flows}


\author{Giuseppe Procopio}
\author{Massimiliano Giona}
\email[]{massimiliano.giona@uniroma1.it}
\affiliation{Dipartimento di Ingegneria Chimica Materiali Ambiente, Sapienza Università di Roma, via Eudossiana 18, Rome 00184, Italy}



\begin{abstract}

We generalize  the  multipole
expansion and  the structure of the Faxén operator  in Stokes flows 
obtained for bodies with no-slip to generic boundary conditions,
 addressing the assumptions 
 under which this generalization is conceivable.
We show that a disturbance field generated by a body immersed in an ambient flow can be expressed as a multipole expansion 
the coefficients of which are  the moments of the volume forces, independently on the boundary conditions.
We find that the dualism between the operator giving the disturbance field of an $n$-th order ambient flow and the $n$-th order Faxén operator, referred to as the Hinch-Kim dualism,  holds only if  the boundary 
conditions  satisfy a property that we call Boundary-Condition reciprocity (BC-reciprocity).
If this property is fulfilled,
the Faxén operators can be expressed  in terms of the $(m,n)$-th order geometrical moments of the volume forces 
(defined in the article). In addition, it is shown that in these cases, the hydromechanics of the
fluid-body system  is completely determined by the entire
set of  the Faxén operators. Finally,  classical boundary  conditions
of hydrodynamic practice are investigated in the light of this property:
 boundary conditions for rigid bodies, Newtonian drops
at the mechanical equilibrium, porous bodies modeled by the Brinkman equations are  BC-reciprocal, while deforming linear elastic bodies, deforming Newtonian drops, non-Newtonian drops and porous bodies modeled by the Darcy equations do not have 
this property.
 For  Navier-slip boundary conditions on a rigid body, we find the analytical expression for low order Faxén operators.

\end{abstract}


\maketitle

\maketitle


\section{Introduction}
The detailed description of the low-Reynolds number hydromechanics
of particles immersed in a fluid  is  highly significant
in rheology and colloidal science \cite{guazzelli-morris,maxey_rev,mewis}, 
fundamental statistical physics \cite{bedeaux-mazur,bian,raizen,pg_fluids}, biological locomotion
and life science \cite{freund,lauga_book,vogel},  microfluidics \cite{undvall, venditti, dicarlo}, etc.

One of the most powerful mathematical tools to investigate  particle-fluid interactions is, when
available, the so called Faxén operator, which is the operator that once  applied to a generic ambient
flow (defined as the flow of the fluid without the disturbance due to the particle inclusion),  
 furnishes forces, torques, stresses and higher-order moments on the particle, without the need
 of solving the specific hydrodynamic problem. 
The introduction of this operator is originally due to Hilding Faxén (hence the name, see the book by Oseen \cite{oseen} or by Happel and Brenner \cite{hb}),
 who found that the force
acting on a sphere with no-slip boundary conditions 
immersed in a generic ambient flow can be expressed in a simple way 
 in terms of  the value of the ambient field and of its Laplacian at the center of the sphere.
Moreover,   the torque is proportional to the vorticity of the ambient field at the center of the sphere. 
The Faxén laws for the no-slip sphere are, essentially, an application 
to the solutions of the Stokes equations \cite{pozri}
of the mean value theorem 
for biharmonic (the velocity solution of the Stokes equations) and  harmonic (the vorticity) functions,
 yielding,
respectively,
 the 
$0$-th and the  asymmetric $1$-st  order  moments of surface traction on the surface of the sphere.

Thereafter, many authors have obtained Faxén operators for several combinations of surface moments, 
shapes of immersed bodies, boundary conditions, fluid regimes.
In the case of the stationary Stokes regime, literature results include  the analytic expressions
for the Faxén operators of lower orders, specifically:
the symmetric $1$-st order  operator
 for a sphere with no-slip boundary conditions \cite{batchelor-green},
 $0$-th and  $1$-st order  
for spheroids \cite{hasimoto,kim85} and, more generally for 
ellipsoids \cite{brenner64,kim86} with no-slip boundary conditions,  $0$-th  and { the Faxén operator for  the Stresslet contribution} for a spherical Newtonian drop \cite{hestroni-haber70,rallison},
 $0$-th,   asymmetric $1$-st  order operators and { the Faxén operator for  the Stresslet contribution}  for a sphere with Navier-slip boundary conditions \cite{
 premlata2021, premlata2022},
 $0$-th and    asymmetric $1$-st  order  operators for porous
spheres using the Darcy model \cite{palaniappan}, and 
 $0$-th and $1$-st order
using the Brinkman model \cite{felderhof78,pad}.
Faxén operators for spheres  has been obtained 
in other flow regimes:
 $0$-th and    asymmetric $1$-st order operators
for a sphere with no-slip boundary conditions
in unsteady Stokes flow \cite{mazur, maxey, yang} and
for  the linearized compressible Navier-Stokes flow
\cite{badeaux}, 
 $0$-th  order operator 
for a spherical Newtonian drop \cite{kaneda},
$0$-th and  asymmetric $1$-st  operators
for a sphere with Navier-slip boundary conditions
in unsteady Stokes flows \cite{felderhof,premlata2020},
 $0$-th and $1$-st order  operators
for 
porous
spheres using the Brinkman model \cite{jones79},
$0$-th order  Faxén counterpart  
for a sphere in  a potential flow \cite{beek}.

By definition, a Faxén operator  is independent of 
 the ambient flow and of the viscosity of the  fluids, and  depends solely  on the geometrical structure of the body and 
on the parameters specifying   the boundary conditions. An 
explicit expression of the Faxén operator for forces and torques has been  given by Brenner \cite{brenner64} for  bodies
with no-slip  conditions and arbitrary shape in terms of an infinite  series of differential operators with polidiadic coefficients.
The Brenner  coefficients  depend only on the geometry of the body and correspond to: i)
 the moments  of the surface traction
associated with the 
solution of the Stokes problem for the translating body in the unbounded fluid (in the case of the Faxén operator for the force),
 and, ii) the moments of the surface traction related the solution of the Stokes problem for the rotating body in the unbounded fluid (in the case of the Faxén operator for the torque).

Hinch \cite{hinch} observed  that the   operator  
applied to the pole of the unbounded Green function of the Stokes flow (usually referred to as the Stokeslet \cite{kim-karrila})
and  returning the disturbance field
generated by a no-slip sphere in a symmetric linear flow 
is exactly the $1$-order symmetric Faxén operator found by Batchelor and Green \cite{batchelor-green}, thus
intuitively  concluding  
 that this is not a  simple coincidence but the consequence of the Lorentz reciprocal theorem for the Stokes flows.
The dualism between the singularity
representation  of the flow generated by an arbitrary body immersed in  a fluid
and the Faxén operators of the body has been 
 proved in a conclusive way by Kim \cite{kim85} by means of the Lorentz reciprocal theorem.
More precisely,  in the case no-slip boundary conditions at the body surface are  assumed, 
the Faxén operator for the force of a body with arbitrary shape
coincides (up to a multiplicative constant $8 \pi \mu$, where $\mu$ is the viscosity of the fluid)
 with the operator that, applied to the pole of
the Stokeslet,
yields the velocity field of the fluid due to the 
translations of the body.
 The extension of the dualism between higher order Faxén operators (torques, Stresslet, etc.) and 
higher order singularity operators (giving the field for rotations, strains, etc.)
 is a straightforward consequence of  the Kim's proof. We refer to this correspondence  as
the {\em Hinch-Kim dualism}.
 
Hinch-Kim dualism implies several  important consequences of theoretical and practical
interest: i) by solving a single hydrodynamic 
problem (either analytically or numerically) it 
is possible to obtain a Faxén operator even for
particles with complex shape (at least  for the leading order terms),
ii)
the flow generated by an immersed particle can
be represented in a compact way by its Faxén operator, the leading-order terms of which
   can be evaluated using the  Brenner
polidiadic expansion even for particles with complex shape,
iii) long range particle-particle and particle-channel interactions can be investigated
taking advantage of this symmetry in order to obtain 
hydrodynamic 
properties of complex systems of particles
  \cite{ batchelor-green,batchelor-green_b,hinch,bossis-brady, mauri},  of
active microswimmers near walls \cite{spagnolie-lauga,kuron,dey}, of microfluidic  flow and separation devices
\cite{brenner-gaydos,swan}, either applying  theoretical approaches or by  means of numerical methods, 
such as  the Stokesian dynamics \cite{brady-bossis}.

The last two decades have seen a  growing interest in generalizing  the nature of the 
 boundary conditions,  
going beyond the  no-slip case, 
and in investigating the interactions with ambient flows more complex then 
purely constant and linear fields. This is mainly due
  to: i) the rise of microfluidics \cite{whitesides}, where  surfaces are chemically treated
and the  properties of the  resulting solid-liquid interfaces exploited, hinging for a more detailed hydrodynamic
description
 \cite{lauga_rev},
 ii) the development of the hydromechanics of biological particles \cite{lauga_book}, where the assumption
of  
 rigid translating and rotating particles equipped with no-slip boundary conditions is evidently too simplified and 
limiting,  and where it has been verified  that 
 the inclusion of only lower order moments, such as forces, torques and Stresslets, 
is not sufficient to explain many interesting
hydrodynamics behaviors 
of biological particles
 \cite{nasouri}.

The scope of this article is to generalize and extend
the results obtained for   the singularity and  Faxén operators 
 and their  mutual relationships
enforcing no-slip at the solid boundaries
to generic boundary conditions and to ambient flows of any order. This  extension
yields several novel results  related to: 
 i) the analytic expression for the singularity operator in terms of
volume moments, ii) the definition of an analytic criterion upon which the Hinch-Kim duality holds,
iii) the application of this criterion to a broad class of boundary conditions of hydrodynamic interest.

The main technical tool in  the present theory is the
 bitensorial
distributional analysis developed in Procopio and Giona\cite{pg_mine}, in which  the moments
with respect to the volume forces  acting on the body -  instead of   the moments associated with  the surface tractions
considered in the  literature - are introduced and  applied in order to express the singularity expansion of 
a disturbance flow.
These two hierarchies of moments  coincide 
in  the no-slip case.
 The advantage of this approach  is that it makes it  possible to obtain
a general expression for the singularity operator
of a disturbance flow in terms of an infinite series of differential operators with  the moments of the  
volume forces as coefficients,  independently of the boundary conditions assumed at the fluid-body interface.

The article is organized as follows. Section \ref{sec2} briefly reviews the bitensorial distributional
theory of hydrodynamic singularities introduced by Procopio and Giona \cite{pg_mine}.
In
Section \ref{sec3} 
we define the $(n,m)$-th order geometrical moments as the $m$-th order moments 
on the body immersed in an 
$n$-th order ambient field,
and  we show that the 
$n$-th order singularity operator of an arbitrary body can be expressed in  series of differential operators with the $(n,m)$-th
 order geometrical moments as coefficients. 
 { In Section \ref{sec4}, we introduce a parity
constraint, referred to as  \textit{boundary condition reciprocity}, for the boundary conditions assumed at the body-fluid interface.
In Section \ref{sec4.1}, we investigate
the Hinch-Kim dualism between $n$-th order singularity operators and $n$-th order Faxén operator for an arbitrary body. 
We show that  the dualism is not a general property deriving from  the Lorentz reciprocity  theorem,  as it applies
solely to a subclass 
of boundary conditions assumed at the surface of the body satisfying the boundary condition reciprocity introduced in Section \ref{sec4}.}
We show that, whenever this dualism
holds (hence reciprocal boundary conditions are considered),  the Brenner expression for  the Faxén operators
can be generalized  by considering  the moments of the volume forces. In addition, 
it is possible to generalize also the property found in the article by Procopio and Giona\cite{pg_mine}, namely that   the hydromechanics of  a
body in Stokes flows  is completely described by
the entire set of its Faxén operators (or
geometrical moments).
 This is a fundamental result for the development of a theory describing the hydrodynamic interaction between many bodies based on the knowledge of the hydrodynamics of each individual component\cite{pg_arxiv}, hence represented by their Faxén operators.

A similar investigation of
the Hinch-Kim dualism has been carried out by Dolata and Zia \cite{dolata-zia} following a  method,
 completely different from  the present approach, based on energetic considerations and  expressing
 the reciprocity between operators instead of  fields. Although their main result
(the conditions under which the dualism hold) can be  mapped into the present theory,  these authors 
have reached some misleading conclusions, such as 
the validity of the dualism 
 for porous particles modeled by the Darcy law. In this
article we show that this is not case. In fact, in the second part 
of this work
(from Section \ref{secIV-A} to Section \ref{sec7})  we analyze a broad class of typical  hydrodynamic
boundary conditions, determining, case by case, whether the dualism holds or not.
 In Section \ref{secIV-A}, we investigate the boundary conditions at the solid-fluid interface,
 finding that the dualism holds for rigid bodies
 with Navier-slip boundary conditions (even with a non uniform  slip length along the surface), but
 not for linear elastic bodies in deformation.
 In Section \ref{sec6}, we analyze the dualism for fluid-fluid boundary conditions finding that
 it  is verified solely  for Newtonian drops at the mechanical equilibrium. 
 In Section \ref{sec7}, we consider the case of   porous bodies, finding that the dualism 
 applies in  the Brinkman model for porous media, but  not for the Darcy model. 
  Finally,
we use  the analytical approach developed in the previous Sections to
{ derive a systematic method for obtaining higher-order Faxén operators for a sphere.
By means of this method, we provide} a closed-form expression for  the $0$-th (already available in literature \cite{premlata2021}), $1$-st and $2$-nd 
(to the best of our knowledge, not yet present in  the literature) order Faxén operators for a sphere 
with Navier-slip boundary conditions,  obtaining, hence, anaclitic expressions for the associated high order flows around the sphere.

\section{Formulation of the problem}
\label{sec2}
Consider a body immersed in a unbounded Stokes fluid. The domain of the body is $D_b \subset \mathbb{R}^3$ with boundaries $\partial D_b$
and the domain of the fluid is $D_f \equiv \mathbb{R}^3/D_b$ with boundaries $\partial D_f \equiv \partial D_b \cup \partial D_\infty $, where $ \partial D_\infty $ is an ideal surface at infinity.
The \textit{ambient flow} of the fluid (i.e. the flow 
of the fluid without the body inclusion)
 is ${\pmb u}({\pmb x})$ with associated pressure
 ${ p}({\pmb x})$ and stress tensor $ {\pmb \pi}({\pmb x}) $,
solution of the Stokes equations
\begin{equation}
\begin{cases}
- \nabla \cdot {\pmb \pi} ({\pmb x}) = \mu \Delta {\pmb u} ({\pmb x})- \nabla p({\pmb x}) = 0 \\
\nabla \cdot  {\pmb u} ({\pmb x}) = 0 \qquad {\pmb x} \in \mathbb{R}^3
\end{cases}
\label{eq1}
\end{equation}
The presence of the body  generates a \textit{disturbance flow} at the boundaries  $\partial D_b$ of the body, 
that we indicate as $ {\pmb w}^S ({\pmb x}) $,
and thus, a disturbance flow ${\pmb w} ({\pmb x})$ in the   whole domain of the fluid
with associated pressure $q({\pmb x})$ and stress tensor ${\pmb \tau }({\pmb x})$
that are solution of the Stokes equations
\begin{equation}
\begin{cases}
- \nabla \cdot {\pmb \tau} ({\pmb x}) = \mu \Delta {\pmb w} ({\pmb x})- \nabla q({\pmb x}) = 0 
\\
\nabla \cdot  {\pmb w} ({\pmb x}) = 0 \qquad {\pmb x} \in D_f
\\
{\pmb w} ({\pmb x})=  {\pmb w}^S ({\pmb x}), \ {\pmb \tau } ({\pmb x})=  {\pmb \tau}^S ({\pmb x}) \qquad {\pmb x} \in \partial D_b
\end{cases}
\label{eq2}
\end{equation}
where ${\pmb \tau}^S ({\pmb x}) $ is the stress tensor of the disturbance flow at the surface of the body.
{ 
The only assumption on ${\pmb w}^S ({\pmb x})$ is that 
 \begin{equation}
 \int_{\partial D_b} {\pmb w}^{(s)}({\pmb x}) \cdot {\pmb n}({\pmb x}) dS = 0
 \label{eq2a}
 \end{equation}
 which is a necessary condition to have a single layer potential expression for ${\pmb w}({\pmb x})$ \cite{lady} and, hence, a multipole expansion.}
The total field
 $({\pmb v}({\pmb x}),{\pmb \sigma}({\pmb x}))=
 ({\pmb u}({\pmb x}),{\pmb \pi}({\pmb x}))+
 ({\pmb w}({\pmb x}),{\pmb \tau}({\pmb x}))$
is  the solution of the Stokes equations
\begin{equation}
\begin{cases}
- \nabla \cdot {\pmb \sigma} ({\pmb x}) = \mu \Delta {\pmb v} ({\pmb x})- \nabla s({\pmb x}) = 0 
\\
\nabla \cdot  {\pmb v} ({\pmb x}) = 0 \qquad {\pmb x} \in D_f
\\
{\pmb v} ({\pmb x})= {\pmb v}^S ({\pmb x}),\ {\pmb \sigma} ({\pmb x})={\pmb \sigma}^S ({\pmb x}) \qquad {\pmb x} \in \partial D_b
\\
{\pmb v} ({\pmb x}) =  {\pmb u} ({\pmb x}) \qquad {\pmb x} \rightarrow \infty
\end{cases}
\label{eq3}
\end{equation}
$s({\pmb x})=p({\pmb x})+q({\pmb x})$ being the total pressure field, $ {\pmb v}^S ({\pmb x})= {\pmb w}^S ({\pmb x})+{\pmb u} ({\pmb x})$ and $ {\pmb \sigma}^S ({\pmb x})={\pmb \tau}^S ({\pmb x})+{\pmb \pi} ({\pmb x})$ the total velocity field and stress tensor, respectively, at the surface of the body.

Making use of the  Ladyzhenskaya volume potential \cite{lady}, 
it is possible to express
the disturbance flow as  the solution  of the non-homogeneous Stokes
equations defined in the whole domain $\mathbb{R}^3$
\begin{equation}
\begin{cases}
- \nabla \cdot {\pmb \tau} ({\pmb x}) = \mu \Delta {\pmb w} ({\pmb x})- \nabla q({\pmb x}) = -{\pmb \psi} ({\pmb x})
\\
\nabla \cdot  {\pmb w} ({\pmb x}) = 0 \qquad
{ {\pmb x} \in \mathbb{R}^3}
\end{cases}
\label{eq4}
\end{equation}
 ${\pmb \psi} ({\pmb x})$ being  any force field distribution, with compact support in 
{the domain of the body } $D_b$, satisfying the
relation
\begin{equation}
\dfrac{1}{8 \pi \mu}
\int_{D_b} \psi_\alpha ({\pmb \xi}) S_{a\, \alpha}({\pmb x},{\pmb \xi}) dV({\pmb \xi})
=
{ w}_a^S({\pmb x})
 \qquad {\pmb x} \in \partial D_b
\label{eq5}
\end{equation}
where $dV({\pmb \xi})$ is the volume element at ${\pmb \xi}$
and $S_{a\, \alpha}({\pmb x},{\pmb \xi})$
 are the entries of the \textit{Oseen bitensor} or \textit{Stokeslet}.
 { Since ${\pmb \psi}({\pmb \xi})$ admits compact support localized on 
$D_b$, the integration  in eq. (\ref{eq5})  can be performed equivalently either
on $D_b$ or on $\mathbb{R}^3$}. The Stokeslet $S_{a\, \alpha}({\pmb x},{\pmb \xi})$ is the Green function of the Stokes equations in the unbounded domain, thus it is solution of the equations
  \cite{pg_mine}
\begin{equation}
\begin{cases}
-\nabla_b \Sigma_{a b \alpha}({\pmb x},{\pmb \xi})
=\Delta  S_{a\, \alpha}({\pmb x},{\pmb \xi})-\nabla_a P_\alpha ({\pmb x},{\pmb \xi})=-8 \pi \delta_{a \alpha}\delta({\pmb x}-{\pmb \xi})
\\
\nabla_a S_{a\, \alpha}({\pmb x},{\pmb \xi})=0
\\
\end{cases}
\label{eq6}
\end{equation}
{ with} $\Sigma_{a b\, \alpha}({\pmb x},{\pmb \xi})$ and 
$P_\alpha ({\pmb x},{\pmb \xi})$  the associated stress tensor and pressure.
{
The explicit solution of eqs. (\ref{eq6}) is \cite{pozri,kim-karrila,lady,pg_meccanica}
\begin{eqnarray}
\nonumber
S_{a \alpha}({\pmb x},{\pmb \xi})&=&
\dfrac{\delta_{a \alpha}}{r}+\dfrac{({\pmb x}-{\pmb \xi})_{a }({\pmb x}-{\pmb \xi})_{\alpha}}{r^3}
\\
[0.5cm]
P_{\alpha}({\pmb x},{\pmb \xi})
&=&
2
\dfrac{({\pmb x}-{\pmb \xi})_{\alpha}}{r^3}
\label{eq6.1}
\\
[0.5cm]
\nonumber
\Sigma_{a b \alpha}({\pmb x},{\pmb \xi})
&=&
6
\dfrac{({\pmb x}-{\pmb \xi})_{a}({\pmb x}-{\pmb \xi})_{b }({\pmb x}-{\pmb \xi})_{\alpha}}{r^5}
\end{eqnarray}
}
{ In eqs. (\ref{eq5})-(\ref{eq6.1})} and 
throughout this article
the Einstein summation convention for  repeated indexes is adopted, as well as 
 the distinction between indexes according to the bitensorial formalism
proposed by Procopio and Giona\cite{pg_mine}. Therefore,
 Greek letters ($\alpha, \beta, ... = 1,2,3$) are used for indexes referred to  field entries 
 at the source point ${\pmb \xi }$, while  Latin letters ($a,b, ... = 1, 2, 3$)  apply for indexes referring to  field 
entries  at the field point ${\pmb x}$. Consistently,
 the operator $\nabla_a$ represents the entries
 of the gradient  with respect to the field-point  coordinate ${\pmb x}$, 
while the operator $\nabla_\alpha$ indicates the entries
 of the gradient  with respect to the source point ${\pmb \xi}$.  
Since we are  analyzing  
bodies 
immersed in a unbounded fluid and since, for  the sake of 
simplicity,
we  consider  the poles of singularities located at a single point, it is always possible to 
express both source and field points in the same Cartesian coordinate system,  and thus the parallel propagator 
\cite{pg_mine,poisson} is  simply $g_{a \alpha}({\pmb x},{\pmb \xi})=\delta_{a \alpha}$.
However, behind the formal correctness in distinguishing entries at different points,
the use of the  bitensorial convention  provides, apart from a higher  notational
clearness,  some practical advantages. Specifically, i) it provides a   direct extension
of the results obtained in simple systems to more complex geometries;
 {  ii) in many cases it is useful to consider poles lying on a manifold 
belonging to the domain of the body, such as for ellipsoids \cite{kim-karrila} in  order to avoid infinite series in the singularity expansion.
In these situations  bitensorial formalism is necessary in order  to 
distinguish the coordinate system describing the pole manifold from 
the coordinates associated with  the  field point;
iii) the points where the operators are applied are naturally specified; iv) different properties and symmetries
between the entries are  clearly highlighted.}

Let us define the $n$-th order moments
\begin{equation}
M_{\alpha\, {\pmb \alpha}_n}({\pmb \xi})=\int_{{ D_b}} 
({\pmb x}-{\pmb \xi})_{{\pmb \alpha}_n}
\psi_{\alpha}({\pmb x}) 
 dV({\pmb x})
\label{eq7}
\end{equation}
where, formally, $ \psi_{\alpha}({\pmb x}) =\delta_{ \alpha a} \psi_{a}({\pmb x}) $ are the entries of the force field distribution at the point ${\pmb x}$ expressed in the coordinate system of the point ${\pmb \xi}$ and where $dV({\pmb x})$ is the volume element at the point ${\pmb x}$.
As shown in the article by Procopio and Giona \cite{pg_mine},
{the classical multipole expansion of a disturbance field for a body with no-slip boundary conditions (expressed in terms of surface moments \cite{kim-karrila}), can be reformulated in terms of volume moments defined by eq. (\ref{eq7}).
Eq. (4.3) in the article by Procopio and Giona\cite{pg_mine} has been  referred to
a body with no-slip boundary conditions. In point of fact,
it is sufficient to exchange $-{\pmb u}({\pmb \xi}) \rightarrow {\pmb w}^s({\pmb \xi})$
 to see  that the distributional analysis developed by Procopio and Giona \cite{pg_mine} 
applies  for generic boundary conditions on the surface of the immersed
body.
 Hence,
 given the set of volume moments $M_{\alpha\, {\pmb \alpha}_n}({\pmb \xi}) $ for a generic body, the disturbance field can be expressed as
\begin{equation}
w_a({\pmb x})=\dfrac{1}{8\pi \mu}\sum_{n=0}^{\infty} 
\dfrac{M_{\alpha\, {\pmb \alpha}_n}({\pmb \xi})}{n!} 
\nabla_{{\pmb \alpha}_n}
S_{a\, \alpha}({\pmb x},{\pmb \xi})
\label{eq8}
\end{equation}
and similarly for the pressure
\begin{equation}
q({\pmb x})=\dfrac{1}{8\pi}\sum_{n=0}^{\infty} 
\dfrac{M_{\alpha\, {\pmb \alpha}_n}({\pmb \xi})}{n!} 
\nabla_{{\pmb \alpha}_n}
P_{ \alpha}({\pmb x},{\pmb \xi})
\label{eq9}
\end{equation}
and  for the stress field
\begin{equation}
\tau_{a b}({\pmb x})=\dfrac{1}{8\pi}\sum_{n=0}^{\infty} 
\dfrac{M_{\alpha\, {\pmb \alpha}_n}({\pmb \xi})}{n!} 
\nabla_{{\pmb \alpha}_n}
\Sigma_{a\, b\, \alpha}({\pmb x},{\pmb \xi})
\label{eq10}
\end{equation}
where ${\pmb \alpha}_n=\alpha_1 ... \alpha_n$ is a multi-index,
$({\pmb x} - {\pmb \xi})_{{\pmb \alpha}_n}=({\pmb x} - {\pmb \xi})_{\alpha_1} ... ({\pmb x} - {\pmb \xi})_{\alpha_n}$
and $ \nabla_{{\pmb \alpha}_n}= \nabla_{\alpha_1} ... \nabla_{ \alpha_n}$.

This is a key point for the general analysis developed below. In fact, while classical  expressions for the coefficients of the multipole expansion in terms of surface integrals are referred
to specific boundary conditions
(for example, the coefficients of multipole expansions reported by many authors \cite{durl,ichiki} are referred to a rigid body with no-slip boundary conditions, while the coefficients of multipole expansion reported in the book by Kim and Karrila \cite[pp. 48-49]{kim-karrila} are referred to rigid particles and drops),
the volume moments in eq. (\ref{eq7}) giving the multipole
expansion in eqs. (\ref{eq8})-(\ref{eq10}) are valid  regardless of the boundary conditions assumed on the surface of the body,
once the integral condition eq. (\ref{eq2a}),
 deriving by eq. (\ref{eq5}) and by the incompressibility of the Stokeslet,
is satisfied.
 As  a consequence, eqs. (\ref{eq8})-(\ref{eq10}) allow us to express disturbance fields around a generic body 
by  means of a differential operator applied at the pole point of the Stokeslet. 
In the the article by Procopio and Giona\cite[Appendix B]{pg_mine}, it is shown that the moments defined by eq. (\ref{eq7}) reduce to the surface moments
defined by several authors \cite{durl,ichiki,kim-karrila} in the case  no-slip boundary conditions are imposed at the surface of the body $\partial D_b$. 
The evaluation of the volume moments directly from their definition eq. (\ref{eq7}) is not straightforward  as the distribution ${\pmb \psi}({\pmb x})$ is in principle not unique.
This issue can be overcome  since these  moments can be expressed 
as surface integrals of the associated total Stokes flows as addressed in the next Section. }

\section{Generalized geometrical moment expansion}
\label{sec3}

Consider a $n$-th order unbounded polynomial ambient Stokes flow, 
centered at a point ${\pmb \xi} \in D_b$ (see the schematic representation in Fig. \ref{fig_gm})
\begin{equation}
u^{(n)}_a({\pmb x},{\pmb \xi})= A_{{a} {\pmb a}_n} ({\pmb x}- {\pmb \xi} )_{{\pmb a}_n}
\label{eq11}
\end{equation}
where  hereafter the superscript $(n)$ indicates  any quantity referred to a $n$-th order ambient flow and { 
$A_{a\, {\pmb a}_n}$ is an intensity constant tensor
with $N_E=3^{n+1}$ entries.

\begin{figure}[h!]
\centering
\includegraphics[scale=0.45]{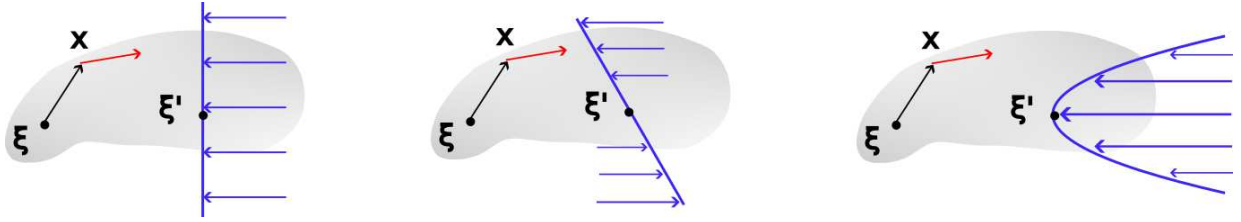}
\caption{
Schematic representation of a body immersed in an
$0$-th, $1$-st and $2$-nd order polynomial ambient flow centered at the point ${\pmb \xi}'$ (blue arrows) as  defined by  eqs. (\ref{eq11})-(\ref{eq11.3}).
Black arrows represent the position vector with respect the point ${\pmb \xi}$ of the force field ${\pmb \psi}({\pmb x})$ (red arrows) by which geometrical moments, defined in eq. (\ref{eq19}), are evaluated.}
\label{fig_gm}
\end{figure}

To ensure that the ambient flow in eq. (\ref{eq11}) satisfies the Stokes eqs. (\ref{eq1}),
the entries of $A_{a {\pmb a}_n}$  should
fulfill the incompressibility constraint 
\begin{equation}
A_{a {\pmb a}_n}\nabla_a ({\pmb x}- {\pmb \xi} )_{{\pmb a}_n}=0
\label{eq11.2}
\end{equation}
and, for $n \geq 3$, also the harmonicity constraint for the vorticity 
\begin{equation}
\varepsilon_{a b c}\, A_{b\, {\pmb a}_n} \Delta
\nabla_c 
({\pmb x}-{\pmb \xi})_{{\pmb a}_n}=0
\label{eq11.3}
\end{equation}
  $\varepsilon_{a b c}$ being the Ricci-Levi Civita symbol.
It can be shown by means of elementary combinatorial analysis \cite{rosen},
that eqs. (\ref{eq11.2}) 
and (\ref{eq11.3})   constitute a system  of 
$$
N_C=
\begin{cases}
\dfrac{(n+2)!}{n!\, 2} \qquad \qquad  \qquad \quad  \; n \leq  2
\vspace{0.5cm}
\\
\dfrac{(n+2)!}{n!\, 2}+\dfrac{n!\, 3}{(n-2)!\, 2}\qquad n\geq 3
\end{cases}
$$ constraints 
for  the $N_E=3^{n+1}$ entries of $A_{a {\pmb a}_n}$. Therefore, 
since $N_E \geq N_C\quad (\forall n \in \mathbb{N})$,
the ambient flow defined by eqs. (\ref{eq11})-(\ref{eq11.3}) exists for all $n \in \mathbb{N}$. 
Specifically, the flows defined by eqs. (\ref{eq11})-(\ref{eq11.3}), 
represent a  more general class of  the {\em external singularities} (singular at infinity) 
 defined by
Chwang and Wu  \cite{chwang-wu}. For example, the fundamental external singularity defined in their work \cite{chwang-wu}, referred to as
the {\em Stokeson}, can be obtained by 
 choosing $A_{a a_1 a_2}=f_\alpha (\delta_{a \alpha}\delta_{a_1 a_2}- \delta_{a_1 \alpha}\delta_{a a_2})$, $f_\alpha$ being the  arbitrary intensities.

The constraint  eq. (\ref{eq11.3}) ensures, by the Schwarz's theorem,
that there exists a scalar field $p^{(n)}({\pmb x})$ (the $n$-th order ambient pressure) such that
\begin{equation}
\nabla_a p^{(n)}({\pmb x},{\pmb \xi}) =\mu A_{a\, {\pmb a}_n} \Delta
({\pmb x}-{\pmb \xi})_{{\pmb a}_n}
\label{eq11.4}
\end{equation}
from which, letting ${\pmb x}_0$  be an arbitrary point at which the the pressure vanishes, we have
\begin{equation}
p^{(n)}({\pmb x},{\pmb \xi}) =\mu A_{a\, {\pmb a}_n}
 \int_{\Gamma( {\pmb x}_0,{\pmb x})}\Delta ({\pmb y}-{\pmb \xi})_{{\pmb a}_n} d{\pmb y}_a
\label{eq11.5}
\end{equation}
where $ \Gamma( {\pmb x}_0,{\pmb x}) $ is any Lipshitz curve, the endpoint of which are  ${\pmb x}_0$ and ${\pmb x}$.

By defining
\begin{equation}
p_{a {\pmb a}_n}({\pmb x},{\pmb \xi}) =
 \int_{\Gamma( {\pmb x}_0,{\pmb x})}\Delta ({\pmb y}-{\pmb \xi})_{{\pmb a}_n} d{\pmb y}_a
\label{eq11.6}
\end{equation} 
 the gradient of which is obviously
\begin{equation}
\nabla_b\, p_{a {\pmb a}_n}({\pmb x},{\pmb \xi}) =\delta_{a b} \Delta
({\pmb x}-{\pmb \xi})_{{\pmb a}_n}
\label{eq11.6.1}
\end{equation}
and 
\begin{equation}
\pi_{b c a {\pmb a}_n}({\pmb x},{\pmb \xi})
 = 
\delta_{b c}p_{a {\pmb a}_n}({\pmb x},{\pmb \xi})-
\left(
\delta_{a c}\nabla_b ({\pmb x} - {\pmb \xi})_{{\pmb a}_n}
+
\delta_{a b}\nabla_c ({\pmb x} - {\pmb \xi})_{{\pmb a}_n}
\right)
\label{eq11.7}
\end{equation}
it is possible to exploit the linearity of the Stokes problem in order
 to express
the pressure $p^{(n)}({\pmb x},{\pmb \xi})$ and the stress tensor $  \pi^{(n)}_{b c}({\pmb x},{\pmb \xi})$ of the $n$-th order ambient flows as
\begin{eqnarray}
& & p^{(n)}({\pmb x},{\pmb \xi}) = \mu A_{a {\pmb a}_n} \, p_{a {\pmb a}_n}({\pmb x},{\pmb \xi}),
\qquad
 \pi^{(n)}_{b c}({\pmb x},{\pmb \xi}) = \mu A_{a {\pmb a}_n} 
\pi_{b c a {\pmb a}_n}({\pmb x},{\pmb \xi}) 
\label{eq12}
\end{eqnarray}
where the dependence on the viscosity $\mu$, the intensity $A_{a {\pmb a}_n}$ and the geometric fields $p_{a {\pmb a}_n}({\pmb x},{\pmb \xi})$ and $\pi_{b c a {\pmb a}_n}({\pmb x},{\pmb \xi})$ has been factorized. 
By using this factorization it is possible to express the volume moments in term of surface integral moments.

To this aim,}  consider a generic total velocity field for a body immersed in an ambient field ${\pmb v}({\pmb x})={\pmb u}({\pmb x})+{\pmb w}({\pmb x})$ in the non-homogeneous form defined in  $\mathbb{R}^3$
\begin{equation}
\begin{cases}
- \nabla \cdot {\pmb \sigma} ({\pmb x}) = \mu \Delta {\pmb v} ({\pmb x})- \nabla s({\pmb x}) = - {\pmb \psi}({\pmb x}) 
\\
\nabla \cdot  {\pmb v} ({\pmb x}) = 0 \qquad {\pmb x} \in \mathbb{R}^3
\\
{\pmb v} ({\pmb x}) =  {\pmb u} ({\pmb x}) \qquad {\pmb x} \rightarrow \infty
\end{cases}
\label{eq14}
\end{equation}
The application of  the Lorentz reciprocal theorem in the differential form \cite{pozri}
to the fields $({\pmb u}^{(n)}({\pmb x},{\pmb \xi}),{\pmb \pi}^{(n)}({\pmb x},{\pmb \xi}))$ and $({\pmb v}({\pmb x}),{\pmb \sigma}({\pmb x}))$
provides
\begin{equation}
u^{(n)}_a({\pmb x},{\pmb \xi}) \nabla_b \sigma_{a b}({\pmb x})- v_a({\pmb x}) \nabla_b \pi^{(n)}_{a b}({\pmb x},{\pmb \xi})=
\nabla_b 
\left(
u^{(n)}_a({\pmb x},{\pmb \xi}) \sigma_{a b}({\pmb x})- v_a({\pmb x}) \pi^{(n)}_{a b}({\pmb x},{\pmb \xi})
\right)
\label{eq15}
\end{equation}
From eq. (\ref{eq15}), considering that $\nabla_b \pi^{(n)}_{b c}({\pmb x},{\pmb \xi}) =0$ and  making
use of (\ref{eq11})-(\ref{eq14}), it follows that
\begin{equation}
\psi_a({\pmb x}) ({\pmb x}-{\pmb \xi})_{{\pmb a}_n}=
\nabla_b 
\left( \sigma_{a b}({\pmb x}) ({\pmb x}-{\pmb \xi})_{{\pmb a}_n}
-\mu v_c({\pmb x})\, \pi_{b c a {\pmb a}_n}({\pmb x},{\pmb \xi})
\right)
\label{eq16}
\end{equation}
Integrating  the latter equation over the volume of the body, using
 the Gauss theorem for the r.h.s of the resulting equation, and enforcing 
the definition eq. (\ref{eq7}),  the 
moments on volume forces  $M_{\alpha {\pmb \alpha}_n}({\pmb \xi})$ can
be expressed as the  surface integrals
\begin{equation}
M_{\alpha {\pmb \alpha}_n}({\pmb \xi})= \int_{\partial D_b} 
\left(
 \sigma_{\alpha b}({\pmb x}) ({\pmb x}-{\pmb \xi})_{{\pmb \alpha}_n}
-\mu v_c({\pmb x})\, \pi_{b c \alpha {\pmb \alpha}_n}({\pmb x},{\pmb \xi})
\right)
n_b({\pmb x}) dS({\pmb x})
\label{eq17}
\end{equation}
where $n_b({\pmb x}) $ are the
entries of the outwardly oriented normal unit vector at point ${\pmb x}$ of $\partial D_b$ and $\sigma_{\alpha b}({\pmb x})= \delta_{\alpha a}\sigma_{a b}({\pmb x})$.
Therefore, by using the expression eq. (\ref{eq12}) for  $\pi_{b c a {\pmb a}_n}({\pmb x},{\pmb \xi}) $, the functional relation connecting the moments  to the values
of  ${\pmb v}({\pmb x})$ and ${\pmb \sigma}({\pmb x})$ assigned at the boundary of the body follows
\begin{eqnarray}
\nonumber
M_{\alpha {\pmb \alpha}_n}({\pmb \xi}) & = & \int_{\partial D_b} 
  ({\pmb x}-{\pmb \xi})_{{\pmb \alpha}_n} \sigma_{\alpha b}({\pmb x})n_b({\pmb x}) dS({\pmb x})
  -\mu  \int_{\partial D_b} p_{\alpha {\pmb \alpha}_n}({\pmb x},{\pmb \xi})
 v_b({\pmb x}) n_b({\pmb x}) dS({\pmb x})
  \\
   &+&
\mu  \int_{\partial D_b} 
 \, 
 \left[
v_\alpha({\pmb x}) n_b({\pmb x})\nabla_b ({\pmb x} - {\pmb \xi})_{{\pmb \alpha}_n}
+
 n_\alpha({\pmb x}) v_c({\pmb x}) 
\nabla_c ({\pmb x} - {\pmb \xi})_{{\pmb \alpha}_n}
\right]
 dS({\pmb x})
\label{eq18}
\end{eqnarray}


We can   introduce  the geometrical moments $m_{\alpha {\pmb \alpha}_m \beta' {\pmb \beta}'_n}({\pmb \xi},{\pmb \xi}')$ as defined  in the article by Procopio and Giona\cite{pg_mine} by the relation
\begin{equation}
 M^{(n)}_{\alpha {\pmb \alpha}_m}({\pmb \xi},{\pmb \xi}')=8 \pi \mu A_{\beta' {\pmb \beta}_n'} m_{\alpha {\pmb \alpha}_m \beta' {\pmb \beta}'_n}({\pmb \xi},{\pmb \xi}')
\label{eq19}
\end{equation}
where $ M^{(n)}_{\alpha {\pmb \alpha}_m}({\pmb \xi},{\pmb \xi}') $ is the $m$-th
order moments on the body  with respect to the point ${\pmb \xi}$ immersed in the $n$-th order ambient 
flow centered at the point ${\pmb \xi}'$ and the index $\beta' {\pmb \beta}'_n$ refers to the 
entries at the point ${\pmb \xi}'$.
{ 
In eq. (\ref{eq19}) we explicit the functional
 dependence of the moments on the pole  ${\pmb \xi}'$ of the ambient flow and 
on its order $n$.
 }

A schematic representation of these special hydrodynamic systems is reported in Fig. \ref{fig_gm}, where a generic body is immersed in $0$-th, $1$-st and $2$-nd order ambient flows. {The relation between the geometrical moments and the entries of the grand-resistance matrix \cite{brenner64a,brenner64b,brenner64c,kim-karrila} is addressed in Appendix \ref{appC}.
}

The $n$-th order disturbance field 
$ ({\pmb w}^{(n)}({\pmb x},{\pmb \xi}'),{\pmb \tau}^{(n)}({\pmb x},{\pmb \xi}')) $ of the ambient field 
$ ({\pmb u}^{(n)}({\pmb x},{\pmb \xi}'),{\pmb \pi}^{(n)}({\pmb x},{\pmb \xi}')) $ 
centered at the point ${\pmb \xi}'\in D_b$ is  the solution of the Stokes equations
\begin{equation}
\begin{cases}
- \nabla \cdot {\pmb \tau}^{(n)} ({\pmb x},{\pmb \xi}') = \mu \Delta {\pmb w}^{(n)} ({\pmb x},{\pmb \xi}')- \nabla q^{(n)}({\pmb x},{\pmb \xi}') = 0 
\\
\nabla \cdot  {\pmb w}^{(n)} ({\pmb x},{\pmb \xi}') = 0 \qquad {\pmb x} \in D_f
\\
{\pmb w}^{(n)} ({\pmb x},{\pmb \xi}')={\pmb w}^{(S,n)} ({\pmb x},{\pmb \xi}'),\ {\pmb \tau}^{(n)} ({\pmb x},{\pmb \xi}')= {\pmb \tau}^{(S,n)} ({\pmb x},{\pmb \xi}') \qquad {\pmb x} \in \partial D_b
\end{cases}
\label{eq20}
\end{equation}
where ${\pmb w}^{(S,n)} ({\pmb x},{\pmb \xi}') $ and ${\pmb \tau}^{(S,n)} ({\pmb x},{\pmb \xi}')$ are the $n$-th order disturbance velocity field and stress tensor at the surface of the body depending on the assigned boundary conditions.
Introducing    the singularity operator
$\mathcal{F}_{\alpha \beta' {\pmb \beta}'_n}$  defined starting from the hierarchy of the
geometrical moments $m_{\alpha {\pmb \alpha}_m \beta' {\pmb \beta}'_n}({\pmb \xi},{\pmb \xi}')$
\begin{equation}
\mathcal{F}_{\alpha \beta' {\pmb \beta}'_n}
=
\sum_{m=0}^{\infty}\dfrac{m_{\alpha {\pmb \alpha}_m \beta' {\pmb \beta}'_n}({\pmb \xi},{\pmb \xi}')}{m!} \nabla_{{\pmb \alpha}_m}
\label{eq21}
\end{equation}
it is possible to express the $n$-th order disturbance field $({\pmb w}^{(n)} ({\pmb x},{\pmb \xi}'),{\pmb \tau}^{(n)} ({\pmb x},{\pmb \xi}'))$
in the form
\begin{equation}
w^{(n)}_a({\pmb x},{\pmb \xi}')=A_{\beta' {\pmb \beta}'_n} {\mathcal{F}_{\alpha \beta' {\pmb \beta}'_n} 
S_{a \alpha}({\pmb x},{\pmb \xi})}
\label{eq22}
\end{equation}
\begin{equation}
q^{(n)}({\pmb x},{\pmb \xi}')= A_{\beta' {\pmb \beta}'_n}{\mathcal{F}_{\alpha \beta' {\pmb \beta}'_n} 
P_{\alpha}({\pmb x},{\pmb \xi})}
\label{eq23}
\end{equation}
 \begin{equation}
\tau^{(n)}_{a b}({\pmb x},{\pmb \xi}')= A_{\beta' {\pmb \beta}'_n} {\mathcal{F}_{\alpha  \beta' {\pmb \beta}'_n} 
\Sigma_{a b \alpha}({\pmb x},{\pmb \xi})}
\label{eq24}
\end{equation}
The $n$-th order total velocity field is $({\pmb v}^{(n)}({\pmb x},{\pmb \xi}'),{\pmb \sigma}^{(n)}({\pmb x},{\pmb \xi}'))=
({\pmb u}^{(n)}({\pmb x},{\pmb \xi}'),{\pmb \pi}^{(n)}({\pmb x},{\pmb \xi}'))+
({\pmb w}^{(n)}({\pmb x},{\pmb \xi}'),{\pmb \tau}^{(n)}({\pmb x},{\pmb \xi}'))
$
and its entries 
 can be expressed by enforcing the linearity of the Stokes flow as
$(v^{(n)}_a ({\pmb x},{\pmb \xi}'),\sigma^{(n)}_{a b}({\pmb x},{\pmb \xi}'))= A_{\beta' {\pmb \beta}'_n} (v_{a \beta' {\pmb \beta}'_n}({\pmb x},{\pmb \xi}'),\mu \sigma_{a b \beta' {\pmb \beta}'_n}({\pmb x},{\pmb \xi}'))$. Therefore, from eqs. (\ref{eq18}) and  (\ref{eq19}), the geometrical moments can be evaluated as  the following surface integrals
\begin{eqnarray}
\nonumber
&& 
8 \pi \,
m_{\alpha {\pmb \alpha}_m \beta' {\pmb \beta}'_n}({\pmb \xi},{\pmb \xi}')
 = 
\\
\nonumber
&& 
  \int_{\partial D_b} 
  ({\pmb x}-{\pmb \xi})_{{\pmb  \alpha}_m} \sigma_{ \alpha b \beta' {\pmb \beta}'_n}({\pmb x},{\pmb \xi}')n_b({\pmb x}) dS({\pmb x})
  -  \int_{\partial D_b} p_{\alpha {\pmb  \alpha}_m}({\pmb x},{\pmb \xi})
 v_{b \beta' {\pmb \beta}'_n}({\pmb x},{\pmb \xi}') n_b({\pmb x}) dS({\pmb x})
  \\
 && +
  \int_{\partial D_b} 
 \, 
 \left(
v_{\alpha \beta' {\pmb \beta}'_n}({\pmb x},{\pmb \xi}') n_b({\pmb x})\nabla_b ({\pmb x} - {\pmb \xi})_{{\pmb  \alpha}_m}
+
 n_\alpha({\pmb x}) v_{c \beta' {\pmb \beta}'_n}({\pmb x},{\pmb \xi}') 
\nabla_c ({\pmb x} - {\pmb \xi})_{{\pmb \alpha}_m}
\right)
 dS({\pmb x})
\label{eq25}
\end{eqnarray}
Without loss of generality, 
we can always  consider ${\pmb \xi}={\pmb \xi}'$ in all the case addressed in the remainder 
since the distinction between these points is  unnecessary. 
For example, as regards the geometrical  moments $m_{\alpha {\pmb \alpha}_m \beta {\pmb \beta}_n}({\pmb \xi},{\pmb \xi}) =
m_{\alpha {\pmb \alpha}_m \beta' {\pmb \beta}'_n}({\pmb \xi},{\pmb \xi}')|_{{\pmb \xi}'={\pmb \xi}}$.

{
For  the sake of notational simplicity, 
we are assuming  an infinite series of singularities centered at a single pole 
for  describing   the disturbance flows, from which eq. (\ref{eq21})
 for
$\mathcal{F}_{\alpha \beta {\pmb \beta}_n }$ follows.
There are   cases, e.g. associated with  ellipsoidal bodies \cite{kim86},
where it is possible to avoid such an infinite series 
by letting the poles lie on an manifold $\Omega \subset D_b$ belonging to the domain of the body and following its symmetries. 
In these cases, the definition of $\mathcal{F}_{\alpha \beta {\pmb \beta}_n } $ can be easily  extended to a continuous set of singularities 
in the form of an  integro-differential linear operator
\begin{equation}
\mathcal{F}_{\alpha \beta' {\pmb \beta}'_n}
=
\dfrac{1}{\text{meas}({ \Omega})}\int d\Omega({\pmb \xi})
\sum_{m=0}^{\infty}\dfrac{m_{\alpha {\pmb \alpha}_m \beta' {\pmb \beta}'_n}({\pmb \xi},{\pmb \xi}')}{m!} \nabla_{{\pmb \alpha}_m}
\label{eq25.1}
\end{equation}
where $d \Omega({\pmb \xi})$ is the measure element at the point ${\pmb \xi}$ and
\begin{equation}
\nonumber
\text{meas}({ \Omega})= \int_\omega d \Omega({\pmb \xi})
\end{equation}
is the Lebesgue measure of the manifold $\Omega$.  This 
extension is  addressed in the article by Procopio and Giona \cite{pg_mine},  and the reader is referred to this
article for the  mathematical details. 

All the results obtained in the remainder  hold regardless of the exact form of $\mathcal{F}_{\alpha \beta {\pmb \beta}_n}$,  and  thus can be 
  extended to the integro-differential representation  eq. (\ref{eq25.1}).
}

\section{Reciprocal boundary conditions}
\label{sec4}
{ Before investigating the relations between the singularity 
operators $ \mathcal{F}_{\alpha \beta {\pmb \beta}_n} $ and the $n$-th order Faxén operators, it is convenient to introduce  the concept of {\em BC-reciprocity}, which as shown in the next Section, is crucial in order to 
 establish a dualism  between the two operators.
 }

For  the sake of { notational conciseness}, we indicate with the symbol $[ \cdot , \cdot ]$ the bilinear operator acting on 
two generic Stokes flows ${\bf v}({\pmb x})$ and ${\bf v}'({\pmb x})$ { at the field point ${\pmb x}$}, corresponding to the surface integral on the body
\begin{equation}
[{\bf v},{\bf v}']=
\int_{\partial D_b}
\left( {\pmb \sigma}\{{\bf v}'({\pmb x})  \}\cdot {\bf v}({\pmb x}) -
 {\pmb \sigma}\{{\bf v}({\pmb x})  \}\cdot {\bf v}'({\pmb x})
\right) \cdot {\pmb n}({\pmb x})dS({\pmb x})
\label{eq26}
\end{equation}
where $ {\pmb \sigma}\{{\bf v}({\pmb x})  \} $ is the stress tensor  related to the field ${\bf v}({\pmb x})$.

It is easy to verify that
\begin{equation}
[{\bf v},{\bf v}']=-[{\bf v}',{\bf v}]
\label{eq27}
\end{equation}
and therefore  the operator $[ \cdot , \cdot ]$ admits odd parity.
Since  the ambient fields are regular homogeneous solutions of the Stokes equations in the domain of the body, given two ambient fields ${\pmb u}({\pmb x})$ and ${\pmb u}'({\pmb x})$, { we can simply apply the Gauss theorem to have}
\begin{equation}
[{\pmb u},{\pmb u}']=0
\label{eq28}
\end{equation}
{ Eq. (\ref{eq28}) is indeed a direct consequence of the Lorentz reciprocal theorem applied to  the Stokes fluid in the domain $D_b$ bounded by $\partial D_b$. }
In the case we consider  two disturbance fields ${\pmb w}({\pmb x})$ and ${\pmb w}'({\pmb x})$ {(not defined in $D_b$)}, 
 applying the Lorentz reciprocal theorem on the surface $\partial D_f \equiv \partial D_b \cup \partial D_\infty$ and considering that the disturbance velocity fields vanish at $\partial D_\infty$ { 
as a Stokeslet ${\pmb w}({\pmb x}) =O(1/|{\pmb x}|)$ and the stress tensor disturbance field as a  Stresslet, i.e.  $O( 1/|{\pmb x}|^2)$ \cite{pozri,kim-karrila}, }
we obtain 
\begin{equation}
[{\pmb w},{\pmb w}']
=
\int_{\partial D_\infty}
\left( {\pmb \sigma}\{{\pmb w}'({\pmb x})  \}\cdot {\pmb w}({\pmb x}) -
 {\pmb \sigma}\{{\pmb w}({\pmb x})  \}\cdot {\pmb w}'({\pmb x})
\right) \cdot {\pmb n}({\pmb x})dS({\pmb x})
=
0
\label{eq29}
\end{equation}
On the other hand, given two total fields
${\pmb v}({\pmb x})={\pmb u}({\pmb x})+{\pmb w}({\pmb x})$  and
${\pmb v}'({\pmb x})={\pmb u}'({\pmb x})+{\pmb w}'({\pmb x})$, the quantity $[{\pmb v},{\pmb v}' ]$ does not vanish in general. 
{ This result can be proved as follows.
By using the bilinear property of the operator $[.\, , .]$, we can expand $[{\pmb v},{\pmb v}']$, hence
\begin{equation}
[{\pmb v},{\pmb v}']
=
[{\pmb u}+{\pmb w},\, {\pmb u}'+{\pmb w}']=
[{\pmb u},{\pmb u}']+
[{\pmb u},{\pmb w}']+
[{\pmb w},{\pmb u}']+
[{\pmb w},{\pmb w}']
\label{eq29.1}
\end{equation}
The first and fourth terms at the r.h.s of eq. (\ref{eq29.1}) vanish because of eqs. (\ref{eq28}) and (\ref{eq29}). By using the property expressed by
 eq. (\ref{eq27}), the following identity holds
\begin{equation}
[{\pmb v},{\pmb v}'] =
[{\pmb u},{\pmb w}']-[{\pmb u}',{\pmb w}]
\label{eq30}
\end{equation}
By applying the Lorentz reciprocal theorem, it is possible to verify that $[{\pmb u},{\pmb w}']- [{\pmb u}',{\pmb w}]$ in general does not vanishes at infinity. For example, consider $[{\pmb u},{\pmb w}']$ and apply the Lorentz reciprocal theorem in the volume of the fluid bounded by $\partial D_f =\partial D_b \cup \partial D_\infty$
\begin{equation}
[{\pmb u},{\pmb w}']
=
\int_{\partial D_\infty}
\left( 
 {\pmb \sigma}\{{\pmb u}({\pmb x})  \}\cdot {\pmb w}'({\pmb x})
-
{\pmb \sigma}\{{\pmb w}'({\pmb x})  \}\cdot {\pmb u}({\pmb x}) 
\right) \cdot {\pmb n}({\pmb x})dS({\pmb x})
\label{eq30.1}
\end{equation}
while for $[{\pmb u}',{\pmb w}]$
\begin{equation}
[{\pmb u}',{\pmb w}]
=
\int_{\partial D_\infty}
\left( 
 {\pmb \sigma}\{{\pmb u}'({\pmb x})  \}\cdot {\pmb w}({\pmb x})
-
{\pmb \sigma}\{{\pmb w}({\pmb x})  \}\cdot {\pmb u}'({\pmb x}) 
\right) \cdot {\pmb n}({\pmb x})dS({\pmb x})
\label{eq30.2}
\end{equation}
Since ${\pmb u}({\pmb x})$ and ${\pmb u}'({\pmb x})$ should not necessarily
decay at infinity, the integrals at the r.h.s of eqs. (\ref{eq30.1}) and (\ref{eq30.2})  are in general different from zero, contrarily to what happens
for disturbance flows.

To show this unambiguously, consider the simple case where ${\pmb u}({\pmb x})={\pmb u}=\mbox{const}$  and ${\pmb u}'({\pmb x})={\pmb A}\cdot ({\pmb x} - {\pmb \xi})$ a linear ambient flow. Let us start  analyzing the integral in eq. (\ref{eq30.1}). 
If the body is a sphere in a linear ambient flow, the leading order Stokeslet term in the multipole expansion of
${\pmb w}'({\pmb x})$
is vanishing due to the spherical symmetries \cite{pozri,kim-karrila}. Therefore, since
${\pmb w}({\pmb x}) =O(1/|{\pmb x}|^2)$ and the associated stress 
${\pmb \sigma}\{ {\pmb w}'({\pmb x})  \}=
O( 1/|{\pmb x}|^3)$ , the integral in eq. (\ref{eq30.1}) vanishes at infinity and $[{\pmb u},{\pmb w}']=0$.
Apart from this specific case possessing spherical symmetry,
the Stokeslet term in the disturbance flow implies ${\pmb w}'({\pmb x}) =O(1/|{\pmb x}|)$ and 
${\pmb \sigma}\{ {\pmb w}'({\pmb x})  \}=
O( 1/|{\pmb x}|^2)$.
 Therefore, integrating over a sphere with radius $R_s$, enclosing the body, and letting the radius go to infinity, we have in general
\begin{equation}
[{\pmb u},{\pmb w}']
\sim
\int_{R_s \rightarrow \infty}
 \dfrac{d r^2}{r^2}  \rightarrow \infty
\label{eq30.3}
\end{equation}
On the other hand, the leading order term in the multipole expansion of 
a disturbance field associated with a body in a constant ambient flow
is a Stokeslet independently of the symmetries of the problem. Therefore, considering $[{\pmb u}',{\pmb w}]$ in eq. (\ref{eq30.2}), where ${\pmb w}({\pmb x}) =O(1/|{\pmb x}|)$ and 
${\pmb \sigma}\{ {\pmb w}({\pmb x})  \}=
O( 1/|{\pmb x}|^2)$,
 we obtain 
 \begin{equation}
[{\pmb u}',{\pmb w}]
\sim
 \int_{R_s \rightarrow \infty}
 \dfrac{d r^2 }{r}  \rightarrow \infty
\label{eq30.4}
\end{equation}
By the different scaling law between eq. (\ref{eq30.3}) and eq. (\ref{eq30.4}), we can also conclude that the difference $[{\pmb u},{\pmb w}']- [{\pmb u}',{\pmb w}]$ does not vanish in general.

The above example shows that it is not possible to determine whether $[{\pmb v},{\pmb v}']=0$ for any pair of total fields ${\pmb v}({\pmb x})$ and ${\pmb v}'({\pmb x})$ by the asymptotic behavior of Stokes flows at infinity and, hence, without considering the specific interactions between the Stokes fluid and the body.
 Therefore, in order to establish  
whether $[{\pmb v},{\pmb v}']=0$,
it is necessary to take into account the boundary conditions assumed at the surface of the body, 
possibly considering the equations governing the internal behavior of the body (if any), for which the Lorentz reciprocal theorem does not hold in principle.}

We call \textit{reciprocal boundary conditions}, any boundary condition for which
\begin{equation}
[{\pmb v},{\pmb v}'] =
0, \qquad \forall \ {\pmb v}({\pmb x}),{\pmb v}'({\pmb x})
\label{eq31}
\end{equation}
Equivalently,  eq. (\ref{eq31}) is  the mathematical definition of the property
referred to as {\em Boundary-Condition reciprocity} (BC-reciprocity, for short).
{
The BC-reciprocity of classical boundary conditions employed to describe the hydrodynamics of Stokes fluid-body interactions is analyzed in Sections \ref{secIV-A}-\ref{sec7}.
In the next Section, we show that, in order to fulfill the Hinch-Kim dualism, it is necessary to assume reciprocal boundary conditions at the interface between the body and the Stokes external fluid.
}
\section{Generalized $n$-th order Faxén operator and  the Hinch-Kim dualism theorem}
\label{sec4.1}

It is possible to express the generic ambient field in the domain of the body by the 
Ladyzhenskaya
 boundary integrals \cite{lady} 
\begin{equation}
u_\alpha({\pmb \xi})=
-\dfrac{[{\pmb S}_{\alpha}({\pmb \xi}),{\pmb u}]}{8\pi \mu}
=
-
\int_{\partial D_b}{ \bigg(}\pi_{a b}({\pmb x}) \dfrac{S_{a \alpha}({\pmb x},{\pmb \xi})}{8 \pi \mu}  - u_a({\pmb x}) \dfrac{\Sigma_{a b \alpha}({\pmb x},{\pmb \xi} )}{8 \pi}  { \bigg)}n_b({\pmb x}) dS({\pmb x})
\label{eq32}
\end{equation}
Applying the operator $A_{\beta {\pmb \beta}_n}\mathcal{F}_{\alpha \beta {\pmb \beta}_n}$, {
acting on the pole point 
${\pmb \xi}$ 
 (as indicated by using Greek subscripts)}, at both  
sides of eq. (\ref{eq32}), and using the relations
(\ref{eq22}) one obtains
\begin{equation}
A_{\beta {\pmb \beta}_n}\mathcal{F}_{\alpha \beta {\pmb \beta}_n}u_\alpha({\pmb \xi}) =-\dfrac{[
A_{\beta {\pmb \beta}_n}\mathcal{F}_{\alpha \beta {\pmb \beta}_n}
{\pmb S}_{\alpha}({\pmb \xi}),{\pmb u}]}{8 \pi \mu}=-\dfrac{[{\pmb w}^{(n)}({\pmb \xi}) ,{\pmb u}]}{8 \pi \mu}
\label{eq33}
\end{equation}
It is possible to add at the r.h.s. of  eq. (\ref{eq33}) the vanishing contribution 
$[{\pmb w}^{(n)}({\pmb \xi}) ,{\pmb w}]$
deriving from   two disturbance fields,
 thus
\begin{equation}
A_{\beta {\pmb \beta}_n}\mathcal{F}_{\alpha \beta {\pmb \beta}_n}u_\alpha({\pmb \xi}) 
 = 
-\dfrac{[{\pmb w}^{(n)}({\pmb \xi}) ,{\pmb u}]+[{\pmb w}^{(n)}({\pmb \xi}),{\pmb w}]}{8\pi\mu}
= -\dfrac{[{\pmb w}^{(n)}({\pmb \xi}) ,{\pmb v}]}{8\pi\mu}
\label{eq34}
\end{equation}
that can be expressed, replacing $ {\pmb w}^{(n)}({\pmb x},{\pmb \xi} ) = 
{\pmb v}^{(n)}({\pmb x},{\pmb \xi})- {\pmb u}^{(n)}({\pmb x},{\pmb \xi}) $,  in the form
\begin{equation}
A_{\beta {\pmb \beta}_n}\mathcal{F}_{\alpha \beta {\pmb \beta}_n}u_\alpha({\pmb \xi}) 
=\dfrac{
[{\pmb u}^{(n)}({\pmb \xi}), {\pmb v}]
-[{\pmb v}^{(n)}({\pmb \xi}) ,{\pmb v}]
}{8\pi\mu}
\label{eq35}
\end{equation}
{  Applying $A_{\alpha {\pmb \alpha}_n}$ on both the sides of eq. (\ref{eq17}) 
\begin{equation}
A_{\alpha {\pmb \alpha}_n} M_{\alpha {\pmb \alpha}_n}({\pmb \xi})=[{\pmb u}^{(n)}({\pmb \xi}), {\pmb v}]
\label{eq35.1}
\end{equation}
}
Finally, comparing eq. (\ref{eq35.1}) with eq. (\ref{eq35}), we obtain
\begin{equation}
M_{\beta {\pmb \beta}_n}({\pmb \xi})=
8\pi\mu \mathcal{F}_{\alpha \beta {\pmb \beta}_n}u_\alpha({\pmb \xi}) 
+[{\pmb v}^{(n)}_{\beta {\pmb \beta}_n}({\pmb \xi}) ,{\pmb v}]
\label{eq36}
\end{equation}
Eq. (\ref{eq36}) is one of the main result of this article, connecting the Hinch-Kim duality to
the condition of BC-reciprocity.
From  eq. (\ref{eq36}) it is possible to state that
 the Hinch-Kim dualism, 
holds
 whenever reciprocal boundary conditions are imposed on the surface of the body, i.e. whenever
\begin{equation}
[{\pmb v}^{(n)}({\pmb \xi}) ,{\pmb v}]
=
\int_{\partial D_b} \big(\sigma_{a b}({\pmb x}) v^{(n)}_a({\pmb x},{\pmb \xi}) -v_a({\pmb x}) \sigma^{(n)}_{a b}({\pmb x},{\pmb \xi}) \big) n_b({\pmb x})  dS({\pmb x})=0, \qquad 
{
 {\forall}\,  {\pmb v}({\pmb x})
}
\label{eq37}
\end{equation}
{ where "$\forall\,  {\pmb v}({\pmb x})$", meaning that ${\pmb v}({\pmb x})$ can be an arbitrary solution of the Stokes equations,  
is a necessary condition for the Faxén operator to be independent of the ambient flow. }

This can be referred to as the Hinch-Kim dualism theorem.
In the BC-reciprocal case,  the $n$-th order singularity operator $ \mathcal{F}_{\alpha \beta {\pmb \beta}_n} $ 
defined by  eq. (\ref{eq21}) furnishes  either the $n$-th order disturbance field, if applied to the pole of the unbounded 
Green function
according eqs. (\ref{eq22})-(\ref{eq24}), or the $n$-th order 
moment on a particle immersed in an ambient field ${\pmb u}({\pmb x})$ according to the relation
\begin{equation}
M_{\beta {\pmb \beta}_n}({\pmb \xi})=8\pi\mu
\mathcal{F}_{\alpha \beta {\pmb \beta}_n}u_\alpha({\pmb \xi}),
\qquad 
{
\forall\, {\pmb u}({\pmb x})
}
\label{eq38}
\end{equation}
Owing to the fact that the operator $\mathcal{F}_{\alpha \beta {\pmb \beta}_n}$ returns 
the $n$-th order moments on the body 
if applied to a generic ambient flow, $\mathcal{F}_{\alpha \beta {\pmb \beta}_n}$ is 
{\em sensu-stricto} a \textit{$n$-th order generalized Faxén operator}.

Let us to show an interesting consequence 
of reciprocal boundary conditions.
If BC-reciprocity holds, by using eq. (\ref{eq8}), the disturbance field
due to the inclusion of the body
 related to a generic ambient field ${\pmb u}({\pmb x})$   can be expressed as
\begin{equation}
w_a({\pmb x})=\dfrac{1}{8\pi \mu}\sum_{n=0}^{\infty} 
\dfrac{
\mathcal{F}_{\beta \alpha {\pmb \alpha}_n}u_\beta({\pmb \xi})
}{n!} 
\nabla_{{\pmb \alpha}_n}
S_{a\, \alpha}({\pmb x},{\pmb \xi})
\label{eq39}
\end{equation}
and, as shown by Procopio and Giona \cite{pg_mine}
and briefly reviewed  in Appendix \ref{appA}, due to the following symmetry of the geometric moments
\begin{equation}
m_{\alpha {\pmb \alpha}_m \beta' {\pmb \beta}'_n}({\pmb \xi},{\pmb \xi}')=m_{\beta' {\pmb \beta}'_n \alpha {\pmb \alpha}_m }({\pmb \xi}',{\pmb \xi})
\label{eq40}
\end{equation}
we obtain an expansion of a generic disturbance
field in terms of  the Faxén operators (i.e. in terms of the geometrical moments)
\begin{equation}
w_a({\pmb x})=\sum_{n=0}^{\infty} 
\dfrac{
\nabla_{{\pmb \beta}_n}
u_\beta({\pmb \xi})
}{n!} 
\mathcal{F}_{\alpha \beta  {\pmb \beta}_n}
S_{a\, \alpha}({\pmb x},{\pmb \xi})
\label{eq41}
\end{equation}
Gathering eqs. (\ref{eq38}) and  (\ref{eq41})  a remarkable property follows, namely
if BC-reciprocity holds, the
hydromechanics (i.e. the motion of the body 
due to the interaction with the fluid and the motion of the fluid due to the interaction with the body)
 of a  fluid-body system in the  Stokes  regime
can be completely described by the knowledge of
the entire set of $(m,n)$-th order
geometrical moments of the body.

In the next Sections we analize  typical  hydrodynamic boundary conditions at the  fluid-body interface 
in order to ascertain  in which cases BC-reciprocity 
i.e. eq. (\ref{eq31}) is fulfilled and $\mathcal{F}_{\alpha \beta {\pmb \beta}_n}$ is a Faxén operator.

\section{Boundary conditions at solid-fluid interfaces}
\label{secIV-A}
BC-reciprocity, i.e. $[{\pmb v},{\pmb v}']=0$,
is straightforwardly verified for  no-slip boundary conditions. { In fact, if
 ${\pmb v}({\pmb x})={\pmb v}'({\pmb x})=0 $ for ${\pmb x} \in \partial D_b$  are assumed
 (as
  by definition of no-slip boundary conditions), the integral in eq. (\ref{eq26}) vanishes trivially.
 This means that the Hinch-Kim dualism holds for any operator $\mathcal{ F}_{\alpha \beta {\pmb \beta}_n}$ when bodies with no-slip boundary conditions are considered.
A typical example is the case of a sphere with radius $R_p$ and with no-slip boundary conditions immersed in a constant ambient flow
(hence ${\pmb v}'({\pmb x}) \equiv {\pmb v}^{(0)}({\pmb x}) $). In this case, the disturbance flow is \cite{pozri}
\begin{equation}
w^{(0)}_a({\pmb x})= -A_\alpha \left( \dfrac{3}{4} R_p +\dfrac{R_p^3 }{8}\Delta_x
\right) S_{a \, \alpha}({\pmb x},{\pmb \xi})
\label{eq41.1}
\end{equation} 
and, by the well known dualism \cite{kim-karrila,pozri}, the force on the same sphere immersed in a generic ambient flow ${\pmb u}({\pmb x})$ is given by 
\begin{equation}
F_a=-M_a= 8 \pi \mu \left( \dfrac{3}{4} R_p +\dfrac{R_p^3 }{8}\Delta_x
\right) u_a({\pmb x})
\label{eq41.2}
\end{equation}
By eq. (\ref{eq41.1}) and (\ref{eq41.2}), it is possible to identify
 $\mathcal{F}_{\alpha \beta} = (3R_p/4+R_p \Delta/8 ) \delta_{\alpha \beta}$ as  the operator acting on the ambient flow in eq. (\ref{eq38}). 

Analogously, the BC-reciprocity is straightforward
considering complete slip 
boundary conditions,
 for which
 ${\pmb \sigma}\{ {\pmb v}({\pmb x}) \} \cdot {\pmb n}({\pmb x})= {\pmb \sigma}\{ {\pmb v}'({\pmb x}) \} \cdot {\pmb n}({\pmb x})=0$ for ${\pmb x} \in \partial D_b$   are assumed in eq. (\ref{eq26}).

It is possible to show that BC-reciprocity holds for any linear relation 
between velocity and traction at the boundary (of which no-slip and complete slip boundary conditions are particular cases) although, in this case, neither ${\pmb v}({\pmb x})$ and ${\pmb v}'({\pmb x})$, nor ${\pmb \sigma}\{ {\pmb v}({\pmb x}) \} \cdot {\pmb n}({\pmb x})$ and $ {\pmb \sigma}\{ {\pmb v}'({\pmb x}) \} \cdot {\pmb n}({\pmb x})$ vanish at the boundary. }
 To this aim, consider the  interfacial mobility matrix ${\pmb {\beta}}({\pmb x})$ \cite{bazant}
defined by the relation
\begin{equation}
{\pmb {\beta}}({\pmb x}) \cdot {\pmb v}({\pmb x})=  {\pmb \sigma}({\pmb x}) \cdot {\pmb n}({\pmb x}),\qquad {\pmb x} \in \partial D_b
\label{eq42}
\end{equation}
Due to its symmetry, 
${\beta}_{a b}({\pmb x})={\beta}_{b a}({\pmb x})$, we have that 
\begin{eqnarray}
\nonumber
&& [{\pmb v}^{(n)}({\pmb \xi}),{\pmb v}]=
\int_{\partial D_b} (\sigma_{a b}({\pmb x}) n_b({\pmb x}) v^{(n)}_a({\pmb x},{\pmb \xi}) -v_a({\pmb x}) \sigma^{(n)}_{a b}({\pmb x},{\pmb \xi}) n_b({\pmb x})  ) dS({\pmb x})
 =  
 \\ 
 &&
\int_{\partial D_b} \big(v_b({\pmb x})
{\beta}_{a b}({\pmb x})
 v^{(n)}_a({\pmb x},{\pmb \xi}) -
  v^{(n)}_b({\pmb x},{\pmb \xi})
{\beta}_{a b}({\pmb x})
 v_a({\pmb x})
\big) dS({\pmb x})
= 0 \qquad 
{
\forall\, {\pmb v}^{(n)}({\pmb x}), {\pmb v}({\pmb x})
}
\label{eq43}
\end{eqnarray}
Consistently,  in this case, { the Hinch-Kim dualism holds and}
$\mathcal{F}_{\alpha \beta {\pmb \beta}_n} $ is a Faxén operator.

Next, let us focus on the case of  Navier-slip boundary conditions.
 Thus, given an ambient field ${\pmb u}({\pmb x})$,
 the total field ${\pmb v}({\pmb x})={\pmb u}({\pmb x})+{\pmb w}({\pmb x})$ satisfies at the boundaries of the body
the relations
\begin{equation}
\begin{cases}
{\pmb v}({\pmb x})\cdot {\pmb n}({\pmb x})=0
\\
{\pmb v}({\pmb x})\cdot {\pmb t}({\pmb x})=-\dfrac{\lambda}{\mu}\, {\pmb h}({\pmb x})\cdot {\pmb t}({\pmb x}), \qquad {\pmb x} \in \partial D_b
\end{cases}
\label{eq47}
\end{equation}
where $\lambda$ is the  slip length of the interface, ${\pmb t}({\pmb x})={\pmb I}-{\pmb n}({\pmb x})\otimes {\pmb n}({\pmb x})$  the unit tangent matrix,
and ${\pmb h}({\pmb x})={\pmb \sigma}({\pmb x})\cdot {\pmb n}({\pmb x})$ 
 the surface traction of the total velocity field.

Navier-slip boundary conditions  eq. (\ref{eq47}) represent a particular 
case where the relation between velocity and traction at the boundary of the body is linear, 
and from what obtained above, BC-reciprocity applies, meaning that
 the Hinch-Kim
dualism  holds. Therefore, given a generic ambient field ${\pmb u}({\pmb x})$,
the moments on the body are given by eq. (\ref{eq38}),
and the disturbance field is expressed by eq. (\ref{eq41}).
The geometrical moments that are needed to explicit the 
 Faxén operators can be obtained by substituting the boundary conditions  eq. (\ref{eq47}) into eq. (\ref{eq25}),
 and by considering the geometrical surface traction of  the body immersed in a $n$-th order ambient field 
$h_{\alpha \beta {\pmb \beta}_n}({\pmb x},{\pmb \xi})=\delta_{\alpha a} \sigma_{a b \beta {\pmb \beta}_n}({\pmb x},{\pmb \xi})n_b({\pmb x})$, thus
\begin{eqnarray}
\nonumber
& & m_{\alpha {\pmb \alpha}_m \beta {\pmb \beta}_n}({\pmb \xi},{\pmb \xi})
  = 
\\ 
\nonumber
&& \int_{\partial D_b} 
\dfrac{h_{\gamma \beta {\pmb \beta}_n}({\pmb x},{\pmb \xi})}{8 \pi }
 \left(
\delta_{\alpha \gamma} ({\pmb x}-{\pmb \xi})_{{\pmb \alpha}_m} -
\lambda ( 
t_{\alpha \gamma}({\pmb x}) n_b({\pmb x})\nabla_b ({\pmb x} - {\pmb \xi})_{{\pmb  \alpha}_m}
+
 n_\alpha({\pmb x}) t_{c \gamma}({\pmb x}) 
\nabla_c ({\pmb x} - {\pmb \xi})_{{\pmb \alpha}_m})
\right)
 dS({\pmb x})
 \\
\label{eq51} 
\end{eqnarray}
The $n$-th order surface traction
$h_{\alpha \beta {\pmb \beta}_n}({\pmb x},{\pmb \xi})$ can be expressed as
\begin{equation}
h_{\alpha \beta {\pmb \beta}_n}({\pmb x},{\pmb \xi})=f_{\alpha \beta {\pmb \beta}_n}({\pmb x},{\pmb \xi})
+
n_{\alpha}({\pmb x})p_{\beta {\pmb \beta}_n}({\pmb x},{\pmb \xi})-
\left(
n_{\beta}({\pmb x})\nabla_\alpha ({\pmb x} - {\pmb \xi})_{{\pmb \beta}_n}
+
\delta_{\beta \alpha}n_\gamma({\pmb x})\nabla_\gamma ({\pmb x} - {\pmb \xi})_{{\pmb \beta}_n}
\right)
\label{eq52}
\end{equation}
where $ f_{\alpha \beta {\pmb \beta}_n}({\pmb x},{\pmb \xi})=\delta_{\alpha a} {\tau}_{a b\beta {\pmb \beta}_n}({\pmb x},{\pmb \xi})  n_b({\pmb x})$ is the surface traction related to the $n$-th order disturbance field.
{ In Section \ref{sec8}, the specific case of a sphere with Navier-slip boundary conditions is addressed in detail and a method, based on this theoretical approach, in order to evaluate all the $n$-th order Faxén operators for the sphere in a systematic way is developed.}

On the other hand, it is easy to  see that $ \mathcal{F}_{\alpha \beta {\pmb \beta}_n} $ is not  a
Faxén operator for a deforming body. In fact,
under the assumption that the body is  a 
linear elastic material solid, the governing equations for the body deformation are \cite{dau7}
\begin{equation}
\begin{cases}
\nabla \cdot {\pmb \sigma }^{[s]}({\pmb x})=-\rho^{[s]} \ddot{\pmb u}^{[s]}({\pmb x}) 
\\
{ \sigma }^{[s]}_{a b}({\pmb x})= 
\delta_{ a b } \, \lambda^{[s]}  \nabla \cdot {\pmb u}^{[s]}({\pmb x}) + \mu^{[s]}
(\nabla_a u^{[s]}_b({\pmb x})+\nabla_b u^{[s]}_a({\pmb x})), \qquad {\pmb x} \in D_b
\end{cases}
\label{eq44}
\end{equation}
where ${\pmb \sigma}^{[s]}({\pmb x})$ is the stress tensor field in the solid, 
${\pmb u}^{[s]}({\pmb x})$ the displacement field of the solid,  $\rho^{[s]}$ the solid
density,  and $\lambda^{[s]} $ and $\mu ^{[s]}$  the 
Lamé coefficients.
In eq. (\ref{eq44})
any upper ``dot'' indicates  the  derivative operation with respect to time.
Enforcing  continuity  conditions at the solid-fluid interface \cite{galdi}
\begin{equation}
\begin{cases}
{\pmb \sigma}({\pmb x}) \cdot {\pmb n}({\pmb x})={\pmb \sigma}^{[s]}({\pmb x}) \cdot {\pmb n}({\pmb x})
\\
{\pmb v}({\pmb x})= \dot{\pmb u}^{[s]}({\pmb x})
\qquad {\pmb x} \in \partial D_b
\end{cases}
\label{eq45}
\end{equation}
and substituting eqs. (\ref{eq45}) in the first integral in eq. (\ref{eq43}), from
 the Maxwell-Betti theorem \cite{maxwell,betti} it follows that
\begin{eqnarray}
\nonumber
 [{\pmb v}^{(n)}({\pmb \xi}),{\pmb v}] & = &
\int_{\partial D_b} \big(\sigma^{[s]}_{a b}({\pmb x}) n_b({\pmb x}) \dot{u}^{[s](n)}_a({\pmb x},{\pmb \xi}) -\dot{u}^{[s]}_a({\pmb x}) \sigma^{[s](n)}_{a b}({\pmb x},{\pmb \xi}) n_b({\pmb x})  \big) dS({\pmb x})
\\
&=& -\rho ^{[s]}\int_{ D_b} \big(
\ddot{u}^{[s]}_a({\pmb x}) \,
\dot{u}^{[s](n)}_a({\pmb x},{\pmb \xi}) -
\dot{u}^{[s]}_a({\pmb x}) \,
\ddot{u}^{[s](n)}_a({\pmb x},{\pmb \xi}) \big) dV({\pmb x})
\label{eq46}
\end{eqnarray}
which does not vanish  in general
for any flow ${\pmb v}({\pmb x})$ and ${\pmb v}^{(n)}({\pmb x},{\pmb \xi})$.
BC-reciprocity
is ensured 
only { when we can consider the body} at the mechanical equilibrium, i.e.  when ${ \nabla \cdot {\pmb \sigma }^{[s]}({\pmb x})=0 }$.
\\
\\

\section{Boundary conditions at fluid-fluid interfaces}
\label{sec6}
In the presence of  a fluid body, the most common linear boundary conditions assumed
at the fluid-fluid interface,
considered incompressible and homogeneous,
 are \cite{brenner_sur,rallison84}
\begin{equation}
\begin{cases}
{\pmb v}({\pmb x})={\pmb v}^{[i]}({\pmb x})\\
{\pmb v}({\pmb x}) \cdot {\pmb n}({\pmb x})
= \dot{r}({\pmb x},t) 
\\
{\pmb \sigma}({\pmb x})\cdot {\pmb n}({\pmb x})={\pmb \sigma}^{[i]}({\pmb x})\cdot {\pmb n}({\pmb x})+\gamma\,  {\pmb n}({\pmb x})\, C({\pmb x}),
\qquad
{\pmb x} \in \partial D_b
\end{cases}
\label{eq103}
\end{equation}
 where ${\pmb v}^{[i]}({\pmb x})$ and $ {\pmb \sigma}^{[i]}({\pmb x})$ are the velocity field and the stress tensor in 
the  disturbing fluid (say a liquid drop or a gas bubble), $C({\pmb x})$  the trace of the curvature tensor of the surface
 and $\gamma$ the surface tension. 

Applying the reciprocity integral  eq. (\ref{eq26})
to the fields ${\pmb v}({\pmb x})$ and ${\pmb v}^{(n)}({\pmb x})$, it follows that 
\begin{eqnarray}
\nonumber
[{\pmb v}^{(n)}({\pmb \xi}),{\pmb v}]=
\int_{\partial D_b} \big( \sigma_{a b}({\pmb x}) n_b({\pmb x}) v^{(n)}_a({\pmb x},{\pmb \xi}) -v_a({\pmb x}) \sigma^{(n)}_{a b}({\pmb x},{\pmb \xi}) n_b({\pmb x})  \big) dS({\pmb x})
& = & \\ 
\nonumber
\int_{\partial D_b} \big( \sigma^{[i]}_{a b}({\pmb x}) n_b({\pmb x}) v^{[i](n)}_a({\pmb x},{\pmb \xi}) -v^{[i]}_a({\pmb x}) \sigma^{[i](n)}_{a b}({\pmb x},{\pmb \xi}) n_b({\pmb x})  \big) dS({\pmb x})
&+& \\
\gamma 
\int_{\partial D_b}  (v^{[i](n)}_a({\pmb x},{\pmb \xi}) -v^{[i]}_a({\pmb x}))C({\pmb x}) n_a({\pmb x}) dS({\pmb x})
&& 
\label{eq104}
\end{eqnarray} 
Since the Lorentz reciprocal theorem is a peculiarity
of Newtonian fluids (and, more generally, of continua characterized by  linear relations between fluxes and thermodynamics forces), the first integral at the r.h.s of eq. (\ref{eq104}) does not vanish, at least in principle, in the
 non-Newtonian case and  consequently
$\mathcal{F}_{\alpha \beta {\pmb \beta}_n}$ cannot be a Faxén operator.
 In the case  the disturbing fluid is  Newtonian, the first integral at the r.h.s of eq. (\ref{eq104}) vanishes
due to the Lorentz reciprocal theorem for Newtonian fluids, but the second integral does not vanish 
until the interface shape does not reach the
equilibrium state. In fact, the velocity at the interface ${\pmb v}({\pmb x})|_{{\pmb x} \in \partial D_b}$ is uniquely determined by the 
Rallison-Acrivos integral equations once the ambient field is assigned \cite{rallison-acrivos,power}.
Therefore, the second integral
at the r.h.s of eq. (\ref{eq104})
does not vanishes for any ambient flows, but solely in the trivial case of ${\pmb u}({\pmb x})={\pmb u}^{(n)}({\pmb x},{\pmb \xi})$.


If the disturbing fluid  is  Newtonian and the shape of the body, say a drop or a bubble,
 is stationary (${\pmb v}({\pmb x}) \cdot {\pmb n}({\pmb x})|_{{\pmb x}\in \partial D_b}=0$), 
 the first integral at the r.h.s of eq. (\ref{eq104}) vanishes 
due to Lorentz's  reciprocity. Furthermore,
since the normal velocity is assumed  to be vanishing at the surface of the body,  also the second integral
at the r.h.s vanishes, independently of the 
shape of the body and of the surface tension.
We obtain $ [{\pmb v}^{(n)}({\pmb \xi}),{\pmb v}]=0$
and consequently  $\mathcal{F}_{\alpha \beta {\pmb \beta}_n} $ is, in this case, a Faxén operator.
This means that both the velocity field in the external fluid and
the moments on the drop  do not depend directly on the surface tension
at the surface, as surface tension has only an indirect influence related to the geometry of the
stationary shape of the drop. 

In this case, in order to evaluate the geometrical moments
providing the Faxén operator, the knowledge either
of the set of $n$-th surface velocity fields ${v}_{a b {\pmb b}_n}({\pmb x},{\pmb \xi})$
 or of   $n$-th external surface traction
$h_{a b {\pmb b}_n}({\pmb x},{\pmb \xi})=\sigma_{a c b {\pmb b}_n}({\pmb x},{\pmb \xi})n_c({\pmb x})$
is required
 since
\begin{eqnarray}
\nonumber
&& 
8 \pi \,
m_{\alpha {\pmb \alpha}_m \beta {\pmb \beta}_n}({\pmb \xi},{\pmb \xi})
 =  \int_{\partial D_b} 
  ({\pmb x}-{\pmb \xi})_{{\pmb  \alpha}_m} h_{ \alpha \beta {\pmb \beta}_n}({\pmb x},{\pmb \xi}) dS({\pmb x})
  \\
 && +
  \int_{\partial D_b} 
 \, 
 \left(
v_{\alpha \beta {\pmb \beta}_n}({\pmb x},{\pmb \xi}) n_b({\pmb x})\nabla_b ({\pmb x} - {\pmb \xi})_{{\pmb  \alpha}_m}
+
 n_\alpha({\pmb x}) v_{c \beta {\pmb \beta}_n}({\pmb x},{\pmb \xi}) 
\nabla_c ({\pmb x} - {\pmb \xi})_{{\pmb \alpha}_m}
\right)
 dS({\pmb x})
\label{eq105}
\end{eqnarray}

\section{Boundary conditions at porous body-fluid interfaces}
\label{sec7}
Next, consider  the case  the inner 
flow $({\pmb v}^{[i]}({\pmb x}),p^{[i]}({\pmb x}))$ inside a porous body is modeled by  means
of the
Darcy equations \cite{darcy,whitaker}
\begin{equation}
\begin{cases}
{\pmb v}^{[i]}({\pmb x})=-\dfrac{k}{\mu }\nabla p^{[i]}({\pmb x})\\
\nabla \cdot {\pmb v}^{[i]}({\pmb x})=0, 
\qquad {\pmb x} \in D_b
\end{cases}
\label{eq108}
\end{equation}
where $k$ is the permeability of the porous medium.
The boundary condition to be imposed at the interface are the
 Beavers-Joseph-Saffman boundary conditions \cite{saffman,jones},  i.e.,
\begin{equation}
\begin{cases}
({\pmb v}({\pmb x})-{\pmb v}^{[i]}({\pmb x}) )\cdot ({\pmb I}-{\pmb n}({\pmb x})\otimes {\pmb n}({\pmb x}) )
=\dfrac{\sqrt{k}}{\alpha}
{\pmb \sigma}({\pmb x})\cdot ({\pmb I}-{\pmb n}({\pmb x})\otimes {\pmb n}({\pmb x}) )
\\
{\pmb v}({\pmb x})\cdot {\pmb n}({\pmb x})={\pmb v}^{[i]}({\pmb x}) \cdot {\pmb n}({\pmb x})\\
p({\pmb x})=p^{[i]}({\pmb x}), 
\qquad {\pmb x} \in \partial D_b
\end{cases}
\label{eq109}
\end{equation}
where $\alpha= \alpha_0 \, \mu$  and $\alpha_0$ 
is a nondimensional constant depending on the geometry and topology of the pore structure.
In this case, BC-reciprocity 
is not satisfied because $[{\pmb v},{\pmb v}']$ does not vanish in general,
since
\begin{eqnarray}
\nonumber
[{\pmb v}^{(n)}({\pmb {\pmb \xi}}),{\pmb v}]=
\int_{\partial D_b} \big( \sigma_{a b}({\pmb x}) n_b({\pmb x}) v^{(n)}_a({\pmb x},{\pmb \xi}) -v_a({\pmb x}) \sigma^{(n)}_{a b}({\pmb x},{\pmb \xi}) n_b({\pmb x})  \big) dS({\pmb x})
& = &\\
- 2 \mu \int_{\partial D_b} \big( e_{a b}\{{\pmb v}({\pmb x})\} v_a^{[i](n)}({\pmb x},{\pmb \xi}) -e_{a b}\{ {\pmb v}^{(n)}({\pmb x},{\pmb \xi})\} v_a^{[i]}({\pmb x})  \big) n_b({\pmb x}) dS({\pmb x})
&\neq &0 
\label{eq110}
\end{eqnarray}
where
\begin{equation}
\nonumber
e_{a b}\{{\pmb v}({\bf x})\}=\dfrac{1}{2}(\nabla_a v_b({\pmb x})+\nabla_b v_a({\pmb x}))
\end{equation}
It is possible to check this result by identifying the singularities in the solution provided 
by Jones \cite{jones}  for the simpler problem of a porous sphere with radius $R_p$
in a constant flow  with components $U_\beta$,
comparing the solution with the Faxén theorem obtained
by Palaniappan \cite{palaniappan} for a Darcy porous sphere in a 
generic ambient flow. 
The disturbance field in the Jones solution is given by the operator applied at the pole of the Stokeslet centered at the center of the sphere ${\pmb \xi}$
\begin{equation}
U_{\beta} \mathcal{F}_{\alpha \beta}=
U_{\beta}\left( \dfrac{R_p A_D }{2}+ \dfrac{R_p^3 B_D}{2} \Delta_{\xi}  \right) \delta_{\alpha \beta}
\label{eq111}
\end{equation}
{ where $\Delta_{\xi}$ is the Laplacian operator acting on the coordinate of the center of the sphere and}
where
\begin{eqnarray}
\nonumber
A_D=-\frac{3 R_p^2 \left(2 \sqrt{k}+\alpha  R_p\right)}{6 k^{3/2}+3 \alpha  k R_p+6 \sqrt{k} R_p^2+2 \alpha  R_p^3}
\\
\nonumber
B_D=-\frac{\alpha  R_p^3}{12 k^{3/2}+6 \alpha  k R_p+12 \sqrt{k} R_p^2+4 \alpha  R_p^3}
\end{eqnarray}
while
the Faxén theorem obtained by Palaniappan \cite{palaniappan} states that the force acting on a sphere in an ambient 
flow ${\pmb u}({\pmb x})$ is
\begin{equation}
F_\alpha= -M_\alpha =-8 \pi \mu 
\left( \dfrac{R_p A_D }{2}+ \dfrac{R_p^3 B_D'}{2} \Delta_\xi \right)
u_\alpha ({\pmb \xi})
\label{eq112}
\end{equation}
where
\begin{equation}
\nonumber
B_D'= B_D-\frac{6 k^{3/2}+3 \alpha  k R_p}{6 k^{3/2}+3 \alpha  k R_p+6 \sqrt{k} R_p^2+2 \alpha  R_p^3}
\end{equation}
The comparison of eqs. (\ref{eq111}) an eq. (\ref{eq112}) shows that the terms proportional
to the Laplacian $\Delta_\xi$ are different in the two expressions (as $B_D \neq B_{D}'$), as it should
be if the Hinch-Kim dualism would not apply. This result, follows almost immediately from
the functional structure of the r.h.s. in eq. (\ref{eq110}).

On the other hand, if the flow of the fluid in the porous medium is modeled by the
Brinkman equations \cite{brinkman} 
\begin{equation}
\begin{cases}
\mu \Delta {\pmb v}^{[i]}({\pmb x})
-\nabla p^{[i]}({\pmb x})=\dfrac{\mu }{k}{\pmb v}^{[i]}({\pmb x})\\
\nabla \cdot {\pmb v}^{[i]}({\pmb x})=0, 
\qquad {\pmb x} \in D_b
\end{cases}
\label{eq113}
\end{equation}
with continuous boundary condition  $({\pmb v}({\pmb x}),{\pmb \sigma}({\pmb x}))=({\pmb v}^{[i]}({\pmb x}),{\pmb \sigma}^{[i]}({\pmb x}))$ at the interface  ${\pmb x} \in \partial D_b $, the reciprocity  of the boundary conditions is fulfilled, since 
 \begin{eqnarray}
\nonumber
&&
[{\pmb v}^{(n)}({\pmb \xi}),{\pmb v}]=
\int_{\partial D_b} \big( \sigma_{a b}({\pmb x}) n_b({\pmb x}) v^{(n)}_a({\pmb x},{\pmb \xi}) -v_a({\pmb x}) \sigma^{(n)}_{a b}({\pmb x},{\pmb \xi}) n_b({\pmb x})  \big) dS({\pmb x}) =\\
\nonumber
&&
\int_{\partial D_b} \big( \sigma_{a b}^{[i]}({\pmb x}) n_b({\pmb x}) v^{[i](n)}_a({\pmb x},{\pmb \xi}) -v_a^{[i]}({\pmb x}) \sigma^{[i](n)}_{a b}({\pmb x},{\pmb \xi}) n_b({\pmb x})  \big) dS({\pmb x})  = \\
&&
-\dfrac{\mu}{k}
\int_{D_b} \big( v_a^{[i]}({\pmb x}) v^{[i](n)}_a({\pmb x},{\pmb \xi}) -v_a^{[i]}({\pmb x})v^{[i](n)}_a({\pmb x},{\pmb \xi}) \big) dV({\pmb x}) =0, \qquad 
{
\forall\, {\pmb v}({\pmb x})}
\label{eq114}
\end{eqnarray}
thus
for Brinkman porous bodies
 $\mathcal{F}_{\alpha \beta {\pmb \beta}_n}$ is a Faxén operator. 

The $0$-th order Faxén operator for this case can be identified by the solutions given by Masliyah and al. \cite{masliyah} or by Yu and Kaloni \cite{yu} for a traslating Brinkman porous sphere in the Stokes flow. We observe that, in this case
 \begin{equation}
U_{\beta} \mathcal{F}_{\alpha \beta}=
U_{\beta}\left( \dfrac{R_p A_B }{2}+ \dfrac{R_p^3 B_B}{2} \Delta_{\xi}  \right) \delta_{\alpha \beta}
\label{eq115}
\end{equation}
where
\begin{eqnarray}
\nonumber
A_B=\frac{3 R_p^2 \left(R_p \cosh \left(\frac{R_p}{\sqrt{k}}\right)-\sqrt{k} \sinh \left(\frac{R_p}{\sqrt{k}}\right)\right)}{6 k^{3/2} \sinh \left(\frac{R_p}{\sqrt{k}}\right)-6 k R_p \cosh \left(\frac{R_p}{\sqrt{k}}\right)-4 R_p^3 \cosh \left(\frac{R_p}{\sqrt{k}}\right)}
\\
[0.5cm]
\nonumber
B_B=\frac{6 k R_p \cosh \left(\frac{R_p}{\sqrt{k}}\right)+R_p^3 \cosh \left(\frac{R_p}{\sqrt{k}}\right)-6 k^{3/2} \sinh \left(\frac{R_p}{\sqrt{k}}\right)-3 \sqrt{k} R_p^2 \sinh \left(\frac{R_p}{\sqrt{k}}\right)}{12 k^{3/2} \sinh \left(\frac{R_p}{\sqrt{k}}\right)-12 k R_p \cosh \left(\frac{R_p}{\sqrt{k}}\right)-8 R_p^3 \cosh \left(\frac{R_p}{\sqrt{k}}\right)}
\end{eqnarray}
The same operator is identifiable in the Faxén theorem found by  
Padmavathi and al. \cite{pad,felderhof78}, according  to which the force on a Brinkman porous sphere with center at ${\pmb \xi}$ immersed in a generic ambient flow 
${\pmb u}({\pmb x})$ is, as expected,
\begin{equation}
F_\alpha= -M_\alpha =-8 \pi \mu 
\left( \dfrac{R_p A_B }{2}+ \dfrac{R_p^3 B_B}{2} \Delta_{\xi} \right)
u_\alpha ({\pmb \xi})
\label{eq116}
\end{equation}
In this case, in order to evaluate  the geometrical moments we 
need to determine the  surface traction and  the velocity at the boundary, since
\begin{eqnarray}
\nonumber
&& 
8 \pi \,
m_{\alpha {\pmb \alpha}_m \beta {\pmb \beta}_n}({\pmb \xi},{\pmb \xi})
 =  \int_{\partial D_b} 
  ({\pmb x}-{\pmb \xi})_{{\pmb  \alpha}_m} h_{ \alpha \beta {\pmb \beta}_n}({\pmb x},{\pmb \xi}) dS({\pmb x})
  -  \int_{\partial D_b} p_{\alpha {\pmb  \alpha}_m}({\pmb x},{\pmb \xi})
 v_{b \beta {\pmb \beta}_n}({\pmb x},{\pmb \xi}) n_b({\pmb x}) dS({\pmb x})
  \\
 && +
  \int_{\partial D_b} 
 \, 
 \left(
v_{\alpha \beta {\pmb \beta}_n}({\pmb x},{\pmb \xi}) n_b({\pmb x})\nabla_b ({\pmb x} - {\pmb \xi})_{{\pmb  \alpha}_m}
+
 n_\alpha({\pmb x}) v_{c \beta {\pmb \beta}_n}({\pmb x},{\pmb \xi}) 
\nabla_c ({\pmb x} - {\pmb \xi})_{{\pmb \alpha}_m}
\right)
 dS({\pmb x})
\label{eq117}
\end{eqnarray}

\section{ Faxén operator for a sphere with Navier-slip boundary conditions }
\label{sec8}

From  eqs. (\ref{eq38}), (\ref{eq41}) and (\ref{eq51}) it follows
that the  hydromechanics of a body in a Stokes fluid with 
Navier-slip boundary conditions
can be determined if the complete set of surface traction is known.
In this Section we develop  an analytic method,
based on the Lorentz reciprocal theorem,
for  determining the 
surface tractions  eq. (\ref{eq52}) 
entering eq. (\ref{eq51}), assuming  Navier-slip boundary conditions on the surface of a spherical object. 
{
The method, here developed for a sphere
with Navier-slip boundary conditions,
can be employed systematically for 
obtaining analytic expressions of $n$-th order Faxén operators of spheres with 
different BC-reciprocal boundary conditions.
To this aim
consider, in the remainder,
a Cartesian coordinate system for the point ${\pmb x}$
with  the origin at the center of the sphere. Also  the entries 
at the source point ${\pmb \xi}$ are expressed in the same Cartesian coordinate system, therefore there is no
substantial
distinction between Greek and Latin indexes.

\subsubsection{$0$-th order Faxén operator}
\label{secIX-1}
In order to determine the  
 $0$-th order
surface traction on a sphere moving in the unbounded Stokes fluid with velocity $-{\pmb U}$,
consider the
disturbance field ${\pmb w}^{(0)}({\pmb x},{\pmb \xi})={\pmb w}^{(0)}({\pmb x})$ due to a sphere with Navier-slip boundary conditions in a constant field ${\pmb u}^{(0)}({\pmb x},{\pmb \xi})={\pmb U}$,  which is
the  solution of the Stokes problem 
\begin{equation}
\begin{cases}
\mu\Delta w^{(0)}_a({\pmb x})-\nabla_a q^{(0)}({\pmb x})=0
\\
\nabla_a w^{(0)}_a({\pmb x})=0, \qquad {\pmb x} \in D_f
\\
w^{(0)}_a(\pmb x)=-U_b
\left(\delta_{a b}+{\lambda}h_{b c}({\pmb x}) t_{a c}({\pmb x}) \right), \qquad {\pmb x} \in \partial D_b 
\end{cases}
\label{eq53}
\end{equation}
and  the Stokeslet
$({\pmb S}_\alpha({\pmb x},{\pmb \xi}), \mu\, {\pmb \Sigma}_\alpha({\pmb x},{\pmb \xi}))$ in eqs. (\ref{eq6.1}).
Applying the Lorentz reciprocal theorem 
to the fields $({\pmb w}^{(0)}({\pmb x}),q^{(0)}({\pmb x}))$ solution of eqs. (\ref{eq53}),
and $({\pmb S}_\alpha({\pmb x},{\pmb \xi}), \mu {\pmb \Sigma}_\alpha({\pmb x},{\pmb \xi}))$
solution of the eqs. (\ref{eq6})
within  the domain of the fluid $D_f$, bounded by the surface $\partial D_b \cup \partial D_\infty$,
and considering that both  fields vanish at infinity
i.e. on $\partial D_\infty$,
 we have
\begin{equation}
[{\pmb w}^{(0)},{\pmb S}_\alpha({\pmb \xi})]=0
 \label{eq54}
\end{equation}
At the surface of the sphere, a Stokeslet
with pole at the center of the sphere (thus
for $r=({\pmb x}-{\pmb \xi})_a({\pmb x}-{\pmb \xi})_a=R_p$ and
 ${\pmb \xi}=(0,0,0)$) reads
\begin{eqnarray}
\nonumber
&& {
S_{a \alpha}({\pmb x},{\pmb \xi})
}
=
\dfrac{\delta_{a \alpha}+n_{a \alpha}({\pmb x})  }{R_p}
\\
[0.5cm]
&& \Sigma_{a b \alpha}({\pmb x},{\pmb \xi})n_b({\pmb x})=6
\dfrac{ \, n_{a \alpha}({\pmb x}) }{R_p^2}
, \qquad r=R_p
\label{eq55}
\end{eqnarray}
where $n_{a a_1 ... a_n}({\pmb x})=n_{a}({\pmb x})n_{a_1}({\pmb x})...n_{a_n}({\pmb x})$
and $n_a({\pmb x})=({\pmb x}-{\pmb \xi})_a/R_p$.

Substituting eqs. (\ref{eq55})  within
eq. (\ref{eq54}) and expliciting the $[\cdot,\cdot]$-operator in eq. (\ref{eq54})
 according  to the  definition eq. (\ref{eq26}),
we have  the following relation between integrals
 \begin{equation}
 -\dfrac{6}{R_p}\int_{r=R_p}
 n_{a \alpha}({\pmb x})
 dS({\pmb x})=
 \int_{r=R_p}
f_{a b}({\pmb x})
(  \delta_{b \alpha}+n_{b \alpha}({\pmb x})  )
 dS({\pmb x})
 \label{eq56}
\end{equation} 
where $f_{a b}({\pmb x})$ is the surface traction related to the disturbance
field  introduced in eq. (\ref{eq52}).
In order to satisfy eq.  (\ref{eq56}),
the surface traction $f_{a b}({\pmb x})$  should  have the generic form
\begin{equation}
f_{a b}({\pmb x})=  a\, \delta_{a b}+ b \, n_{a b}(\pmb x)
\label{eq57.0}
\end{equation}
where $a$ and $b$ are constant to be determined.
Alternatively, expressed in terms of normal and tangential components,
\begin{equation}
f_{a b}({\pmb x})=f^n \, n_{a b}(\pmb x)+f^t (\delta_{a b}-n_{a b}(\pmb x))
\label{eq57}
\end{equation}
where $f^n=a+b$ and $f^t= a$.

Substituting eq. (\ref{eq57}) into eq. (\ref{eq56}), a first relation between $f^n$ and $f^t$ is obtained
 \begin{equation}
 -\dfrac{6}{R_p}\int_{r=R_p}
 n_{a \alpha}({\pmb x})
 dS({\pmb x}) =
 2 f^n
\int_{r=R_p }
n_{a \alpha}(\pmb x)
 dS({\pmb x})
+
 f^t
 \int_{r=R_p }
 (\delta_{a \alpha}-n_{a \alpha}(\pmb x))
 dS({\pmb x})
 \label{eq58}
\end{equation} 
that, solving the  surface integrals,  attains the simple form
\begin{equation}
f^n+f^t=-\dfrac{3}{R_p}
\label{eq59}
\end{equation}
To determine $f^n$ and $f^t$, a further independent  
relation between $f^t$ and $f^n$ is required. To this aim,
we can consider 
the lowest order potential Stokes singularity, i.e. the so called
source doublet
$(-\Delta_\xi{\pmb S}_\alpha({\pmb x},{\pmb \xi})/2, -\mu \Delta_\xi{\pmb \Sigma}_\alpha({\pmb x},{\pmb \xi})/2)$
reported in the supplementary material \cite{suppl}. Also in this case, applying the Lorentz reciprocal theorem, we obtain
\begin{equation}
[{\pmb w}^{(0)},\Delta_\xi{\pmb S}_\alpha({\pmb \xi})]=0
 \label{eq60}
\end{equation}
where $\Delta_\xi$ is the Laplacian operator 
acting on  the coordinates of the pole.
At the surface of the sphere, the Source Doublet with pole at the center of the sphere is
\begin{eqnarray}
\nonumber
&& -\dfrac{
{\Delta_{\xi}\, S_{a \alpha}({\pmb x},{\pmb \xi})}
}{2}
=
\dfrac{-\delta_{a \alpha}+3\, n_{a \alpha}({\pmb x})  }{R_p^3}
\\
[0.5cm]
&&-\dfrac{
\left(
\Delta_{\xi}\, \Sigma_{a b \alpha}({\pmb x},{\pmb \xi})
\right)n_b({\pmb x})}{2}
=
6\, 
 \dfrac{-\delta_{a \alpha}+3\, n_{a \alpha}({\pmb x}) }{R_p^4}
, \qquad r=R_p
\label{eq61}
\end{eqnarray}
Consequently, the second relation for $f^n$ and $f^t$  stemming from eq. (\ref{eq60}) is 
  \begin{eqnarray}
 \nonumber 
 \dfrac{6}{R_p}\int_{r=R_p}(
\delta_{a \alpha}-3 n_{a \alpha}({\pmb x}) 
 )dS({\pmb x})
& + &
  \dfrac{6 \lambda f^t}{R_p}\int_{r=R_p}(
\delta_{a \alpha}- n_{a \alpha} ({\pmb x})
 )dS({\pmb x})
  = \\
 2 f^n
\int_{r=R_p }
n_{a \alpha}(\pmb x)
 dS({\pmb x})
&-&
 f^t
 \int_{r=R_p }
 (\delta_{a b}-n_{a b}(\pmb x))
 dS({\pmb x})
 \label{eq62}
\end{eqnarray} 
that, upon explicit integration, simplifies as
\begin{equation}
\left( 
6\hat{\lambda} + 1
\right)
f^t-f^n=0,\qquad \hat{\lambda}=\lambda/R_p
\label{eq63}
\end{equation}
The solution of the linear system eqs. (\ref{eq59}) and (\ref{eq63}) provides
\begin{eqnarray}
f^n=-\dfrac{3(1+6\hat{\lambda})}{2 R_p(1+3\hat{\lambda})}, \qquad
f^t=-\dfrac{3}{2 R_p(1+3\hat{\lambda})}
\label{eq64}
\end{eqnarray}
and, thus, the total $0$-th order geometrical surface traction is
\begin{equation}
h_{a b}(\pmb x)=
f_{a b}({\pmb x})=-\dfrac{3}{2 R_p}\left( 
\dfrac{\delta_{a b} +6 \hat{\lambda} n_{a b}(\pmb x)}{1+3\hat{\lambda}}
\right)
\label{eq65}
\end{equation}
In Appendix  \ref{appB}, we determine the analytical expression for the
geometrical moments $m_{\alpha {\pmb \alpha}_m \beta {\pmb \beta}_n}({\pmb \xi},{\pmb \xi})$
 useful 
to evaluate  the Faxén operators of the $0$-th,
$1$-st and $2$-nd order, by means of the surface traction obtained in this Section. 
To express the $0$-th order Faxén operator according the eq. (\ref{eq21}), we need  the geometrical moments for $n=0$.
Geometrical moments for
$n=0$ and $m=0,1,2$
are obtained in eqs. (\ref{eqB1})-(\ref{eqB4}). Due to the symmetry of the sphere,
the moments for $n=0$ and $m=1$ in eq. (\ref{eqB2}) vanish, and equivalently also  the 
moments for $n=0$ and $m=3,5,7, ...$. It is easy to see that the
 moments for $n=0$ and $m=4, 6, 8, ...$ contribute to the Faxén operator in eq. (\ref{eq58})
 by introducing terms proportional  either to the  divergence operators $\nabla_\beta$ and $\nabla_\alpha $ or  
to the bilaplacian operator $\Delta_\xi \Delta_\xi$. Since Stokes fields are both divergence free and biharmonic, their
action is immaterial. 
Therefore,
in agreement with the result obtained by Premlata and Wei \cite{premlata2021},
 the $0$-th order Faxén operator for a sphere
with Navier-slip boundary conditions is  
\begin{equation}
\mathcal{F}_{\beta \alpha }=
-
\left(
\dfrac{1+2 \hat{\lambda}}{1+ 3 \hat{\lambda}}
\right)
\left( 
\dfrac{3}{4}R_p  + \dfrac{1}{8} \dfrac{R_p^3}{(1+2  \hat{\lambda})} \Delta_\xi
\right)\delta_{ \alpha \beta }
\label{eq66}
\end{equation}
 where $\Delta_\xi$ is the Laplacian respect to the coordinates of the center of the sphere ${\pmb \xi}$.

Since the force ${F}_\alpha$ exerted by the fluid on the particle 
is the $0$-th order moment with reverse sign,  for a sphere immersed in the ambient flow ${\pmb u}({\pmb x})$ 
we have
\begin{equation}
F_\alpha=-M_\alpha({\pmb \xi})
=
8\pi\mu
\left(
\dfrac{1+2 \hat{\lambda}}{1+ 3 \hat{\lambda}}
\right)
\left( 
\dfrac{3}{4}R_p  + \dfrac{1}{8} \dfrac{R_p^3}{(1+2  \hat{\lambda})} \Delta_\xi 
\right)
u_\alpha({\pmb \xi})
\end{equation}

\subsubsection{$1$-st order Faxén operator }
In order to evaluate the surface traction on a sphere immersed in a $1$-st order ambient field, consider the 
disturbance field
\begin{equation}
\begin{cases}
\mu\Delta w_a^{(1)}({\pmb x},{\pmb \xi})-\nabla_a q^{(1)}({\pmb x},{\pmb \xi})=0
\\
\nabla_a w_a^{(1)}({\pmb x},{\pmb \xi})=0, \qquad {\pmb x} \in D_f
\\
w_a^{(1)}({\pmb x},{\pmb \xi})=-A_{b b_1}
\left(({\pmb x}-{\pmb \xi})_{b_1} \delta_{a b}+{\lambda}h_{c b b_1}({\pmb x},{\pmb \xi}) t_{a c}({\pmb x}) \right), \qquad {\pmb x} \in \partial D_b 
\end{cases}
\label{eq67}
\end{equation}
with ${\pmb \xi}=(0,0,0)$ at the center of the sphere. From eq. (\ref{eq52}) we have
\begin{equation}
h_{a b { b}_1}({\pmb x})=f_{a b {b}_1}({\pmb x})
-
\left(
n_{b}({\pmb x})\, 
\delta_{a b_1} 
+
\delta_{a b}\, n_{b_1}({\pmb x})
\right)
\label{eq68}
\end{equation}
The procedure to obtain
$ f_{a b {b}_1}({\pmb x})$ (and thus
 $ h_{a b { b}_1}({\pmb x})$) is equivalent to that followed
in eqs. (\ref{eq54})-(\ref{eq65}) and is reported in detail in the supplementary materials \cite{suppl}. In this case, the most general form for $ f_{a b b_1}({\pmb x}) $
in order to satisfy reciprocity relations with singularities centered at the center of the sphere is
\begin{equation}
f_{a b b_1}({\pmb x})=
a\, \delta_{a b} n_{b_1}({\pmb x})
+b\,  \delta_{a b_1} n_b({\pmb x})
+c\, n_{a b b_1}({\pmb x})
+ d\, \delta_{b b_1} n_{ a1 }({\pmb x})
\label{eq72}
\end{equation}
where $a$, $b$, $c$ and $d$ are constant to be determined.
Since,  according  to the definition  eq. (\ref{eq11}), $A_{b b_1}=0$ due to the imcompressibility of the ambient 
flow,
the last term in eq. (\ref{eq72}) does not contribute 
to the surface traction, and we can set $d=0$.
If we apply the Lorentz reciprocal theorem to the 
solution $({\pmb w}^{(1)}({\pmb x},{\pmb \xi}),{\pmb \tau}^{(1)}({\pmb x},{\pmb \xi}))$ of eqs. (\ref{eq67}) 
and to the Stokeslet
or the Source doublet, 
as in the previous paragraph,
we obtain that all the integrals on the surface of the sphere vanish due to the spherical symmetry.
For this reason,  
it is necessary to
 consider the Stokes doublet
 $(\nabla_\beta {\pmb S}_\alpha({\pmb x},{\pmb \xi}), \mu \nabla_\beta {\pmb \Sigma}_\alpha({\pmb x},{\pmb \xi}))$
 with pole at the center of the sphere, and therefore
\begin{equation}
[{\pmb w}^{(1)}({\pmb \xi}),\nabla_\beta {\pmb S}_\alpha({\pmb \xi})]=0
\label{eq70}
\end{equation}
from which one obtains a system of two linear equations in the coefficients
$a$, $b$, $c$,
\begin{equation}
\begin{cases}
 -(8+6 \hat{\lambda }  )a + (2+24 \hat{\lambda } ) b -3c=24+18\hat{\lambda } 
\\
(2+24 \hat{\lambda } ) a - (8+6 \hat{\lambda }  ) b -3c=-6+18\hat{\lambda } 
\end{cases}
\label{eq75}
\end{equation}
In order to solve the system we need another linearly independent equation. To this aim, it is possible to apply the Lorentz reciprocal theorem between the 
Source Quadrupole  
$( \Delta_{\xi} \nabla_\beta {\pmb S}_\alpha({\pmb x},{\pmb \xi}), \mu\, \Delta_{\xi}  \nabla_\beta {\pmb \Sigma}_\alpha({\pmb x},{\pmb \xi}))$
and $({\pmb w}^{(1)}({\pmb x}),{\pmb \tau^{(1)}}({\pmb x}))$
.
Thus, applying the relation
\begin{equation}
[{\pmb w}^{(1)}({\pmb \xi}), \Delta_\xi \nabla_\beta {\pmb S}_\alpha({\pmb \xi})]=0
\label{eq77}
\end{equation}
 one obtains another independent relation amongst $a$, $b$ and $c$, 
\begin{equation}
c-8\hat{\lambda}(a+b)=-16\hat{\lambda}
\label{eq79}
\end{equation}
Solving eqs. (\ref{eq75}) and (\ref{eq79}) one finally gets
\begin{eqnarray}
a =\frac{-3-7 \hat{\lambda }  +15 \hat{\lambda }  ^2}{(1+5 \hat{\lambda } )(1+3\hat{\lambda } )}, \qquad
b = \frac{\hat{\lambda }   (8+15 \hat{\lambda } )}{(1+5 \hat{\lambda } )(1+3\hat{\lambda } )}, \qquad
c=-\frac{40 \hat{\lambda }  }{1+5 \hat{\lambda } }
\label{eq80}
\end{eqnarray}
The surface tensor in eq. (\ref{eq52})
is obtained by substituting the constant defined by eq. (\ref{eq80}) into eq. (\ref{eq72}),
\begin{equation}
h_{a b b_1}({\pmb x})=
-\dfrac{
(4+15\hat{\lambda } )\, \delta_{a b} n_{b_1}({\pmb x})
+\,  \delta_{a b_1} n_b({\pmb x})
+40\hat{\lambda }  (1+3 \hat{\lambda } )\, n_{a b b_1}({\pmb x})
}{ (1+5 \hat{\lambda } )(1+3\hat{\lambda } )}
\label{eq81}
\end{equation}
To determine the Faxén operator $\mathcal{F}_{\alpha \beta \beta_1}$, the 
moments $m_{\alpha {\pmb \alpha}_m \beta {\pmb \beta}_n}({\pmb \xi},{\pmb \xi})$ at the center of the sphere for $n=1$ 
should be evaluated  using the surface traction defined by eq. (\ref{eq81}).
Since geometrical moments  with even $m=0,2, ...$ vanish due to the symmetry of the sphere, only the geometrical
moments for odd $m=1, 3, ...$ are needed. Geometrical moments for $m=1$ and $m=3$ are 
given in eq. (\ref{eqB5}) and eq. (\ref{eqB6}) respectively. 
Higher order geometrical moments contribute to the Faxén operator by divergence and bilaplacian operators and
thus, they can be neglected as their action is immaterial. Thus,
according to eq. (\ref{eq58}),
 the $1$-st order Faxén operator is given by
\begin{eqnarray}
\nonumber
&&
\mathcal{F}_{\alpha \beta \beta_1 }=
-\dfrac{R_p^3}{6(1+5  \hat{\lambda})(1+3 \hat{\lambda})}
\bigg\{
\left[
(4 +20  \hat{\lambda} +15  \hat{\lambda}^2)
\delta_{\alpha \beta} \nabla_{\beta_1}
+
(1 +5 \hat{\lambda} +15  \hat{\lambda}^2)
\delta_{\alpha \beta_1} \nabla_{\beta}
\right]
\\
&&
+\dfrac{R_p^2}{10}
\left[
(4 +12  \hat{\lambda} - 15  \hat{\lambda}^2)
\Delta_\xi \nabla_{\beta_1}\delta_{\alpha \beta}
+
(1 +3 \hat{\lambda} +15  \hat{\lambda}^2)
\Delta_{{\xi}} \nabla_{\beta}\delta_{\alpha \beta_1}
\right]
\bigg\}
\label{eq82}
\end{eqnarray}
The Faxén operator
$\mathcal{T}_{\gamma \alpha}$,
 yielding the torque
 $T_\alpha= \varepsilon_{\alpha \beta \beta_1}M_{\beta \beta_1}({\pmb \xi})$
  on the body 
if  an ambient flow is applied, 
or the velocity field due to the body rotation if applied to the Stokeslet, can be defined by
the antisymmetric part of the first order Faxén operator, i.e., 
\begin{equation}
\mathcal{T}_{\gamma \alpha}=\varepsilon_{\alpha \beta  \beta_1} \mathcal{F}_{\gamma \beta \beta_1}
\label{eq83}
\end{equation}
Considering that  $\varepsilon_{\gamma \alpha \alpha_1 } \Delta \nabla_{\alpha_1} u_\alpha({\pmb x})=0$ for any ambient flow,
terms containing the third order derivatives
 in eq. (\ref{eq82}) 
are immaterial and may be neglected, so we obtain 
\begin{equation}
\mathcal{T}_{\gamma \alpha }=
\dfrac{\varepsilon_{\alpha \gamma \gamma_1} R_p^3 \nabla_{\gamma_1}}{2(1+3  \hat{\lambda})}
\label{eq84}
\end{equation}
 in agreement with \cite{premlata2021}.
 
Thus, the torque on the sphere in the ambient flow $u_\alpha({\pmb x})$ is give by
\begin{equation}
T_{\alpha }=
\dfrac{4 \pi \mu \, \varepsilon_{\alpha \gamma \gamma_1} R_p^3 \nabla_{\gamma_1}
u_\gamma({\pmb \xi})
}{(1+3  \hat{\lambda})}
\label{eq84a}
\end{equation}

{
As shown in Appendix \ref{appC}, 
the off-diagonal elements of the symmetric part of $M_{\beta \beta_1}({\pmb \xi})$  defined by  eq. (\ref{eq7})
represent, with reversed sign,
the  Stresslet
$\mathsf{S}_{\alpha \alpha_1}=-\gamma_{\alpha \alpha_1 \beta \beta_1}M_{\beta \beta_1}({\pmb \xi})$,
where
\begin{equation}
\nonumber
\gamma_{\alpha \alpha_1 \beta \beta_1}=\dfrac{
\delta_{\alpha \beta} \delta_{\alpha_1 \beta_1}+
\delta_{\alpha \beta_1} \delta_{\alpha_1 \beta}}{2}-
\dfrac{\delta_{\alpha \alpha_1} \delta_{\beta \beta_1}}{3}
\end{equation}
which physically corresponds to
the additional contribution of any single particle to the effective stress of a diluted suspension \cite{batchelor-green_b,kim-karrila,premlata2022}.

Therefore, the associated Faxén operator $ \mathcal{E}_{\alpha \beta \beta_1}$ providing the Stresslet on the body immersed in an ambient flow ${\pmb u}({\pmb x})$ according to
\begin{equation}
 \mathsf{S}_{\beta \beta_1}
 = 8\pi \mu \mathcal{E}_{\alpha \beta \beta_1} u_\alpha({\pmb \xi})
\end{equation}
is generically expressed by
\begin{equation}
\mathcal{E}_{\alpha \beta \beta_1} =-
\dfrac{\mathcal{F}_{\alpha \beta \beta_1}+\mathcal{F}_{\alpha \beta_1 \beta}}{2}+\dfrac{\mathcal{F}_{\alpha \gamma \gamma} \delta_{\beta \beta_1}}{3}
\end{equation}
 Since,  $\mathcal{F}_{\alpha \beta \beta}u_\alpha({\pmb x})$ in eq. (\ref{eq82}) vanishes due to the incompressibility, in the case of a spherical body with Navier-slip boundary conditions, the Faxén operator giving the Stresslet contribution is simply the
 symmetric part of the $1$-st order Faxén operator with opposite sign. Therefore,  in agreement with \cite{premlata2022}, we obtain
\begin{equation}
 \mathcal{E}_{\alpha \beta \beta_1}
 =
-\dfrac{\mathcal{F}_{\alpha \beta \beta_1}+\mathcal{F}_{\alpha \beta_1 \beta}}{2}
=\left( 
\dfrac{5+10 \hat{\lambda}}{6+30 \hat{\lambda}}+
\dfrac{\Delta_\xi }{12(1+5  \hat{\lambda})}
\right)
\left(
\dfrac{\nabla_\beta \delta_{ \alpha \beta_1 }+\nabla_\beta \delta_{\alpha \beta_1}}{2}
\right)
\label{eq85}
\end{equation} 
}

According  to eq. (\ref{eq22}), the Faxén operator in  eq. (\ref{eq82}) determines the flow around a sphere in the linear ambient
flow $u_a({\pmb x})=\delta_{ a\, 3 }\,\delta_{ b\, 1 }\, x_{b}=\delta_{ a\, 3 }\, x_1$. In fact, by choosing
$A_{\beta \beta_1}= \delta_{ \beta\, 3 } \delta_{ \beta_1\, 1 }$ in eq. (\ref{eq22})
the disturbance field reads
\begin{eqnarray}
\nonumber
&& w^{(1)}_a({\pmb x},{\pmb \xi})=
-\dfrac{R_p^3}{6(1+5 \hat{\lambda})(1+3 \hat{\lambda})} \left\{
(4+20  \hat{\lambda} + 15  \hat{\lambda}^2 )S_{a\, 3, 1}({\pmb x},{\pmb \xi})
+
(1+5 \hat{\lambda} + 15  \hat{\lambda}^2 )S_{a\, 1, 3}({\pmb x},{\pmb \xi})
\right.
\\
&& \left.
\dfrac{R_p^2}{10}
\left[
(4+12 \hat{\lambda} - 15  \hat{\lambda}^2 )\Delta_\xi S_{a\, 3, 1}({\pmb x},{\pmb \xi})
+
(1+3 \hat{\lambda} + 15  \hat{\lambda}^2 )
\Delta_\xi S_{a\, 1, 3}({\pmb x},{\pmb \xi})
\right]
\right\}
\end{eqnarray}
Fig. \ref{xfig1}  depicts the streamlines for the total
flow in the case of 
 no-slip, complete slip,  and for two different values of  
$\hat{\lambda}=1,\,  10$. 

\begin{figure}
\centering
\includegraphics[scale=0.25]{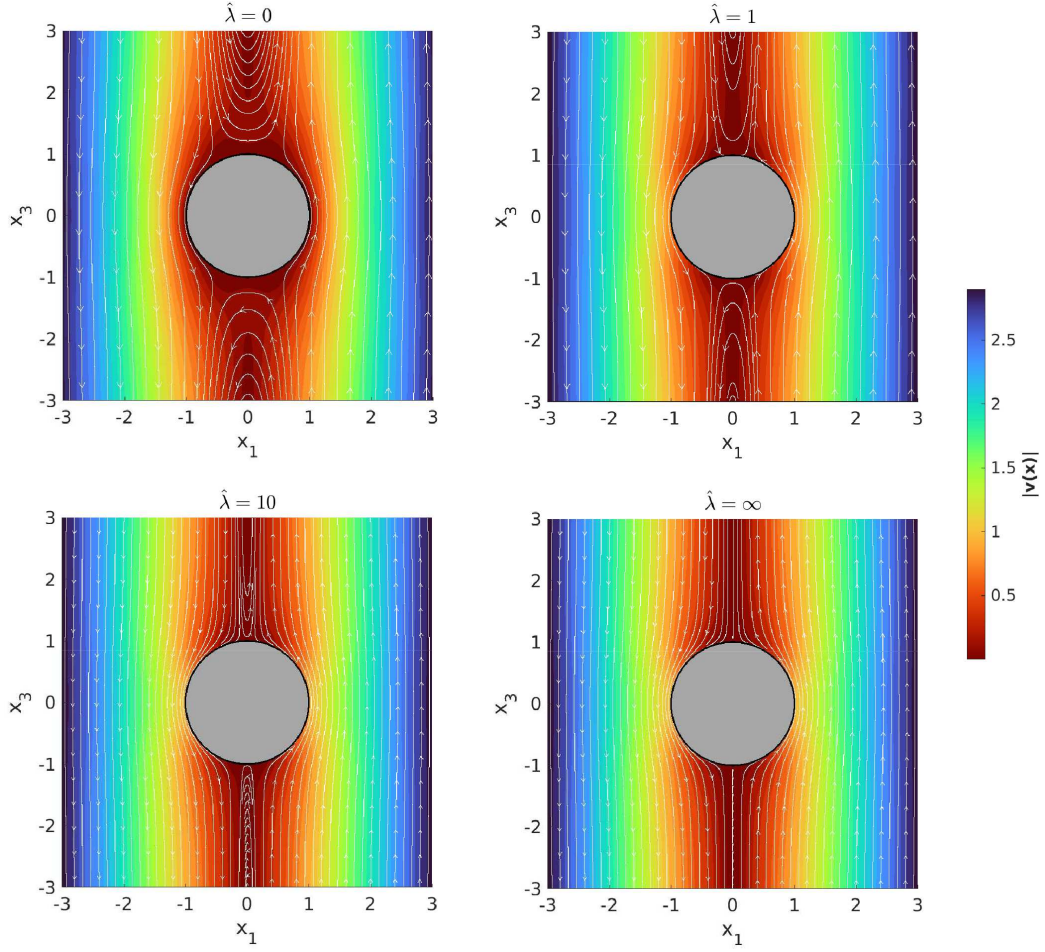}
\caption{Streamlines on the plane $x_2=0$ of the fluid around a sphere
in a linear ambient flow $u_a({\pmb x})=\delta_{ a\, 3 }\, x_1$ for different values of dimensionless slip length at the surface of the sphere.
}
\label{xfig1}
\end{figure}

\subsubsection{$2$-nd order Faxén operator }
In order to  evaluate the surface traction on a sphere immersed in a $2$-nd order ambient field, consider the disturbance field 
\begin{equation}
\begin{cases}
\mu\Delta w_a^{(2)}({\pmb x},{\pmb \xi})-\nabla_a q^{(2)}({\pmb x},{\pmb \xi})=0
\\
\nabla_a w_a^{(2)}({\pmb x},{\pmb \xi})=0, \qquad {\pmb x} \in D_f
\\
w_a^{(2)}({\pmb x},{\pmb \xi})=-A_{b b_1 b_2}
\left(
({\pmb x}-{\pmb \xi})_{b_1 b_2} \delta_{a b}+{\lambda}h_{c b b_1 b_2}({\pmb x},{\pmb \xi}) t_{a c}({\pmb x}) \right), \qquad {\pmb x} \in \partial D_b 
\end{cases}
\label{eq86}
\end{equation}
In this case, the ambient pressure is $p_{b b_1 b_2}({\pmb x},{\pmb \xi})=({\pmb x} - {\pmb \xi})_b \Delta ({\pmb x}- {\pmb \xi})_{b_1 b_2}$. Therefore, from eq. (\ref{eq52}) with ${\pmb \xi}=(0,0,0)$ we can set 
\begin{equation}
h_{a b  b_1 b_2}({\pmb x})=f_{a b { b}_1 b_2}({\pmb x})
-R_p
\left(
n_{b b_1}({\pmb x}) \delta_{a b_2} 
+n_{b b_2}({\pmb x}) \delta_{a b_1} 
+2(\delta_{a b}n_{b_1 b_2}({\pmb x})
-\delta_{b_1 b_2}n_{a b}({\pmb x}))
\right)
\label{eq87}
\end{equation}
Considering that $ f_{a b { b}_1 b_2}({\pmb x}) $ must be symmetric with respect  to the indexes $b_1$ and $b_2$, and that 
$A_{b b}=0$, the most general form  for this traction is
\begin{equation}
f_{a b { b}_1 b_2}({\pmb x})=
a\, \delta_{a b} n_{b_1 b_2} ({\pmb x})+
b\, \delta_{a b} \delta_{b_1 b_2}+
c\, (\delta_{a b_1} n_{b b_2}({\pmb x})+ \delta_{a b_2} n_{b b_1}({\pmb x}))+
d\, \delta_{b_1 b_2} n_{a b}({\pmb x}) +
e\, n_{a b b_1 b_2}({\pmb x})
\label{eq88}
\end{equation}
Following the same procedure used for  the geometrical surface tractions of lower orders, we consider the Stokes Quadrupole
 $(\nabla_\gamma \nabla_\beta {\pmb S}_\alpha({\pmb x},{\pmb \xi}), \mu \nabla_\gamma  \nabla_\beta {\pmb \Sigma}_\alpha({\pmb x},{\pmb \xi}))$.
Applying the Lorentz reciprocal theorem, we have
\begin{equation}
[{\pmb w}^{(2)}({\pmb \xi}),\nabla_\gamma \nabla_\beta {\pmb S}_\alpha({\pmb \xi})]=0
\label{eq90}
\end{equation}
Solving the integrals
{(the detailed procedure is reported in the supplementary material \cite{suppl})}, we obtain four linear equations for  the five unknown
\begin{equation}
\begin{cases}
-2 (3+4) a
+2(1+20 \hat{\lambda}) c 
-2 e=(25+24 \hat{\lambda}) R_p 
\\
(-5 +12 \hat{\lambda})a 
+84 \hat{\lambda} b
- 3 (1-8  \hat{\lambda}) c 
-14 d 
-4 e = 
3 (5+16\hat{\lambda})R_p
\\
(1+20 \hat{\lambda}) a 
-(5-12 \hat{\lambda}) c 
- 2 e=
-(3-52 \hat{\lambda}) R_p
\\
-4(1+6 \hat{\lambda}) a 
+42 \hat{\lambda} b 
+ 6 (1 -\hat{\lambda})c
-7 d+ 
e=
6 (2-9 \hat{\lambda}) R_p 
\label{eq92}
\end{cases}
\end{equation}
out of which  only three are linearly independent. In fact, by
summing the fourth equation multiplied by $2$ to the third equation multiplied by $3$, we obtain the second equation.
To obtain a further equation, we 
consider the Source Hexapole 
$(\Delta_\xi \nabla_\gamma \nabla_\beta {\pmb S}_\alpha({\pmb x},{\pmb \xi}), \mu \Delta_\xi \nabla_\gamma  \nabla_\beta {\pmb \Sigma}_\alpha({\pmb x},{\pmb \xi}))$.
As in the previous cases, the application of the the Lorentz reciprocal theorem provides
\begin{equation}
[{\pmb w}^{(2)}({\pmb \xi}),\Delta_\xi \nabla_\gamma \nabla_\beta {\pmb S}_\alpha({\pmb \xi})]=0
\label{eq94}
\end{equation}
from which it follows that
\begin{equation}
-10  \hat{\lambda} a - 20  \hat{\lambda} c +e = - 40  \hat{\lambda}  
\label{eq96}
\end{equation}
We need another equation, linearly independent of eqs. (\ref{eq96}) and of the three linearly independent
eqs. (\ref{eq92}) that can obtained by applying the Lorentz reciprocal theorem to the Stokeslet
\begin{equation}
[{\pmb w}^{(2)}({\pmb \xi}), {\pmb S}_\alpha({\pmb \xi})]=0
\label{eq97}
\end{equation}
resulting the relation
\begin{equation}
3 a +10 b+ c+5 d +e=-3 R_p
\label{eq99}
\end{equation}
By solving the linear system formed by the first,  third, and fourth equations in  eq. (\ref{eq92}), 
eq. (\ref{eq96}) and eq. (\ref{eq99}), that possesses a non-vanishing determinant for any $ \hat{\lambda} \geq 0$, 
the constant $a$, $b$, $c$, $d$, $e$ entering 
eq. (\ref{eq88}) are determined
\begin{eqnarray}
\nonumber
\dfrac{a}{R_p}=
\dfrac{-17 -52  \hat{\lambda} +244  \hat{\lambda}^2}{4(1+4 \hat{\lambda})(1+ 7 \hat{\lambda})}\qquad 
&&
\dfrac{b}{ R_p}=
\dfrac{3+17  \hat{\lambda}}{
4(1+3 \hat{\lambda})(1+4 \hat{\lambda})(1+ 7 \hat{\lambda})
}
\qquad
\dfrac{e}{R_p}=-
\dfrac{175  \hat{\lambda}}{2+14  \hat{\lambda}}
\\
[10pt]
\dfrac{c}{R_p}=
\dfrac{-1 +44 \hat{\lambda}+ 112  \hat{\lambda}^2}{4(1+4 \hat{\lambda})(1+ 7 \hat{\lambda})}\qquad 
&&
\dfrac{d}{R_p}=
\dfrac{1+28 \hat{\lambda}+127 \hat{\lambda}^2+84  \hat{\lambda}^3}{
2(1+3 \hat{\lambda})(1+4 \hat{\lambda})(1+ 7 \hat{\lambda})
}
\label{eq100}
\end{eqnarray}
and therefore, the geometrical surface traction in eq. (\ref{eq87}) is given by
\begin{eqnarray}
\nonumber
&&
h_{a b { b}_1 b_2}({\pmb x})=\dfrac{R_p}{
4(1+4 \hat{\lambda})(1+ 7 \hat{\lambda})
}
\left[
-5(5+28  \hat{\lambda}) \delta_{a b} n_{b_1 b_2} ({\pmb x})
\right.
\\
\nonumber
&+&
\left.
\dfrac{3+17  \hat{\lambda}}{1+3  \hat{\lambda}}\, \delta_{a b} \delta_{b_1 b_2}
-5
\, (\delta_{a b_1} n_{b b_2}({\pmb x})+ \delta_{a b_2} n_{b b_1}({\pmb x}))
\right.
\\
&+&
\left.
\dfrac{2(5+84 \lambda +371 \lambda ^2+420 \lambda ^3)}
{1+3  \hat{\lambda}}
\, \delta_{b_1 b_2} n_{a b}({\pmb x}) 
-350  \hat{\lambda} (1+4  \hat{\lambda})
\, n_{a b b_1 b_2}({\pmb x}))
\right]
\label{eq101}
\end{eqnarray}
through which it is possible  to obtain the
geometrical moments at the center of the sphere $m_{\alpha \beta {\pmb \beta}_2}({\pmb \xi},{\pmb \xi})$, $m_{\alpha {\pmb \alpha}_2 \beta {\pmb \beta}_2}({\pmb \xi},{\pmb \xi})$, $m_{\alpha {\pmb \alpha}_4 \beta {\pmb \beta}_2}({\pmb \xi},{\pmb \xi})$.
Their analytic expression is reported  in Appendix \ref{appB},  eqs. 
(\ref{eqB7}), (\ref{eqB8}), (\ref{eqB9}). Using these results it is possible to express analytically the  $2$-nd 
order Faxén operator 
\begin{eqnarray}
\nonumber
&&\mathcal{F}_{\alpha \beta \beta_1 \beta_2}=
-\dfrac{R_p^3}{4(1+4 \hat{\lambda})(1+7 \hat{\lambda})}
\left\{
\dfrac{(1+4  \hat{\lambda})(1+7 \hat{\lambda})}{1+3 \hat{\lambda}}
\left[
\delta_{\alpha \beta} \delta_{\beta_1 \beta_2}
+
\hat{\lambda}(\delta_{\beta \beta_1}\delta_{\alpha \beta_2}
+
\delta_{\beta\beta_2} \delta_{\alpha \beta_1} 
)
\right]
\right.
\\
\nonumber
&&
+\dfrac{R_p^2}{6}
\bigg[
-4 \hat{\lambda}^2 \left( \dfrac{4+21 \hat{\lambda}}{1+3 \hat{\lambda}} \right)\Delta_\xi \delta_{\beta_1 \beta_2} \delta_{ \alpha \beta }
+
5(1+6 \hat{\lambda})\nabla_{\beta_1 \beta_2}\delta_{ \alpha \beta }+
\\
\nonumber
&&
\left.
(1+6 \hat{\lambda} + 28 \hat{\lambda}^2)(\nabla_{\beta \beta_1}\delta_{\alpha \beta_2}
+
\nabla_{\beta \beta_2} \delta_{\alpha \beta_1}
)+
(1+12 \hat{\lambda}+56 \hat{\lambda}^2)(
\delta_{ \alpha \beta_1 }\delta_{ \beta \beta_2 }
+\delta_{ \alpha \beta_2 } \delta_{ \beta \beta_1 }
)
\Delta_\xi
\bigg]
\right.
\\
&&
\left. 
+\dfrac{R_p^4}{84}
\left[
(5+20 \hat{\lambda}-56 \hat{\lambda}^2)\nabla_{\beta_1 \beta_2}\delta_{ \alpha \beta }
+(1+4 \hat{\lambda}+28  \hat{\lambda}^2)(\nabla_{\beta \beta_1}\delta_{\alpha \beta_2}
+
\nabla_{\beta \beta_2}\delta_{\alpha \beta_1}
)
\right]\Delta_\xi
\right\}
\label{eq102}
\end{eqnarray}
The Faxén operator in  eq. (\ref{eq102}) can be used to obtain the flow around a sphere in the unbounded Poiseuille ambient
flow $$ u^{(2)}_a({\pmb x})=\delta_{ a\, 3 }(\delta_{ a_1 a_2 }-\delta_{ a_1\, 3}\delta_{ a_2\, 3 })\, ({\pmb x}-{\pmb \xi})_{a_1 a_2}=\delta_{ a\, 3 }\, (({\pmb x}-{\pmb \xi})_1^2+({\pmb x}-{\pmb \xi})_{2}^2)$$ ${\pmb \xi}$ being the center of the sphere. 
Choosing
$A_{\beta \beta_1 \beta_2}= 
\delta_{ \beta\, 3 }(\delta_{ \beta_1 \beta_2 }-\delta_{ \beta_1\, 3}\delta_{ \beta_2\, 3 })
$ in eq. (\ref{eq22})
the disturbance field reads
\begin{eqnarray}
\nonumber
&& w^{(2)}_a({\pmb x},{\pmb \xi})=
-\dfrac{R_p^3}{24(1+7\hat{\lambda})} \left\{
12\left(\dfrac{
1+7 \hat{\lambda}}{1+3 \hat{\lambda}}
\right)
S_{a\, 3}({\pmb x},{\pmb \xi}) +
\right.
\\
\nonumber
&& \left.
{R_p^2}
\left[\left(
\dfrac{5+25 \hat{\lambda}-42 \hat{\lambda}^2}{1+3 \hat{\lambda}}
\right)
\Delta_\xi S_{a\, 3}({\pmb x},{\pmb \xi})
-
7(1+2 \hat{\lambda})
 S_{a\, 3, 3 \, 3}({\pmb x},{\pmb \xi})
\right]-R_p^4\dfrac{
\Delta_\xi S_{a\, 3, 3 \, 3}({\pmb x},{\pmb \xi})}{2}
\right\}\\
\label{eq102.1}
\end{eqnarray}
The streamlines obtained 
using  eq. (\ref{eq102.1}) in the case of no-slip, complete slip, Navier-slip with 
$\hat{\lambda}=1$ and  $ \hat{\lambda}=10$ are reported in Fig. \ref{xfig2}.

\begin{figure}[h!]
\centering
\includegraphics[scale=0.2]{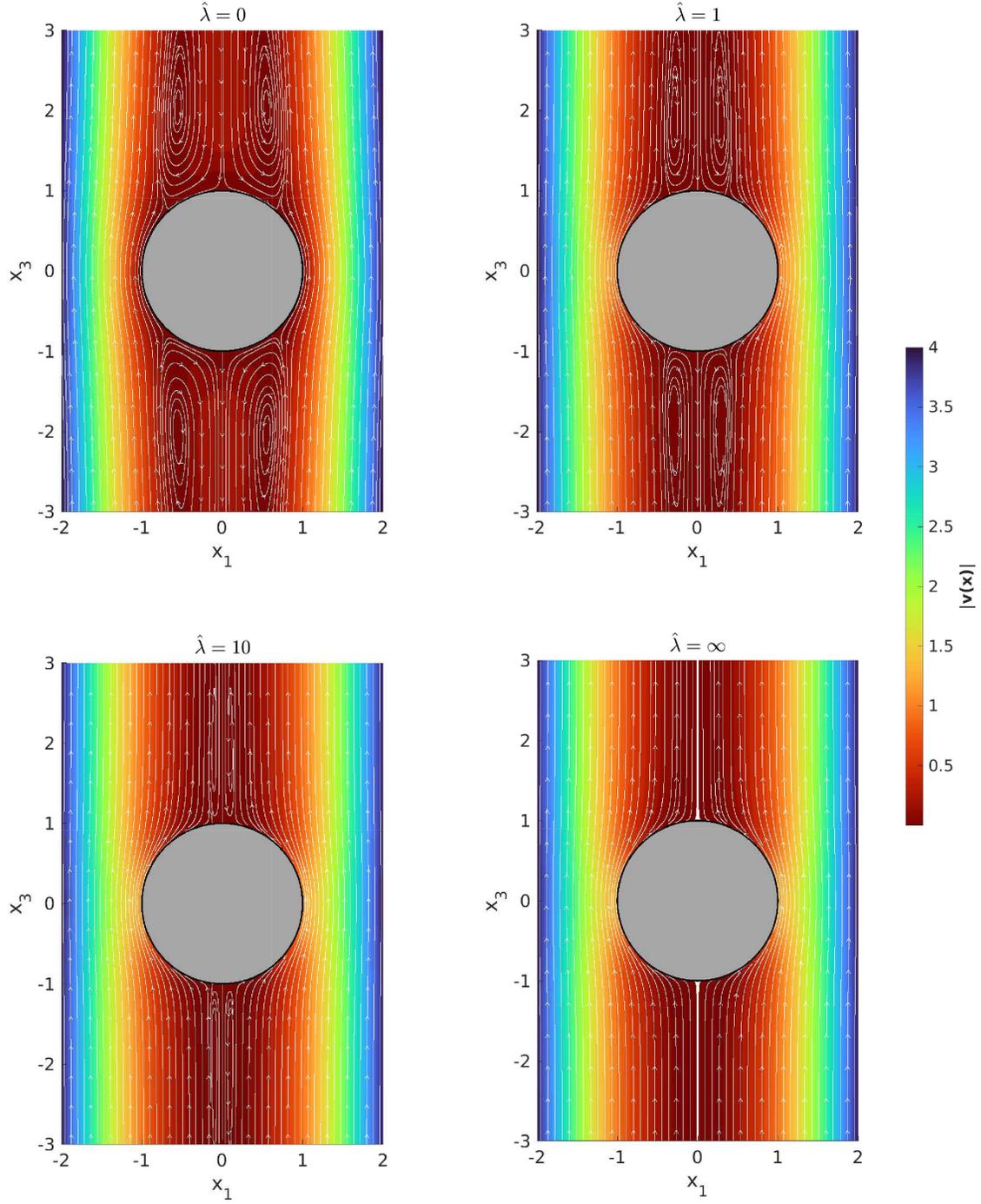}
\caption{Streamlines on the plane $x_2=0$ of the fluid around a sphere
in the unbounded Poiseuille ambient flow $u_a({\pmb x})=\delta_{ a\, 3 }\, (x_1^2+x_2^2)$ for different values of dimensionless slip length, at the surface of the sphere.}
\label{xfig2}
\end{figure}

\newpage

\section{Conclusions}
This work has investigated the operators 
describing the hydromechanics of bodies immersed in Stokes flow:
the singularity operator giving the disturbance
flow due to the body immersed in the fluid and the Faxén operator providing the moments on the body
due to the fluid flowing around it.

By considering moments of the volume forces on the
body, we provided a general expression for the 
singularity operator. Specifically, a generic
$n$-th order singularity operator can be obtained by defining ($m$, $n$)-th order geometrical moments
 as the $m$-th order moments of volume forces 
 on the body immersed in an $n$-th order ambient 
 flow singular at infinity.
{ Furthermore, it is found that
the same operator is a Faxén operator only if the  boundary  conditions assumed at the surface of the body are BC-reciprocal.} This shows that the Hinch-Kim's dualism
is not an intrinsic property of the Stokes flow
 due to the reciprocity of its governing equations,
 but it depends also on the nature
 of the interaction with the body. This is not
 a trivial conclusion, since other properties
 following from the reciprocity between
 thermodynamic forces and fluxes of the governing
 equations, such as the symmetry of the resistance matrix (following from the reciprocity of the Stokes equations and, firstly, from that of the Onsager relations \cite{dau5,dau6}), are independent
 on the nature of the interaction with the body
(this is also shown in Appendix \ref{appA} for the resistance matrix of the Stokes flow). 

{The results reported in Section \ref{sec4.1} show that bodies for which the Hinch-Kim's dualism holds are an interesting class of bodies
since their hydromechanic interaction with the fluid can be completely described 
by collecting all the associated geometrical moments.}
This makes this analysis a fundamental tool for the development of a theory describing the hydrodynamic interactions between many bodies in confined geometries starting from the knowledge of the hydrodynamics of single bodies in unbounded flows and of the confinement Green function \cite{pg_arxiv}.
{The strength} of this method is that the operators have an explicit expression that allow us to investigate case by case the dualism, so we find that Hinch-Kim's dualism
 hold only for particular (but of great interest) boundary conditions (such as for rigid particles,
drop at the mechanical equilibrium or porous bodies
modeled by the Birkman equations), but it does not hold for other systems (elastic deforming body, deforming drop, non-Newtonian drops, porous bodies modeled by the Darcy law).
It is interesting that in the case where Navier-slip boundary conditions are considered (for which we found in Section
\ref{secIV-A} that Hinch-Kim's dualism holds)
 the $n$-th order Faxén operator is determined
 by the $n$-th order surface traction on the body
(i.e. the surface traction on the body immersed in an $n$-th order ambient field). Therefore, by using 
a method based on the
Lorentz's reciprocal theorem, 
we have obtain the $0$-th, $1$-st and $2$-nd order surface traction on a sphere with Navier-slip boundary condition
and, then, { applying the expressions developed by enforcing the representation in terms of volume geometrical moments}, the $0$-th, $1$-st, $2$-nd
Faxén operator giving both the moments, up to the second order, on the sphere in a generic ambient flow and the disturbance field due to the sphere in an ambient flow up to the $2$-nd order, such as the Poiseuille flow.

\appendix

\section{Symmetry of  the geometrical moments}
\label{appA}
To show the symmetry in eq. (\ref{eq40}), if BC-reciprocity applies,
 consider the moments $ M^{(n)}_{\alpha {\pmb \alpha}_m}({\pmb \xi},{\pmb \xi}') $ 
and their expression eq. (\ref{eq17}). Using the notation 
developed in Section \ref{sec4}
\begin{equation}
A_{\alpha {\pmb \alpha}_m} M^{(n)}_{\alpha {\pmb \alpha}_m}({\pmb \xi},{\pmb \xi}')=[{\pmb u}^{(m)}({\pmb \xi}),{\pmb v}^{(n)}({\pmb \xi}')]
\label{eqA1}
\end{equation}
and analogously
\begin{equation}
A_{\beta' {\pmb \beta}'_n} M^{(m)}_{\beta' {\pmb \beta}'_n}({\pmb \xi}',{\pmb \xi})=[{\pmb u}^{(n)}({\pmb \xi}'),{\pmb v}^{(m)}({\pmb \xi})]
\label{eqA2}
\end{equation}
Therefore, their difference can be  expressed  as
\begin{eqnarray}
\nonumber
&& A_{\alpha {\pmb \alpha}_m} M^{(n)}_{\alpha {\pmb \alpha}_m}({\pmb \xi},{\pmb \xi}')
- 
A_{\beta' {\pmb \beta}'_n} M^{(m)}_{\beta' {\pmb \beta}'_n}({\pmb \xi}',{\pmb \xi})
=
[{\pmb u}^{(m)}({\pmb \xi}),{\pmb v}^{(n)}({\pmb \xi}')]
-
[{\pmb u}^{(n)}({\pmb \xi}'),{\pmb v}^{(m)}({\pmb \xi})]  \nonumber \\
&& =
[{\pmb u}^{(m)}({\pmb \xi}),{\pmb u}^{(n)}({\pmb \xi}')]
+
[{\pmb u}^{(m)}({\pmb \xi}),{\pmb w}^{(n)}({\pmb \xi}')] 
-
[{\pmb u}^{(n)}({\pmb \xi}'),{\pmb u}^{(m)}({\pmb \xi})]
-
[{\pmb u}^{(n)}({\pmb \xi}'),{\pmb w}^{(m)}({\pmb \xi})] \nonumber \\
& & =
[{\pmb u}^{(m)}({\pmb \xi}),{\pmb w}^{(n)}({\pmb \xi}')] 
-
[{\pmb u}^{(n)}({\pmb \xi}'),{\pmb w}^{(m)}({\pmb \xi})]
\label{eqA3}
\end{eqnarray}
Under the hypothesis of reciprocal boundary conditions,
from the identity  eq. (\ref{eq30}) it follows that
\begin{equation}
[{\pmb u}^{(m)}({\pmb \xi}),{\pmb w}^{(n)}({\pmb \xi}')] 
-[{\pmb u}^{(n)}({\pmb \xi}'),{\pmb w}^{(m)}({\pmb \xi})]
=
[{\pmb v}^{(m)}({\pmb \xi}),{\pmb v}^{(n)}({\pmb \xi}')] = 0
\label{eqA4}
\end{equation}
i.e.,
\begin{equation}
A_{\alpha {\pmb \alpha}_m} M^{(n)}_{\alpha {\pmb \alpha}_m}({\pmb \xi},{\pmb \xi}')
=
A_{\beta' {\pmb \beta}'_n} M^{(m)}_{\beta' {\pmb \beta}'_n}({\pmb \xi}',{\pmb \xi})
\label{eqA5}
\end{equation}
Therefore, from  the definition of geometrical moments  eq. (\ref{eq19}), since
$A_{\beta' {\pmb \beta}_n'}$ could be in principle arbitrary, we have
\begin{equation}
m_{\alpha {\pmb \alpha}_m \beta' {\pmb \beta}'_n}({\pmb \xi},{\pmb \xi}')=m_{\beta' {\pmb \beta}'_n \alpha {\pmb \alpha}_m }({\pmb \xi}',{\pmb \xi})
\label{eqA6}
\end{equation}
corresponding to eq. (\ref{eq40}).
If the boundary conditions are not reciprocal,  the r.h.s. of
eq. (\ref{eqA3}) possesses the property 
\begin{equation}
[{\pmb u}^{(m)}({\pmb \xi}),{\pmb w}^{(n)}({\pmb \xi}')] 
-[{\pmb u}^{(n)}({\pmb \xi}'),{\pmb w}^{(m)}({\pmb \xi})]
\; \, \begin{cases}
=0, \quad n=m
\\
\neq 0, \quad n \neq m
\end{cases}
\label{eqA7}
\end{equation}
For $n=m=0$ 
eq. (\ref{eqA7}) expresses the thermodynamic condition of symmetry of the resistance matrix, 
$
m_{\alpha \beta}=m_{\beta  \alpha  }
$, independently on the boundary conditions and on the nature of the immersed body \cite{dau6}.
This result represents a purely mechanical proof of the symmetry of the resistance matrix
independently of the boundary conditions. The thermodynamic proof has been given by Landau
\cite{dau5,dau6}, while the mechanical proof by Brenner uses specific (no-slip) boundary conditions  see \cite[p. 166]{hb}
and the discussion therein.

As seen in Section \ref{sec3}, a consequence of the
symmetry of the geometrical moments, in the case the Hinch-Kim dualism holds, is the equivalence between eq. (\ref{eq8}) and
eq. (\ref{eq41}). This equivalence can be proved 
by substituting  the Hinch-Kim theorem expressed by eq. (\ref{eq38}) in eq. (\ref{eq8}), hence
\begin{equation}
w_a({\pmb x})=\dfrac{1}{8\pi \mu}\sum_{m=0}^{\infty} 
\dfrac{M_{\alpha\, {\pmb \alpha}_m}({\pmb \xi})}{m!} 
\nabla_{{\pmb \alpha}_m}
S_{a\, \alpha}({\pmb x},{\pmb \xi})
=\sum_{m=0}^{\infty} 
\dfrac{\mathcal{F}_{\beta' \alpha {\pmb \alpha}_m}u_{\beta'}({\pmb \xi}') }{m!} 
\nabla_{{\pmb \alpha}_m}
S_{a\, \alpha}({\pmb x},{\pmb \xi})
\label{eqA8}
\end{equation}
substituting the expression  eq. (\ref{eq21}) for the  Faxén  operator 
and using the symmetry eq. (\ref{eqA6})
\begin{eqnarray}
\nonumber
w_a({\pmb x})
&&=\sum_{m=0}^{\infty}
 \sum_{n=0}^{\infty}
\dfrac{
m_{\beta' {\pmb \beta}'_n
\alpha {\pmb \alpha}_m 
}({\pmb \xi}',{\pmb \xi})\nabla_{{\pmb \beta'}_n
}
u_{\beta'}({\pmb \xi}') }{m!n!} 
\nabla_{{\pmb \alpha}_m}
S_{a\, \alpha}({\pmb x},{\pmb \xi})
\\
&&
=
\sum_{m=0}^{\infty}
 \sum_{n=0}^{\infty}
\dfrac{
m_{\alpha {\pmb \alpha}_m 
\beta' {\pmb \beta}'_n
}({\pmb \xi},{\pmb \xi}')\nabla_{{\pmb \beta'}_n
}
u_{\beta'}({\pmb \xi}') }{m!n!} 
\nabla_{{\pmb \alpha}_m}
S_{a\, \alpha}({\pmb x},{\pmb \xi})
\label{eqA9}
\end{eqnarray}
 in which a new representation of the  Faxén operator acting on the point ${\pmb \xi}'$ is introduced
\begin{equation}
\nonumber
\mathcal{F}_{\alpha \beta' {\pmb \beta}'_n}=
\sum_{m=0}^{\infty}
\dfrac{
m_{\alpha {\pmb \alpha}_m 
\beta' {\pmb \beta}'_n
}({\pmb \xi},{\pmb \xi}')
\nabla_{{\pmb \alpha}_m}}{m!}
\end{equation}
from which it follows that
\begin{equation}
w_a({\pmb x})=
 \sum_{n=0}^{\infty}
 \dfrac{
 \nabla_{{\pmb \beta}_n'
}
u_{\beta'}({\pmb \xi}') 
 }{n!}
 \mathcal{F}_{\alpha \beta' {\pmb \beta}_n'}
 S_{a\, \alpha}({\pmb x},{\pmb \xi})
 \label{eqA10}
\end{equation}
that becomes eq. (\ref{eq41}) for ${\pmb \xi}'={\pmb \xi}$.

\section{Geometrical moments for a sphere with Navier-slip  conditions}
\label{appB}
In this Appendix, we provide the analytical expression for the Cartesian  entries
$m_{\alpha {\pmb \alpha}_m \beta {\pmb \beta}_n}({\pmb \xi},{\pmb \xi})$
 of the geometrical moments, 
 for a sphere with Navier-slip boundary conditions, 
 {${\pmb \xi}$ being the center of the sphere}.

\subsubsection{ $(m,n)$-th order geometrical moments  with $n=0$}

In order to evaluate the geometrical moment with $n=0$,  eqs. (\ref{eq51})-(\ref{eq52})  can be applied 
using for the geometrical surface traction  eq. (\ref{eq65}).

In the case $m=0$ we obtain the well-known
Basset  term \cite{basset} 
\begin{equation}
m_{\alpha \beta}=-\dfrac{3}{16 \pi R_p}\int_{r=R_p}
\left[ 
\dfrac{\delta_{\alpha \beta} +6 \hat{\lambda} n_{\alpha \beta}(\pmb x)}{1+3\hat{\lambda}}
\right]
 dS(\pmb x)=-\dfrac{3}{4} 
 \left[ 
\dfrac{1 +2 \hat{\lambda}}{1+3\hat{\lambda}}
\right]R_p \, \delta_{\alpha \beta}
\label{eqB1}
\end{equation}
For $m=1$  the geometrical moments at the center of the sphere ${\pmb \xi}=(0,0,0)$ vanish, in fact
\begin{eqnarray}
\nonumber
&& m_{\alpha { \alpha}_1 \beta}
  = 
  m_{\beta \alpha { \alpha}_1 }
  = 
\\ 
\nonumber
 && 
 -\dfrac{3}{16 \pi }\int_{r=R_p}
 \left[ 
\dfrac{\delta_{\gamma \beta} +6 \hat{\lambda} n_{\gamma \beta}(\pmb x)}{1+3\hat{\lambda}}
\right]
 \left[
\delta_{\alpha \gamma} n_{{ \alpha}_1}({\pmb x}) -
\hat{\lambda }( 
t_{\alpha \gamma}({\pmb x}) n_{\alpha_1}({\pmb x})
+
 n_\alpha({\pmb x}) t_{\alpha_1 \gamma}({\pmb x}) )
\right]
 dS({\pmb x})=0
 \\
\label{eqB2}
\end{eqnarray}
and, due to the symmetry of the sphere,
\begin{equation}
 m_{\alpha {\pmb \alpha}_m \beta}=0,
 \qquad m=1,3,5\, ...
\label{eqB3}
\end{equation}
for any odd value of $m$.
For $m=2$
\begin{eqnarray}
\nonumber
&& m_{\alpha {\pmb \alpha}_2 \beta }
  =  
m_{ \beta \alpha {\pmb \alpha}_2 }=
  -\dfrac{3 R_p}{16 \pi}\int_{r=R_p}
 \left[ 
\dfrac{\delta_{\gamma \beta} +6 \hat{\lambda} n_{\gamma \beta}(\pmb x)}{1+3\hat{\lambda}}
\right] \times  
\\
\nonumber
& \times & \bigg[
\delta_{\alpha \gamma} n_{\alpha_1 \alpha_2}({\pmb x}) 
-
\hat{\lambda} (2 t_{\alpha \gamma} n_{\alpha_1 \alpha_2}(\pmb x) 
+ t_{\alpha_1 \gamma} n_{\alpha \alpha_2}(\pmb x) 
+
t_{\alpha_2 \gamma} n_{\alpha \alpha_1}(\pmb x) 
 )
\bigg]
dS({\pmb x})= 
 \\
&&
-\dfrac{ R_p^3}{4(1+3 \hat{\lambda})}
\bigg[
\delta_{\alpha \beta} \delta_{\alpha_1 \alpha_2}
+
\hat{\lambda}(\delta_{\alpha \alpha_1}\delta_{\beta \alpha_2}
+
\delta_{\alpha\alpha_2} \delta_{\beta \alpha_1} 
)
\bigg]
\label{eqB4}
\end{eqnarray}

\subsubsection{(m,n)-th order geometrical moments with $n=1$}

For the geometrical moments with  $n=1$,  eqs. (\ref{eq51})-(\ref{eq52})  can be used 
with the geometrical surface traction expressed by eq.  (\ref{eq101}).
By the symmetry expressed by eq.  (\ref{eqB2}), the moment with $m=0$ vanishes.
For $m=1$,
\begin{eqnarray}
\nonumber
&& 
m_{\alpha {\alpha}_1 \beta \beta_1}
  =  
m_{ \beta \beta_1 \alpha {\alpha}_1}
=
-R_p
  \int_{r=R_p}
 \left[ 
\dfrac{
(4+15\hat{\lambda } )\, \delta_{\gamma \beta} n_{\beta_1}({\pmb x})
+\,  \delta_{\gamma \beta_1} n_\beta({\pmb x})
+40\hat{\lambda }  (1+3 \hat{\lambda } )\, n_{\gamma \beta \beta_1}({\pmb x})
}{ 8 \pi (1+5 \hat{\lambda } )(1+3\hat{\lambda } )}
\right] 
\\
\nonumber
&\times & \bigg[
\delta_{\alpha \gamma} n_{\alpha_1}({\pmb x}) 
-
\hat{\lambda} (
t_{\alpha \gamma}({\pmb x})n_{\alpha_1}({\pmb x})
+t_{\alpha_1 \gamma}({\pmb x})n_{\alpha}({\pmb x})
 )
\bigg]
dS({\pmb x})= 
 \\
 \nonumber
 &&
 -\dfrac{R_p^3}{6 (1+5 \hat{\lambda } ) (1+3 \hat{\lambda})}
\bigg[
\left( 4+20 \hat{\lambda } +15 \hat{\lambda }  ^2\right) \delta _{\alpha  \beta } \delta _{\alpha_1 \beta_1}
+
\left(1+5 \hat{\lambda }+15 \hat{\lambda} ^2\right) 
\delta _{\alpha \beta_1} \delta _{\alpha_1 \beta }
+10 \hat{\lambda }   ( 1+3 \hat{\lambda } ) \delta _{\alpha  \alpha_1} \delta _{\beta \beta_1}
\bigg]\\
\label{eqB5}
\end{eqnarray}
Due to symmetry of the sphere, for odd $n$ and even $m$  the geometrical moments vanish, therefore 
 $m_{\alpha {\pmb \alpha}_2 \beta \beta_1} = m_{\beta \beta_1 \alpha {\pmb \alpha}_2}= 0$.
 For $m=3$,
\begin{eqnarray}
\nonumber
&& 
m_{\alpha { \pmb \alpha}_3  \beta \beta_1}
  =  
m_{ \beta \beta_1 \alpha {\pmb \alpha}_3}=
\\
\nonumber
&-&R_p^3
  \int_{r=R_p}
 \left[ 
\dfrac{
(4+15\hat{\lambda } )\, \delta_{\gamma \beta} n_{\beta_1}({\pmb x})
+\,  \delta_{\gamma \beta_1} n_\beta({\pmb x})
+40\hat{\lambda }  (1+3 \hat{\lambda } )\, n_{\gamma \beta \beta_1}({\pmb x})
}{ 8 \pi (1+5 \hat{\lambda } )(1+3\hat{\lambda } )}
\right] 
 \times  
\\
\nonumber
&& \bigg[
\delta_{\alpha \gamma} n_{\alpha_1 \alpha_2 \alpha_3}({\pmb x}) 
-
\hat{\lambda} (
3 t_{\alpha \gamma}({\pmb x})n_{\alpha_1 \alpha_2 \alpha_3}({\pmb x})
+t_{\alpha_1 \gamma}({\pmb x})n_{\alpha \alpha_2 \alpha_3}({\pmb x})
+t_{\alpha_2 \gamma}({\pmb x})n_{\alpha_1 \alpha \alpha_3}({\pmb x})
+t_{\alpha_3 \gamma}({\pmb x})n_{\alpha_1 \alpha_2 \alpha}({\pmb x})
 )
\bigg]
dS({\pmb x})
 \\
 \nonumber
 &=&
 -\dfrac{ R_p^5}{30(1+5 \hat{\lambda } ) (1+3 \hat{\lambda})}
\bigg[
\left( 4+12\hat{\lambda }-15 \hat{\lambda }  ^2\right) \delta _{\alpha  \beta } \eta_{ \beta_1 {\pmb \alpha}_3}
+
\left(1+3\hat{\lambda }+15 \hat{\lambda} ^2\right) 
\delta _{\alpha  \beta_1} \eta _{ \beta {\pmb \alpha}_3}
\\
&&
+5 \hat{\lambda }   ( 1+3 \hat{\lambda } ) 
(\delta _{\alpha \alpha_1} \eta _{\beta \beta_1  \alpha_2 \alpha_3}+
\delta _{\alpha \alpha_2} \eta _{\beta \beta_1 \alpha_1 \alpha_3}+
\delta _{\alpha \alpha_3} \eta _{\beta \beta_1  \alpha_1 \alpha_2}
)
\bigg]
\label{eqB6}
\end{eqnarray}
where
$\eta_{\alpha \beta \gamma \delta}=
\delta_{\alpha \beta} \delta_{\gamma \delta}+
\delta_{\alpha \delta} \delta_{\beta\gamma }+
\delta_{\alpha \gamma} \delta_{\beta \delta}
$.

\subsubsection{(m,n)-th order geometrical moments with n=2}

It is possible to obtain the geometrical moments
 for $n=2$ and $m=0$ by symmetry from eq. (\ref{eqB4}),
thus
\begin{equation}
m_{\alpha  \beta {\pmb \beta}_2}=
-
\dfrac{R_p^3}{4(1+3 \hat{\lambda})}
\bigg[
\delta_{\alpha \beta} \delta_{\beta_1 \beta_2}
+
\hat{\lambda}(\delta_{\beta \beta_1}\delta_{\alpha \beta_2}
+
\delta_{\beta\beta_2} \delta_{\alpha \beta_1} 
)
\bigg]
\label{eqB7}
\end{equation}

For $m=2$
\begin{eqnarray}
\nonumber
&& m_{\alpha {\pmb \alpha}_2 \beta {\pmb \beta}_2}
  =  
m_{ \beta {\pmb \beta}_2 \alpha {\pmb \alpha}_2 }=
\\
\nonumber
&&
R_p^2
\int_{r=R_p}
\dfrac{h_{\alpha \beta {\pmb \beta}_2}({\pmb x})}{8\pi}
 \bigg[
\delta_{\alpha \gamma} n_{\alpha_1 \alpha_2}({\pmb x}) 
-
\hat{\lambda} (2 t_{\alpha \gamma} n_{\alpha_1 \alpha_2}(\pmb x) 
+ t_{\alpha_1 \gamma} n_{\alpha \alpha_2}(\pmb x) 
+
t_{\alpha_2 \gamma} n_{\alpha \alpha_1}(\pmb x) 
 )
\bigg]
dS({\pmb x})= 
 \\
 \nonumber
&&
-\dfrac{R_p^5}{
24(1+4 \hat{\lambda})(1+ 7 \hat{\lambda})
}
\left\{
\delta_{\alpha \beta}
\left[-8  \hat{\lambda}^2
\left(
\dfrac{
4+21  \hat{\lambda}
}{1+3  \hat{\lambda}} \right)
 \delta_{\alpha_1 \alpha_2} \delta_{\beta_1 \beta_2}
 +5(1+6 \hat{\lambda})(
  \delta_{\alpha_1 \beta_2} \delta_{\beta_1 \alpha_2}+
   \delta_{\alpha_1 \beta_1 } \delta_{\alpha_2 \beta_2}
 )
 \right]
 \right.
 \\
 \nonumber
&&
+
(1+6 \hat{\lambda}+28 \hat{\lambda}^2)
(
\delta_{\alpha \beta_1 }
(\delta_{\alpha_1 \beta_2}\delta_{ \alpha_2 \beta}
+
\delta_{\alpha_1 \beta}\delta_{ \alpha_2 \beta_2}
)
+
\delta_{\alpha \beta_2}
(\delta_{\alpha_1 \beta_1 }\delta_{ \alpha_2 \beta}
+
\delta_{\alpha_1 \beta}\delta_{\alpha_2 \beta_1}
)
)
\\
\nonumber
&&
+
(1+12  \hat{\lambda}+56  \hat{\lambda}^2)
(\delta_{ \alpha \beta_1 }\delta_{ \beta \beta_2 }
+
\delta_{ \alpha \beta_2 }\delta_{ \beta \beta_1 })\delta_{ \alpha_1 \alpha_2 }
+(\text{terms giving vanishing contribution to}\,  \mathcal{F}_{\alpha \beta {\pmb \beta}_2})\\
\label{eqB8}
\end{eqnarray}
For $m=4$
\begin{eqnarray}
\nonumber
&& m_{\alpha {\pmb \alpha}_4 \beta {\pmb \beta}_2}
  =  
m_{ \beta {\pmb \beta}_2 \alpha {\pmb \alpha}_4 }=
\\
\nonumber
&&
R_p^4 \int_{r=R_p}
\dfrac{h_{\alpha \beta {\pmb \beta}_2}({\pmb x})}{8\pi}
 \bigg[
\delta_{\alpha \gamma} n_{{\pmb \alpha}_4}({\pmb x}) 
-
\hat{\lambda} (4 t_{\alpha \gamma} n_{{\pmb \alpha}_4}(\pmb x) 
+ t_{\alpha_1 \gamma} n_{\alpha \alpha_2 \alpha_3 \alpha_4}(\pmb x) 
+
t_{\alpha_2 \gamma} n_{\alpha \alpha_1 \alpha_3 \alpha_4}(\pmb x)  
  \\
 \nonumber
 &&+ t_{\alpha_3 \gamma} n_{\alpha \alpha_1 \alpha_2 \alpha_4}(\pmb x) 
+
t_{\alpha_4 \gamma} n_{\alpha \alpha_1 \alpha_2 \alpha_3}(\pmb x) 
 ) 
\bigg]
dS({\pmb x})= 
 \\
 \nonumber
&&
-\dfrac{R_p^7}{
168(1+4 \hat{\lambda})(1+ 7 \hat{\lambda})
}
\bigg[
(5+20 \hat{\lambda}-56 \hat{\lambda}^2)
\delta_{\alpha \beta} \mathrm{H}_{{\pmb \beta}_2 {\pmb \alpha}_4}
+(1+4 \hat{\lambda}+28 \hat{\lambda}^2)
(\delta_{\alpha \beta_1} \mathrm{H}_{\beta \beta_2 {\pmb \alpha}_4}+
\delta_{\alpha \beta_2} \mathrm{H}_{\beta \beta_1 {\pmb \alpha}_4}
)\bigg]
\\
&&
+
(\text{terms giving vanishing contribution to}\, \mathcal{F}_{\alpha \beta {\pmb \beta}_2})
\label{eqB9}
\end{eqnarray}
where
\begin{equation}
\nonumber
\mathrm{H}_{{\pmb \beta}_2 {\pmb \alpha}_4}=
\mathrm{H}_{\beta_1 \beta_2 \alpha_1 \alpha_2 \alpha_3 \alpha_4}=
\delta_{\beta_1 \alpha_1} 
\eta_{\beta_2 \alpha_2 \alpha_3 \alpha_4}+
\delta_{\beta_1 \alpha_2} 
\eta_{\beta_2 \alpha_1 \alpha_3 \alpha_4}+
\delta_{\beta_1 \alpha_3} 
\eta_{\beta_2 \alpha_1 \alpha_2 \alpha_4}+
\delta_{\beta_1 \alpha_4} 
\eta_{\beta_2 \alpha_1 \alpha_2 \alpha_3}
\end{equation}
{

\section{Extended grand-resistance matrix}
\label{appC}
As can be observed by the definition eq. (\ref{eq19}), geometrical moments are strictly related to the entries of the grand-resistance
matrix \cite{brenner64a,brenner64b,brenner64c,kim-karrila},
 which provides the linear relation between the hydrodynamic
forces ${\pmb F}$, torques ${\pmb T}$ and Stresslets $\mathsf{\pmb S}$ exerted by the fluid onto a body   
 and the 
translation velocity ${\pmb U}$, the rotation angular velocity ${\pmb \Omega}$ and strain intensity ${\pmb E}$ of the body, according to the relation
\begin{equation}
\left (
\begin{array}{ccccc}
{\pmb F} \\
{\pmb T} \\
\mathsf{\pmb S}
\end{array}
\right )
=-
\left (
\begin{array}{ccccc}
{\pmb R}^{(F U)} &  & {\pmb R}^{(F \Omega)} &  & {\pmb R}^{(F E)} \\
{\pmb R}^{(T U)} & & {\pmb R}^{(T \Omega)} &  & {\pmb R}^{(T E)} \\
{\pmb R}^{(S U)} & & {\pmb R}^{(S \Omega)} &  & {\pmb R}^{(S E)}
\end{array}
\right ) 
\left (
\begin{array}{ccccc}
{\pmb U} \\
{\pmb \Omega} \\
{\pmb E}
\end{array}
\right )
\label{eqC1}
\end{equation}
where
${\pmb R}^{(F U)}  $, $ {\pmb R}^{(F \Omega)}$, ${\pmb R}^{(T U)} $ and ${\pmb R}^{(T \Omega)} $ are constant tensors with rank $2$,
$ {\pmb R}^{(F E)}$, $ {\pmb R}^{(T E)}$, $ {\pmb R}^{(S U)}$ and $ {\pmb R}^{(S \Omega)}$ are constant tensors with rank $3$
and  ${\pmb R}^{(S E)} $ is a constant tensor with rank $4$. Componentwise, we have
\begin{eqnarray}
\nonumber
F_\alpha &=& 
-
 {R}^{(F U)}_{\alpha \beta} U_\beta
-
 {R}^{(F \Omega)}_{\alpha \beta} \Omega_\beta
-
 {R}^{(F E)}_{\alpha \beta \beta_1} E_{\beta \beta_1}
\\
T_\alpha &=&
-
 {R}^{(T U)}_{\alpha \beta} U_\beta
-
 {R}^{(T \Omega)}_{\alpha \beta} \Omega_\beta
-
 {R}^{(T E)}_{\alpha \beta \beta_1} E_{\beta \beta_1}
\\
\nonumber
\mathsf{S}_{\alpha \alpha_1} &=&
-
 {R}^{(S U)}_{\alpha \alpha_1 \beta} U_\beta
-
 {R}^{(S \Omega)}_{\alpha \alpha_1 \beta} \Omega_\beta
-
 {R}^{(S E)}_{\alpha \alpha_1 \beta \beta_1} E_{\beta \beta_1}
\label{eqC2}
\end{eqnarray}
Specifically, the hydrodynamic force exerted by the fluid onto the body
is given by
\begin{equation}
F_\alpha= -\int_{\partial D_b}{ \sigma}_{\alpha \beta}({\pmb x})\cdot {n}_\beta({\pmb x})\,  dS
\end{equation}
the torque reads
\begin{equation}
T_\alpha= -
\varepsilon_{\alpha \gamma_1 \gamma}
\int_{\partial D_b}({\pmb x}-{\pmb \xi})_{\gamma_1}\, { \sigma}_{\gamma \beta}({\pmb x})\cdot {n}_\beta({\pmb x})\,  dS
\label{eqC4}
\end{equation}
and the so called {Stresslet}, giving the contribution of the single body to the effective stress of the bulk fluid, is defined as \cite{batchelor-green_b} 
\begin{eqnarray}
\nonumber
&& \mathsf{S}_{\alpha \alpha_1} =-  \dfrac{1}{2}\int_{\partial D_b} 
\left(
  ({\pmb x}-{\pmb \xi})_{ \alpha_1} \sigma_{\alpha \beta}({\pmb x})n_\beta({\pmb x})
  +
   ({\pmb x}-{\pmb \xi})_{ \alpha} \sigma_{\alpha_1 \beta}({\pmb x})n_\beta({\pmb x})  
\right)dS({\pmb x})
\\
[10pt]
&&
+  \dfrac{1}{3}  \int_{\partial D_b}  ({\pmb x}-{\pmb \xi})_{ \alpha} \sigma_{\alpha \beta}({\pmb x})n_\beta({\pmb x})  dS({\pmb x})
 -
 \int_{\partial D_b}  
\left(
\mu 
v_\alpha ({\pmb x}) n_{\alpha_1}({\pmb x})
+
 n_\alpha({\pmb x}) v_{\alpha_1}({\pmb x})
 \right)
 dS({\pmb x})
\label{eqC5}
\end{eqnarray}
On the other hand, the $0$-th order moment defined eq. (\ref{eq7}) 
\begin{equation}
M_{\alpha }=\int_{{ D_b}} 
\psi_{\alpha}({\pmb x}) 
 dV({\pmb x})=
 \int_{\partial D_b}{ \sigma}_{\alpha \beta}({\pmb x})\cdot {n}_\beta({\pmb x})\,  dS
\end{equation}
represents the force exerted by the body onto the fluid, hence $M_\alpha=-F_\alpha$. By comparing the relation providing the force onto a translating sphere $F_\alpha = -R_{\alpha \beta}^{(F U)}\, U_\beta$
with
the relation
$ M_{\alpha }^{(0)}=8 \pi \mu\, m_{\alpha \beta}\, A_\beta$
 yielding the $0$-th order moment of the body in a constant ambient flow, with $A_a=-U_a$, it easy to obtain 
\begin{equation}
R_{\alpha \beta}^{(F U)}=- 8 \pi \mu\, m_{\alpha \beta}
\end{equation}
 Since $R_{\alpha \beta}^{(F U)}$ is a positive semi-definite matrix \cite{hb},
$m_{\alpha \beta}$ necessarily is a negative semi-definite matrix.

The torque exerted by the body onto the fluid can be obtained by applying the Levi-Civita symbol to the $1$-st order moment, thus
\begin{equation}
\varepsilon_{\alpha \beta_1 \beta}
M_{\beta \beta_1}({\pmb \xi})=
\varepsilon_{\alpha \beta_1 \beta}
\int_{{ D_b}} 
({\pmb x}-{\pmb \xi})_{\beta_1}
\psi_{\beta}({\pmb x}) 
 dV({\pmb x})=
 \varepsilon_{\alpha \beta_1 \beta}
 \int_{\partial D_b}
 ({\pmb x}-{\pmb \xi})_{\beta_1}
 { \sigma}_{\beta \gamma}({\pmb x})\cdot {n}_\gamma({\pmb x})\,  dS
 \label{eqC7}
\end{equation}
By eqs. (\ref{eqC7}) and (\ref{eqC4}), the relation between the 
torque acting on the body and the $1$-st order volume moment is $T_\alpha= \varepsilon_{\alpha \beta \beta_1}
M_{\beta \beta_1}({\pmb \xi})$.

By comparing the relation giving the torque on the rotating body
$T_\alpha=- {R}_{\alpha \beta}^{(T \Omega)} \Omega_\beta$
with the relation
$\varepsilon_{\alpha \beta \beta_1}
M_{\beta \beta_1}^{(1)}({\pmb \xi})= 8 \pi \mu \varepsilon_{\alpha \gamma \gamma_1} m_{\gamma \gamma_1 \delta \delta_1}({\pmb \xi,{\pmb \xi}})A_{\delta \delta_1}$  yielding the momentum
of a body immersed in a 
rotating flow with angular velocity $-\Omega_\beta$, hence with $A_{\delta \delta_1}=-\varepsilon_{\beta \delta \delta_1}\Omega_\beta$, we obtain
\begin{equation}
{R}_{\alpha \beta}^{(T \Omega)}=
- 8 \pi \mu \,\varepsilon_{\alpha \gamma \gamma_1} \varepsilon_{\beta \delta \delta_1} m_{\gamma \gamma_1 \delta \delta_1}({\pmb \xi},{\pmb \xi})
\end{equation}

Let us consider the off-diagonal elements of the symmetric 
part of $1$-st order volume moment $\gamma_{\alpha \alpha_1 \beta \beta_1}M_{\beta \beta_1}({\pmb \xi})$, where
\begin{equation}
\nonumber
\gamma_{\alpha \alpha_1 \beta \beta_1}=\dfrac{
\delta_{\alpha \beta} \delta_{\alpha_1 \beta_1}+
\delta_{\alpha \beta_1} \delta_{\alpha_1 \beta}}{2}-
\dfrac{\delta_{\alpha \alpha_1} \delta_{\beta \beta_1}}{3}
\end{equation}
By eq. (\ref{eq18}), we obtain
\begin{eqnarray}
\nonumber
&& \gamma_{\alpha \alpha_1 \beta \beta_1}M_{\beta \beta_1}({\pmb \xi}) = \dfrac{1}{2}\int_{\partial D_b} 
\left(
  ({\pmb x}-{\pmb \xi})_{ \alpha_1} \sigma_{\alpha \beta}({\pmb x})n_\beta({\pmb x})
  +
   ({\pmb x}-{\pmb \xi})_{ \alpha} \sigma_{\alpha_1 \beta}({\pmb x})n_\beta({\pmb x})  
\right)dS({\pmb x})
\\
[10pt]
&&
- \dfrac{1}{3}  \int_{\partial D_b}  ({\pmb x}-{\pmb \xi})_{ \alpha} \sigma_{\alpha \beta}({\pmb x})n_\beta({\pmb x})  dS({\pmb x})
 +
 \int_{\partial D_b}  
\left(
\mu 
v_\alpha ({\pmb x}) n_{\alpha_1}({\pmb x})
+
 n_\alpha({\pmb x}) v_{\alpha_1}({\pmb x})
 \right)
 dS({\pmb x})
\nonumber 
 \\
\label{eqC10}
\end{eqnarray}
Eq. (\ref{eqC10}) expresses the same integrals in eq. (\ref{eqC4}) with reversed sign, hence
we have $\mathsf{S}_{\alpha \alpha_1}= - \gamma_{\alpha \alpha_1 \beta \beta_1} M_{\beta \beta_1}({\pmb \xi})$.
Comparing the relation $\mathsf{S}_{\alpha \alpha_1}=-
 {R}^{(S E)}_{\alpha \beta \gamma \delta}\, E_{\gamma \delta}$ with
the relation $ \gamma_{\alpha \alpha_1 \beta \beta_1} M_{\beta \beta_1}({\pmb \xi})= 8 \pi \mu
\gamma_{\alpha \alpha_1 \beta \beta_1}
 m_{\beta \beta_1 \delta \delta_1}({\pmb \xi},{\pmb \xi}) A_{\delta \delta_1}$ giving the off-diagonal entries of the symmetric part of
 $ M_{\beta \beta_1}({\pmb \xi})$
for a body in a linear ambient flow  
$A_{\delta \delta_1}=-E_{\delta \delta_1} $, we obtain
\begin{equation}
{R}^{(S E)}_{\alpha \alpha_1 \beta \beta_1}=
-
8 \pi \mu\,
\gamma_{\alpha \alpha_1 \delta \delta_1}
 m_{\delta \delta_1 \beta \beta_1}({\pmb \xi},{\pmb \xi})
\label{eqC11}
\end{equation}
Since $E_{\delta \delta}=0$, by incompressibility, it is also possible to express eq. (\ref{eqC11}) in the more symmetric form
\begin{equation}
{R}^{(S E)}_{\alpha \alpha_1 \beta \beta_1}=
-
8 \pi \mu\,
\gamma_{\alpha \alpha_1 \delta \delta_1}
\gamma_{ \beta \beta_1 \gamma \gamma_1}
 m_{\delta \delta_1 \gamma \gamma_1}({\pmb \xi},{\pmb \xi})
\label{eqC11}
\end{equation}

In order to obtain the expression for the coupling tensor ${R}_{\alpha \beta}^{(F \Omega)}$, it is possible to compare the expression  $F_\alpha= -{R}_{\alpha \beta}^{(F \Omega)} \Omega_\beta$, giving the force exerted by the  fluid onto  a rotating body with velocity $\Omega_\beta$, with the expression   $ M_{\alpha }^{(0)}=8 \pi \mu\, m_{\alpha \beta \beta_1}\, A_{\beta \beta_1}$
giving the $0$-th order moment of the body in a rotating ambient flow
 with angular velocity $-\Omega_\beta$, hence with $A_{\delta \delta_1}=-\varepsilon_{\beta \delta \delta_1}\Omega_\beta$, to obtain
\begin{equation}
R_{\alpha \beta}^{(F\,\Omega)}=  8 \pi \mu\, \varepsilon_{\beta \gamma \gamma_1} m_{\alpha \gamma \gamma_1}({\pmb \xi},{\pmb \xi}) 
\label{eqC12}
\end{equation} 
and analogously, by iterating the same procedure,
\begin{equation}
R_{\alpha \beta}^{(T\,U)}=
 8 \pi \mu\, \varepsilon_{\alpha \gamma \gamma_1} m_{ \gamma \gamma_1 \beta}({\pmb \xi},{\pmb \xi}) 
\label{eqC13}
\end{equation} 
Since $ m_{\alpha \gamma \gamma_1}({\pmb \xi},{\pmb \xi}) = m_{ \gamma \gamma_1 \alpha}({\pmb \xi},{\pmb \xi}) $
in the case BC-reciprocity holds, the well known  symmetry $R_{\alpha \beta}^{(F\,\Omega)}= R_{\beta \alpha}^{(T\,U)}$
is straightforwardly obtained by eqs. (\ref{eqC12}) and (\ref{eqC13}) \cite{hb}.

The coupling tensor between force and strain is expressed by
\begin{equation}
R_{\alpha \beta \beta_1}^{(F\,E)}=
-
8 \pi \mu\, \gamma_{\beta \beta_1 \gamma \gamma_1} m_{\alpha \gamma \gamma_1}({\pmb \xi},{\pmb \xi}) 
\label{eqC14}
\end{equation}
while the coupling tensor between stresslet and translation reads
\begin{equation}
R_{\alpha \alpha_1 \beta}^{(S\,U)}=
-
8 \pi \mu\, \gamma_{\alpha \alpha_1 \delta \delta_1} m_{ \delta \delta_1 \beta}({\pmb \xi},{\pmb \xi}) 
\label{eqC15}
\end{equation}

Finally ,
the coupling tensor between torque and strain \cite{premlata2022} is given by
\begin{equation}
R_{\alpha \beta \beta_1}^{(T\,E)}=
 8 \pi \mu\, \varepsilon_{\alpha \gamma \gamma_1} 
 \gamma_{\beta \beta_1 \delta \delta_1}
 m_{ \gamma \gamma_1 \delta \delta_1}({\pmb \xi},{\pmb \xi}) 
\label{eqC16}
\end{equation}
and 
the coupling tensor between stresslet and rotation reads
\begin{equation}
R_{\alpha \alpha_1 \beta}^{(S\, \Omega)}=
 8 \pi \mu\, \varepsilon_{\beta \gamma \gamma_1} 
 \gamma_{\alpha \alpha_1 \delta \delta_1}
 m_{\delta \delta_1 \gamma \gamma_1 }({\pmb \xi},{\pmb \xi}) 
\label{eqC17}
\end{equation}
}


\begin{thebibliography}{50}

\bibitem{guazzelli-morris}
E. Guazzelli and J. F. Morris, \emph{A physical introduction to suspension dynamics.} (Cambridge University Press, New York, 2012).

\bibitem{maxey_rev}
M. Maxey, Simulation methods for particulate flows and concentrated suspensions, Annu. Rev. Fluid Mech., {\bf 49}, 171 (2017).

\bibitem{mewis}
J. Mewis and N. J. Wagner, \emph{Colloidal suspension rheology}, (Cambridge university press, 2012).

\bibitem{bedeaux-mazur}
D. Bedeaux and P. Mazur, Brownian motion and fluctuating hydrodynamics, Physica {\bf 76}, 247 (1974).

\bibitem{bian}
X. Bian, C. Kim and G. E. Karniadakis, 111 years of Brownian motion, Soft Matter, {\bf 12}, 6331 (2016).

\bibitem{raizen}
J. Mo and M. G. Raizen, Highly resolved Brownian motion in space and in time, Annu. Rev. Fluid Mech., {\bf 51}, 403 (2019).

\bibitem{pg_fluids}
G. Procopio and M. Giona, Stochastic Modeling of Particle Transport in Confined Geometries: Problems and Peculiarities, Fluids {\bf 7}, 105 (2022).

\bibitem{lauga_book}
E. Lauga, \emph{The fluid dynamics of cell motility} (Cambridge University Press, 2020).

\bibitem{vogel}
S. Vogel, \emph{Life in Moving Fluids: The Physical Biology of Flow}, (Princeton University Press, 1994).

\bibitem{freund}
J. B. Freund, Numerical simulation of flowing blood cells, Annu. Rev. Fluid Mech., {\bf 46}, 67 (2014).

\bibitem{venditti}
C. Venditti, S. Cerbelli, G. Procopio, and A. Adrover, Comparison between one-and two-way coupling approaches for estimating effective transport properties of suspended particles undergoing Brownian sieving hydrodynamic chromatography, Phys. Fluids, {\bf 34}, 042010 (2022).

\bibitem{undvall}
E. Undvall, F. Garofalo, G. Procopio, W. Qiu, A. Lenshof, T. Laurell, and T. Baasch, Inertia-Induced Breakdown of Acoustic Sorting Efficiency at High Flow Rates, Phys. Rev. Appl., {\bf 17}, 034014 (2022).

\bibitem{dicarlo}
D. Di Carlo, Inertial microfluidics, Lab Chip {\bf 9}, 3038,(2009).

\bibitem{oseen}
C. W. Oseen, \emph{Neuere methoden und ergebnisse in der hydrodynamik.}  (Akademische Verlagsgesellschaft mb H., Leipzig, 1927).

\bibitem{hb}
J. Happel and H. Brenner,
\emph{Low Reynolds number hydrodynamics: with special applications to particulate media},
(Martinus Nijhoff, The Hague (Ne), 1983).

\bibitem{pozri}
    C. Pozrikidis,
     \emph{Boundary integral and singularity methods for linearized viscous flow},
     (Cambridge University Press,Cambridge, 1992).

\bibitem{batchelor-green}
G. K. Batchelor and J. T. Green, The hydrodynamic interaction of two small freely-moving spheres in a linear flow field, J. Fluid Mech., {\bf 56}, 375 (1972).



\bibitem{hasimoto}
H. Hasimoto, An Extension of Faxén's Law to the Ellipsoid of Revolution, J. Phys. Soc. Jpn. {\bf 52}, 3294 (1983).


\bibitem{kim85}
S. Kim, A note on Faxén laws for nonspherical particles,  Int. J. Multiph. Flow, {\bf 11}, 713 (1985).


\bibitem{brenner64}
H. Brenner, The Stokes resistance of an arbitrary particle—IV arbitrary fields of flow, Chem. Eng. Sci. {\bf 19}, 703 (1964).





\bibitem{kim86}
S. Kim, Singularity solutions for ellipsoids in low-Reynolds-number flows: with applications to the calculation of hydrodynamic interactions in suspensions of ellipsoids, Int. J. Multiph. Flow, {\bf 12}, 469 (1986).


\bibitem{hestroni-haber70}
G. Hetsroni and S. Haber. The flow in and around a droplet or bubble submerged in an unbound arbitrary velocity field. Rheol. Acta {\bf 9}, 488 (1970).

\bibitem{rallison}
J. M. Rallison, Note on the Faxén relations for a particle in Stokes flow, J. Fluid Mech. {\bf 88}, 529 (1978).


\bibitem{premlata2021}
A. R. Premlata  and H. H. Wei, Coupled Faxén relations for non-uniform slip Janus spheres, Phys. Fluids {\bf 33}, 112003 (2021).


\bibitem{premlata2022}
{ A. R. Premlata  and H. H. Wei, Anisotropic stresslet and rheology of stick–slip Janus spheres, J. Fluid Mech. {\bf 945}, A1 (2022).
}



\bibitem{palaniappan}
D. Palaniappan, Arbitrary Stokes flow past a porous sphere, {Mech. Res. Commun.}, {\bf 20}, 309 (1993).
 
\bibitem{pad}
B. S. Padmavathi, and T. Amaranath, Stokes flow past a composite porous spherical shell with a solid core, Arch. Mech. {\bf 48}, 311 (1996).

\bibitem{felderhof78}
B. U. Felderhof and R. B. Jones, Faxén theorems for spherically symmetric polymers in solution, Physica A, {\bf 93}, 457 (1978).

\bibitem{mazur}
P. Mazur and D. Bedeaux, A generalization of Faxén's theorem to nonsteady motion of a sphere through an incompressible fluid in arbitrary flow, Physica {\bf 76}, 235 (1974).

\bibitem{maxey}
M. R. Maxey and J.J. Riley, Equation of motion for a small rigid sphere in a nonuniform flow, Phys. Fluids {\bf 26}, 883 (1983).

\bibitem{yang}
S-M. Yang, Motions of a sphere in a time-dependent stokes flow: A generalization of Faxén’s law, Korean J. Chem. Eng., {\bf 4}, 15 (1987).

\bibitem{badeaux}
Bedeaux, D., and P. Mazur. A generalization of Faxén's theorem to nonsteady motion of a sphere through a compressible fluid in arbitrary flow. Physica, { \bf 78}, 505 (1974).

\bibitem{kaneda}
Y. Kaneda, A generalization of Faxén's theorem to nonsteady motion of an almost spherical drop in an arbitrary flow of a compressible fluid, Physica A, { \bf 101}, 407 (1980).

\bibitem{felderhof}
B. U. Felderhof, Force density induced on a sphere in linear hydrodynamics: II. Moving sphere, mixed boundary conditions, Physica A, {\bf 84}, 569 (1976).

\bibitem{premlata2020}
A. R. Premlata and H. H. Wei, Atypical non-Basset particle dynamics due to hydrodynamic slip, {Phys.  Fluids} {\bf 32},097109 (2020).

\bibitem{jones79}
R. B. Jones, Faxén theorems for a spherically symmetric polymer in time dependent compressible flow, Physica A, {\bf 95}, 104 (1979).

\bibitem{beek}
P. Van Beek, A counterpart of Faxén's formula in potential flow, Int. J. Multiph. Flow., {\bf 11},873 (1985).

\bibitem{hinch}
 E. J. Hinch, An averaged-equation approach to particle interactions in a fluid suspension. J. Fluid Mech., {\bf 83}, 695 (1977).

\bibitem{kim-karrila}
    S. Kim and S. J. Karrila,
     \emph{Microhydrodynamics: principles and selected applications},
    (Dover Publications Inc., Mineola (NY), 2005).

\bibitem{batchelor-green_b}
G. K. Batchelor and J. T. Green, The determination of the bulk stress in a suspension of spherical particles to order $c^2$, J. Fluid Mech. {\bf 56}, 401 (1972).

\bibitem{bossis-brady}
G. Bossis and J. F. Brady, The rheology of Brownian suspensions, J. Chem. Phys. {\bf 91}, 1866 (1989).

\bibitem{mauri}
R. Mauri, A new application of the reciprocity relations to the study of fluid flows through fixed beds, J. Eng. Math. { \bf 33}, 103 (1998).

\bibitem{spagnolie-lauga}
S. E. Spagnolie and E. Lauga, Hydrodynamics of self-propulsion near a boundary: predictions and accuracy of far-field approximations, J. Fluid Mech. {\bf 700}, 105 (2012).

\bibitem{kuron}
M. Kuron, P. Stärk, C. Holm, and J. De Graaf, Hydrodynamic mobility reversal of squirmers near flat and curved surfaces, Soft Matter {\bf 15}, 5908 (2019).

\bibitem{dey}
R. Dey, C. M. Buness, B. V. Hokmabad, C. Jin, and C. C. Maass, Oscillatory rheotaxis of artificial swimmers in microchannels, Nat. Commun. {\bf 13}, 1 (2022).

\bibitem{brenner-gaydos}    
H. Brenner and L. J. Gaydos, The constrained Brownian movement of spherical particles in cylindrical pores of comparable radius: models of the diffusive and convective transport of solute molecules in membranes and porous media, J. Colloid Interface Sci., {\bf 58}, 312 (1977).   
    
\bibitem{swan}   
J. W. Swan and J. F. Brady, Particle motion between parallel walls: Hydrodynamics and simulation, Phys. Fluids, {\bf 22}, 103310 (2010).

\bibitem{brady-bossis}
J. F. Brady and G. Bossis, Stokesian dynamics, Annu. Rev. Fluid Mech. {\bf 20}, 111 (1988).

\bibitem{whitesides}
G. M. Whitesides, The origins and the future of microfluidics, Nature {\bf 442}, 368 (2006).

\bibitem{lauga_rev}
E. Lauga, M. P. Brenner and H. A. Stone,  Microfluidics: the no-slip boundary condition, arXiv preprint cond-mat/0501557 (2005).

\bibitem{nasouri}
B. Nasouri and G. J. Elfring, Higher-order force moments of active particles, { Phys. Rev. Fluids.} {\bf 25}, 044101 (2018).

\bibitem{pg_mine}
G. Procopio and M. Giona, Bitensorial formulation of the singularity method for Stokes flows,
{ Math. Eng.}, {\bf 5}, 1-34 (2023).

\bibitem{pg_arxiv}
G. Procopio and M. Giona,
On the theory of body motion in confined Stokesian fluids, arXiv preprint (2309.03527).



\bibitem{dolata-zia}
B. E. Dolata and R. N. Zia, Faxén formulas for particles of arbitrary shape and material composition, J. Fluid Mech., {\bf 910} (2021).

\bibitem{lady}
O. A. Ladyzhenskaya, 
\emph{The mathematical theory of viscous incompressible flow},
(Martino Publishing, Mansfield Centre (CT), 2014).

\bibitem{pg_meccanica}
{ M. Giona, G. Procopio and R. Mauri, Hydrodynamic Green functions: paradoxes in unsteady Stokes conditions and infinite propagation velocity in incompressible viscous models. Meccanica {\bf 57}, 1055 (2022).}

\bibitem{poisson}
E. Poisson, A. Pound and I. Vega, The motion of point particles in curved spacetime. Living Rev. Relativ., {\bf 14}, 1 (2011).

\bibitem{durl}
L. Durlofsky, J. F. Brady and G. Bossis, Dynamic simulation of hydrodynamically interacting particles, J. Fluid Mech., {\bf 180}, 21 (1987).

\bibitem{ichiki}
K. Ichiki, Improvement of the Stokesian dynamics method for systems with a finite number of particles, J. Fluid Mech., {\bf 452}, 231 (2002).

\bibitem{rosen}
K. H. Rosen, \emph{ Handbook of discrete and combinatorial mathematics.} (CRC press, 1999).

\bibitem{chwang-wu}
A. T. Chwang and T. Y. T. Wu, Hydromechanics of low-Reynolds-number flow. Part 2. Singularity method for Stokes flows, J. Fluid Mech., {\bf 67}, 787 (1975).

\bibitem{brenner64a}
{ H. Brenner, The Stokes resistance of an arbitrary particle—II: An extension, Chem. Eng. Sci. {\bf 18}, 1 (1963).
}

\bibitem{brenner64b}
{ H. Brenner, The Stokes resistance of an arbitrary particle—II: An extension, Chem. Eng. Sci. {\bf 19}, 599 (1964).
}

\bibitem{brenner64c}
{ H. Brenner, The Stokes resistance of an arbitrary particle—III: Shear fields, Chem. Eng. Sci. {\bf 19}, 631 (1964).
}

\bibitem{bazant}
M. Z. Bazant and O. I. Vinogradova, Tensorial hydrodynamic slip, J. Fluid Mech. {\bf 613}, 125 (2008).

\bibitem{dau7}
L. D. Landau and E. M. Lifshitz, \emph{Theory of elasticity}, Vol. 7, 2nd ed., Course of Theoretical Physics (Elsevier, Oxford, 2012).

\bibitem{galdi}
G. P. Galdi and R. Rannacher, \emph{Fundamental trends in fluid-structure interaction}, Vol.1, (World Scientific, Singapore, 2010).

\bibitem{maxwell}
J. C. Maxwell,  On the calculation of the equilibrium and stiffness of frames, Lond. Edinb. Dublin Philos. Mag. J. Sci., {\bf 27}, 294 (1864).

\bibitem{betti}
E. Betti, Teoria della elasticità, Il Nuovo Cimento, {\bf 7}, 69 (1872).


\bibitem{brenner_sur}
D. A. Edwards, H. Brenner, D. T. Wasan, and A. M. Kraynik, \emph{Interfacial Transport Processes and Rheology}, (Butterworth-Heinemann, Stonheam (MA), 1993).

\bibitem{rallison84}
J. M. Rallison, {The deformation of small viscous drops and bubbles in shear flows}, Annu. Rev. Fluid Mech. {\bf 16}, 45 (1984).

\bibitem{rallison-acrivos}
J. M. Rallison and A. Acrivos, { A numerical study of the deformation and burst of a viscous drop in an extensional flow}, J. Fluid Mech. {\bf 89}, 191 (1978).

\bibitem{power}
H. Power, On the Rallison and Acrivos solution for the deformation and burst of a viscous drop in an extensional flow, J. Fluid Mech., {\bf 185}, 547 (1987).

\bibitem{darcy}
H. Darcy, \emph{ Les fontaines publiques de la ville de Dijon: Exposition et application des principes à suivre et des formules à employer dans les questions de distribution d'eau}. (V. Dalmont, 1856).

\bibitem{whitaker}
S. Whitaker, Flow in porous media I: A theoretical derivation of Darcy's law, Transp. Porous Media {\bf 1}, 3 (1986).

\bibitem{saffman}
P. G. Saffman, { On the boundary condition at the surface of a porous medium}, Stud. Appl. Math., {\bf 50}, 93 (1971).

\bibitem{jones}
I. P. Jones, { Low Reynolds number flow past a porous spherical shell},Math. Proc. Camb. Philos.,{\bf 73}, 231 (1973).

\bibitem{brinkman}
H. C. Brinkman, { A calculation of the viscous force exerted by a flowing fluid on a dense swarm of particles.} Flow Turbul Combust., { \bf 1},27 (1949).

\bibitem{masliyah}
J. H. Masliyah, G. Neale, K. Malysa, and T. G. M. Van De Ven, { Creeping flow over a composite sphere: solid core with porous shell}, Chem. Eng. Sci. {\bf 42}, 245 (1987).

\bibitem{yu}
Q. Yu, and P. N. Kaloni, { A Cartesian-tensor solution of the Brinkman equation}, J. Eng. Math. {\bf 22}, 177 (1988).

\bibitem{suppl}
G. Procopio and M. Giona,
{Supplementary materials to this article}, Phys. Rev. Fluids (2023).

\bibitem{basset}
A. B. Basset, \emph{ A treatise on hydrodynamics: with numerous examples}, Vol. 2, (Bell and Company, Deighton, 1888).

\bibitem{dau5}
L. D. Landau and E. M. Lifshitz, \emph{Statistical Physics}, Vol. 5, 2nd ed., Course of Theoretical Physics (Elsevier, Oxford, 1987).

\bibitem{dau6}
L. D. Landau and E. M. Lifshitz, \emph{Fluid Mechanics}, Vol. 6, 2nd ed., Course of Theoretical Physics (Elsevier, Oxford, 1987).




\end{thebibliography}
\end{document}